\documentclass[12pt, a4paper]{article}
\usepackage[dvipdfmx]{graphicx}
\usepackage{amssymb}
\usepackage{amsmath}
\usepackage{bm}
\usepackage{color}
\usepackage{cite}
\usepackage{slashed}
\usepackage{subfigure}
\usepackage{epstopdf}            
\usepackage{epsfig}
\usepackage{caption}

\renewcommand{\thefootnote}{\fnsymbol{footnote}}
\newcommand{\beq}{\begin{equation}}
\newcommand{\eeq}{\end{equation}}
\newcommand{\ov}{\overline}

\newcommand{\eps}{\epsilon}
\newcommand{\veps}{\varepsilon}
\newcommand{\ol}[1]{\overline{#1}}
\newcommand{\wt}[1]{\widetilde{#1}}
\newcommand{\abs}[1]{\left|#1\right|}
\newcommand{\order}[1]{\mathcal{O}\left(#1\right) }

\newcommand{\la}{\lambda}

\newcommand{\Lcal}{\mathcal{L}}
\newcommand{\kap}{\kappa}
\newcommand{\tkap}{\tilde{\kappa}}
\newcommand{\tH}{\tilde{H}}

\newcommand{\sig}{\sigma}

\newcommand{\VEV}[1]{\left\langle{#1}\right\rangle}

\newcounter{num} 

\newcommand{\rnum}[1]{\mathrm{\setcounter{num}{#1} \roman{num}}}

\newcommand{\lp}{\ell} 
\newcommand{\fst}{A} 

\DeclareMathOperator{\arccot}{arccot}

\usepackage[colorlinks=true, linkcolor=black, citecolor=black,
urlcolor=black]{hyperref}

\begin{document}

\begin{titlepage}

\begin{center}

{\Large
{\bf 
Current status and muon $g-2$ explanation\\
of lepton portal dark matter
}
}

\vskip 2cm

Junichiro Kawamura$^{1,2}$\footnote{kawamura.14@osu.edu}, 
Shohei Okawa$^{3}$\footnote{okawa@uvic.ca}
and
Yuji Omura$^{4}$\footnote{yomura@phys.kindai.ac.jp}

\vskip 0.5cm

{\it $^1$Department of Physics,
The Ohio State University, 
Columbus, OH 43210, USA}\\[3pt]

{\it $^2$Department of Physics, 
Keio University, Yokohama 223-8522,  Japan} \\[3pt]

{\it $^3$ Department of Physics and Astronomy, University of Victoria, \\
Victoria, BC V8P 5C2, Canada}

{\it $^4$
Department of Physics, Kindai University, Higashi-Osaka, Osaka 577-8502, Japan}\\[3pt]

\vskip 1.5cm

\begin{abstract}
In this paper, we summarize phenomenology in lepton portal dark matter (DM) models, where DM couples to leptons and extra leptons/sleptons.
There are several possible setups: complex/real scalar DM and Dirac/Majorana fermion DM. 
In addition, there are choices for the lepton chirality that couples to DM. 
We discuss the prediction of each model and compare it with the latest experimental constraints from the DM, the LHC, and the flavor experiments. 
We also propose a simple setup to achieve the discrepancy in the anomalous magnetic moment of muon.
\end{abstract}

\end{center}
\end{titlepage}

\tableofcontents 

\renewcommand{\thefootnote}{\#\arabic{footnote}}
\setcounter{footnote}{0}

\section{Introduction}

Dark matter (DM) is one of the biggest mysteries in our universe. While evidences of the existence have been built up since the first suggestion in 30's, these are obtained via gravitational force and the fundamental nature of DM, e.g. spin and mass, is known very little. This has motivated theorists to propose a lot of possibilities for particle DM, and encouraged many attempts to reveal the physical characters.

As a good DM candidate, a massive elementary particle that weakly interacts with particles in the Standard Model (SM) has been discussed. If the size of the interaction is compatible with the weak coupling strength and DM mass ranges from a few GeV to a few TeV, the thermal relic density of DM is surprisingly in agreement with the observed value. This fascinating candidate is called weakly interacting massive particle (WIMP) DM, and significant attention has been paid to this kind of DM. 
In fact, there are many possible extensions of the SM predicting WIMP DM candidates, 
e.g. wino, bino and higgsino in supersymmetric models~\cite{Jungman:1995df,Ellis:2003cw}, Kaluza-Klein particles in extra dimension models~\cite{Servant:2002aq,Cheng:2002ej}, inert scalars in extended Higgs models~\cite{Arhrib:2013ela,Goudelis:2013uca,Ilnicka:2015jba,Diaz:2015pyv,Belyaev:2016lok}, and so on. 
In these extended models, DM is often accompanied with extra charged or colored particles, which mediates interaction between DM and SM. Such extra particles provide good opportunities to test DM models at the LHC experiment.

Many of those extended models are originally motivated by theoretical problems of the SM, such as the gauge hierarchy problem, rather than by explaining DM itself. However, compatibility of the models with theoretical problems and experimental results considerably restrict the parameter space where the DM problem is resolved. 
In contrast, once going away from the original motivation and working only on the DM problem, one can take a more effective way using simplified models, in which an electromagnetically (EM) neutral stable particle is simply added as a DM candidate, and interaction of DM to the SM sector is introduced in an {\em ad hoc} manner. An additional symmetry may also be imposed in another way to guarantee the DM stability. Given a specific setup in the framework of simplified models, one can study DM physics, such as thermal relic density and DM signatures, with a limited number of parameters, and examine the viability of the setup. 
The studies based on simplified models can cover many theoretically motivated models.

To examine simplified models, it is helpful to classify DM candidates according to spin and interaction with SM particles\footnote{In some cases, it is also important whether DM is self-conjugate or not, as discussed later.}. 
As an example, let us assume that DM is not charged under the electroweak (EW) symmetry and a (complex) scalar field. The simplest possible interaction between this DM candidate and the SM sector is via a scalar quartic coupling, $\la_X|X|^2|H|^2$, where $X$ is the DM field and $H$ the SM doublet Higgs field. 
The quartic coupling is responsible for DM thermal production and, once it is fixed by the observed value, 
one can unambiguously predict signatures of this DM candidate at high-energy collider, direct and indirect detection experiments. 
This setup is known as a Higgs portal DM model and has been studied well~\cite{Kanemura:2010sh,Djouadi:2011aa,Djouadi:2012zc,Escudero:2016gzx,Balazs:2017ple,Athron:2018ipf}.
In another example, the scalar DM can be coupled directly to SM quarks/leptons by newly introducing vectorlike charged (or colored) fermions. Yukawa couplings $\la_f^i X\ol{F}f^i$ involving the DM $X$, the vectorlike fermions $F$ and quarks/leptons $f^i$ ($i$:~flavor) are crucial to DM physics. 
This model makes different DM signatures from that of the Higgs portal model. The vectorlike fermions can be produced at high energy collider, and can be tested directly. Besides, the non-trivial flavor structure of the Yukawa couplings $\la_f^i$ can induce flavor changing processes, which may bring other constraints into play. These models are called fermion portal models~\cite{Chang:2013oia,Bai:2013iqa} and what we study in this paper. 

In the latter example, DM annihilation is induced by $t$-channel exchange of new particle. 
We can consider similar setups: DM can be either of scalar or fermion field; it may be self-conjugate or not; SM fermions coupled to DM are either of quarks or leptons, which are further $SU(2)_L$ doublets or singlets. 
A certain new fermion or sfermion (scalar) field is introduced,  
so that the portal Yukawa coupling is allowed and the $t$-channel annihilations are turned on.
There are many works on fermion portal models, but most of them focused on only limited setups or particular phenomena. More importantly, all these studies were done several years ago. 
In the present paper, therefore, we examine all possible setups in fermion portal models by taking into account the latest data from LHC searches, flavor physics and DM physics, including higher-order corrections to DM searches that have not been studied well. Then, we update the current status of each setup. In particular, we concentrate on namely lepton portal DM models~\cite{Bai:2014osa,Chang:2014tea}, given no positive results of direct searches at the LHC and direct detection experiments. 
For quark portal models, see e.g. Refs.~\cite{Chang:2013oia,Bai:2013iqa,Garny:2014waa,Ibarra:2015fqa,Bhattacharya:2015xha,Abe:2016wck,Baek:2016lnv,Baek:2017ykw,Blanke:2017tnb,Kawamura:2017ecz,Colucci:2018vxz,Kawamura:2018kut} 
and references therein. 

In the present paper, we also study impact of lepton portal DM models on precision observables. 
There is an explicit correlation between these observables and DM thermal relic density, since both of them are induced via DM-lepton Yukawa couplings. 
Once DM density is fixed, some observables are predicted.  
Of these, we especially discuss the anomalous magnetic moment ($g-2$) for muon.
There is a longstanding discrepancy in the muon $g-2$~\cite{Tanabashi:2018oca}:
\begin{align}
\label{eq-Delamu}
 \Delta a_\mu \equiv a_\mu^\mathrm{exp}-a_\mu^\mathrm{SM} = \left(268\pm76\right)\times 10^{-11},
\end{align}
where $a_\mu^\mathrm{exp}$ and $a_\mu^\mathrm{SM}$ are the experimental result and the SM prediction 
of $a_\mu=(g-2)/2$ for muon.\footnote{Recently, QED NLO corrections to a pion form factor have been calculated in Ref.~\cite{Campanario:2019mjh}, where a possibility of the corrections accommodating the discrepancy is examined. However, the authors have shown that the corrections are too small to diminish the existing discrepancy in the SM. For BSM interpretation of the discrepancy, see e.g. Ref.~\cite{Lindner:2016bgg}.}
It is well-known that the size of the discrepancy is the same order as that of the EW corrections. 
On the other hand, in the thermal freeze-out scenario, 
an EW scale mass and coupling are required to produce the correct DM abundance. 
This coincidence suggests that the DM problem and the discrepancy in $\Delta a_\mu$ 
can be resolved in this class of models. 
This issue has been addressed in the literature working on lepton portal DM models~\cite{Kopp:2014tsa,Agrawal:2014ufa,Kowalska:2017iqv,Calibbi:2018rzv}. 
It was observed, however, that due to strong chirality suppressions, 
$\Delta a_\mu$ cannot be accommodated in the minimal models, 
where either an $SU(2)_L$ singlet or doublet leptonic mediator is introduced. 
We point out in this paper that if we modify the minimal models and introduce both mediators, 
the suppression is removed, and the discrepancy in $\Delta a_\mu$ can be resolved. 
This possibility is studied in the scalar DM model in Refs.~\cite{Fukushima:2013efa,Kowalska:2017iqv}. 
In this paper, we examine both scalar and fermion DM models, and improve the analysis of the DM direct detection taking into account the loop correction. 
To our knowledge, we identify for the first time which setup and which parameter space can achieve the discrepancy in the fermion DM models.

This paper is organized as follows. 
In Sec.~\ref{sec;setup}, we classify lepton portal DM models, based on the spin of DM, 
gauge invariance and renormalizability. In our study, DM is assumed to be either of scalar or fermion. 
The EW and color neutrality of DM is also required there. 
Then, we study the phenomenology in each model. 
In Sec.~\ref{sec-LHC}, the LHC constraint on each model is reviewed. 
In Sec.~\ref{sec-minimalDM}, we study the DM physics, i.e. relic density, direct detection and indirect detection, within the minimal models, and summarize the updated status of each model. Contributions to $\Delta a_\mu$ in the minimal models are also calculated there, and we reconfirm difficulty in inducing a large $\Delta a_\mu$. 
In Sec.~\ref{sec;DMLR}, we examine extended models that is specialized in explaining the discrepancy in $\Delta a_\mu$. 
Sec.~\ref{sec;summary} is devoted to summary. 
In Appendix, the detail of our calculation concerned with DM physics and renormalization group (RG) are shown.

\section{Lepton portal DM models} 
\label{sec;setup}
\begin{table}[t]
\centering 
\begin{tabular}{c|ccc}\hline
Name        & self-conjugate  & ~~spin of DM~~   & ~~spin of mediator~~       \\ \hline\hline 
real           & Yes                     & 0 & 1/2            \\ 
complex    & No                       &0        & 1/2          \\
Majorana  & Yes                     & 1/2 & 0             \\
Dirac         & No                       & 1/2        & 0             \\
\hline 
\end{tabular}
\caption{\label{tab-DMtypes} 
Classification of DM models based on spins of DMs and mediators. 
}
\end{table} 

We study DM that is a singlet under the SM gauge group and has spin 0 or $1/2$. 
The scalar DM is denoted by $X$ and the fermion DM is denoted by $\chi$.  
The DM field is also classified into whether it is self-conjugate or not.  
For the self-conjugate scalar (fermion) DM, $X^\dag = X$ ($\chi^c = \chi$). 
The self-conjugation nature is important in DM physics.  
We further introduce a leptonic mediator whose spin is $1/2$ or $0$ for the scalar or fermion DM, respectively.  
The classification of DM is shown in Table~\ref{tab-DMtypes}. 
In the following, we shall introduce scalar and fermion DM models separately.

\subsection{Scalar DM models} 
\label{sec-modelscal}

\begin{table}[t]
\centering 
\begin{tabular}{c|cccc}\hline
fields &~~spin~~   & ~~$\text{SU}(3)_c$~~ & ~~$\text{SU}(2)_L$~~  & ~~$\text{U}(1)_Y$~~ \\ \hline\hline 
   $L_R$   & 1/2  & ${\bf 1}$        &${\bf2}$      &         $-1/2$           \\
    $L_L$   &  1/2  & ${\bf 1}$        &${\bf2}$      &         $-1/2$           \\ \hline
     $E_R$  &  1/2  & ${\bf 1}$        &${\bf 1}$      &         $-1$              \\
    $E_L$   &  1/2 & ${\bf 1}$        &${\bf1}$      &         $-1$               \\  \hline
   $ X$  &   0  & ${\bf1}$         &${\bf 1}$      &            $0$         \\
    \hline 
\end{tabular}
\caption{\label{tab-scalarDM} 
Matter content in the scalar DM model. 
The subscripts $L,R$ for the fermions represent their chiralities. 
$L$ and $E$, are vectorlike fermions to respect the anomaly-free condition. 
}
\end{table}

We shall set up scalar DM models in which DM is an EM and color neutral real or complex scalar.
To couple DM to SM leptons at renormalizable level, extra leptons are introduced, which are $(\mathbf{2}, -1/2)$ or $(\mathbf{1}, -1)$ under the EW gauge symmetry $SU(2)_L\times U(1)_Y$. 
Then, the DM particle has Yukawa couplings with these leptons. 
The extra leptons should be vectorlike to make the models anomaly-free. 
The matter content in the scalar DM models is shown in Table~\ref{tab-scalarDM}. 
In order to stabilize DM, 
a parity symmetry $Z_2$ or global $U(1)$ symmetry is imposed on the models, under which DM and the extra leptons non-trivially transform while all the SM fields are trivial. 
This symmetry distinguishes the extra leptons $L_L$ ($E_R$) 
from the corresponding SM ones $\lp_L^i$ ($e_R^i$).  
Note that the Higgs portal coupling $|X|^2 |H|^2$ is always allowed in the scalar models, 
and it could have some impacts on DM annihilation and direct detection. 
In this study, to highlight phenomenology that lepton portal couplings induce, we assume that the Higgs portal couplings is negligibly small, and 
do not consider any combinational effects with it.

We introduce relevant part of Lagrangian in the scalar DM models. 
The mass terms for the vectorlike leptons are given by 
\begin{align}
- \Lcal_{S, \mathrm{mass}} =  m_L \ol{L}_L {L}_R+ m_E \ol{E}_L E_R.    
\end{align}
The Yukawa interactions involving the vectorlike leptons are given by 
\begin{align}
\label{eq-yukS}
- \Lcal_{S, \mathrm{Yukawa}} =  
 \la_L^i  \, \ol{\lp}^i_L X L_R + \la_R^i  \, \ol{E}_L \, X^* e^i_R 
+
 \kappa \, \ol{L}_L \, H \,  E_{R}+ \tilde{\kappa} \, \ol{E}_L \, \tilde{H} \,  L_{R}  
+h.c., 
\end{align}
where $\lp_L^i$ and $e_R^i$ are the $SU(2)_L$ doublet and singlet leptons in the SM. 
The index $i=1,2,3$ runs over the SM three generations.   
Here, $\tilde{H} \equiv i \sigma_2 H^\dag$.  
The first two terms are the portal couplings of DM to the SM leptons. 
The latter two terms in Eq.~\eqref{eq-yukS} generate a mass mixing between the vectorlike charged leptons after the EW symmetry breaking. The mass matrix for them is given by 
\begin{align}
\label{eq-VLmasmat}
\begin{pmatrix}
 \ol{E}^\prime_L  & \ol{E}_L
\end{pmatrix}
\begin{pmatrix}
 m_L &  \tkap \, v_H \\
  \kap \, v_H & m_E 
\end{pmatrix}
\begin{pmatrix}
 E^\prime_R  \\ E_R
\end{pmatrix}, 
\end{align}
where $v_H \equiv \langle H_0\rangle$ is the Higgs vacuum expectation value (VEV). 
The primed field $E^\prime_{L}$ ($E^\prime_R$) 
is the charged component in the doublet vectorlike lepton $L_{L}$ ($L_R$).  
We define the mass eigenstates as 
\begin{align}
\label{eq-VLLmix}
  \begin{pmatrix}
    E'_R \\ E_R 
  \end{pmatrix}
 = 
\begin{pmatrix}
 c_R & s_R \\ -s_R & c_R
\end{pmatrix}
   \begin{pmatrix}
    E_{R_1} \\ E_{R_2}
   \end{pmatrix},
\quad 
  \begin{pmatrix}
    E'_L \\ E_L 
  \end{pmatrix}
 = 
\begin{pmatrix}
 c_L & s_L \\ -s_L & c_L
\end{pmatrix}
   \begin{pmatrix}
    E_{L_1} \\ E_{L_2}
   \end{pmatrix},
\end{align}
where $c_X$, $s_X$ ($X=L,R$) satisfy $c_X^2+s_X^2=1$.  
The left- and right-handed fermions can be combined to Dirac fermions as 
$E_a \equiv (E_{L_a},E_{R_a})$, where $a=1,2$. 
Their masses are denoted by $m_{E_1}$ and $m_{E_2}$. 
In the mass base, 
the Yukawa couplings involving the DM and vectorlike leptons $E_{1,2}$ are given by 
\begin{align}
\label{eq-muyuk}
\Lcal_{S,\mathrm{Yukawa}} \supset X \, \ol{e}_i \left[ \left(  \la_L^i P_R c_R - \la^{i \,*}_RP_L s_L  \right) E_1+
                                            \left(  \la_L^i P_R s_R + \la^{i \,*}_R P_L c_L  \right) E_2  \right] + h.c.,   
\end{align}
where $e_i$ is a Dirac fermion for a SM charged lepton in the $i$-th generation. 

The model has chance to resolve the discrepancy in the muon anomalous magnetic moment. 
The new physics contribution to $\Delta a_\mu$ in the scalar DM models is given by~\cite{Dermisek:2013gta,Jegerlehner:2009ry,Lynch:2001zs} 
\begin{align}
\label{eq-DelamuS}
 \Delta a_\mu =  \frac{m_\mu}{16\pi^2 m_X^2} 
                  \Bigl[&   \left( c_R^2 |\la^\mu_L|^2 + s^2_L |\la^{\mu}_R|^2 \right) m_\mu F_f(x_1)
                        + c_R s_L \text{Re} \left( \la^\mu_L \la^\mu_R \right) m_{E_1} G_f(x_1)    \\ \notag 
           &\quad              +    \left( s_R^2 |\la^\mu_L|^2 + c^2_L |\la^{\mu}_R|^2 \right) m_\mu F_f(x_2)
                        - c_L s_R \text{Re} \left( \la^\mu_L \la^\mu_R \right) m_{E_2} G_f(x_2)\Bigr], 
\end{align}
where $x_i = m_{E_i}^2/m_X^2$. The loop functions are defined as 
\begin{align}
 F_f(x) &\ = \frac{2+3x-6x^2+x^3+6x\log x}{6(1-x)^4},\quad 
 G_f(x) = \frac{3-4x+x^2+2\log x}{(1-x)^3}.   
\end{align}
When $m_X \sim \order{100\ \mathrm{GeV}}$, 
the current discrepancy can be explained 
only if chirality-flip effect proportional to the vectorlike lepton mass $m_{E_{1,2}}$ is sizable. 
This arises only in the model with a non-vanishing mass mixing, i.e. $s_L\neq0$ or $s_R\neq0$. 

A discrepancy in the electron anomalous magnetic moment $\Delta a_e$ 
is also reported~\cite{Davoudiasl:2018fbb} recently, although it is less significant than that of the muon. 
We expect that $\Delta a_e$ can be explained in the lepton portal models instead of $\Delta a_\mu$. 
However, these cannot be explained simultaneously,   
since the lepton flavor violating (LFV) decay $\mu\to e\gamma$ is induced 
just as the portal Yukawa coupling to the electron is turned on~\footnote{
Similar conclusion is obtained in models which $\Delta a_{e,\mu}$ 
is explained by the loop corrections involving vectorlike leptons 
and $Z^\prime$ boson~\cite{Kawamura:2019rth,CarcamoHernandez:2019ydc,Kawamura:2019hxp}.  
}. 
We do not pursue this possibility in this paper. 
The simultaneous explanation for both anomalies are studied in Refs.~\cite{
Giudice:2012ms,Crivellin:2018qmi,Liu:2018xkx,Dutta:2018fge,Parker:2018vye,Han:2018znu,Endo:2019bcj,Abdullah:2019ofw,Bauer:2019gfk,Cornella:2019uxs}.  

Throughout this paper, the ''minimal'' setups refer to the models in which either doublet or singlet leptonic mediator exists, and the other is decoupled. There is no mass mixing between them. 
In the model Lagrangian introduced above, the minimal setups are therefore obtained by simply neglecting either $L$ or $E$. 
As alluded above, the non-minimal setup that involves both the vectorlike leptons in the model 
is interesting because it can accommodate the discrepancy in the muon $g-2$.

\subsection{Fermion DM models}   
\begin{table}[t]
\centering 
\begin{tabular}{c|cccc}\hline
fields &~~spin~~   & ~~$\text{SU}(3)_c$~~ & ~~$\text{SU}(2)_L$~~  & ~~$\text{U}(1)_Y$~~ \\ \hline\hline 
   $\wt{L}$   & 0  & ${\bf 1}$        &${\bf2}$      &         $-1/2$           \\
    $\wt{E}$ &  0  & ${\bf 1}$        &${\bf 1}$      &         $-1$              \\ \hline 
   $\chi$  &   1/2  & ${\bf1}$         &${\bf 1}$      &            $0$         \\
    \hline 
\end{tabular}
\caption{\label{tab-fermiDM} 
Matter content in the fermion DM model.  
}
\end{table} 

The fermion DM can be Majorana or Dirac. 
In the fermion DM models, mediators are complex scalars 
and are $(\mathbf{2}, -1/2)$ or $(\mathbf{1}, -1)$ under the EW symmetry. 
The $SU(2)_L$ doublet (singlet) mediator, named slepton, is denoted by $\wt{L}$ ($\wt{E}$). 
The matter content is shown in Table~\ref{tab-fermiDM}. 
As in the scalar DM models, the minimal setups are obtained by neglecting either of the sleptons.

The slepton masses and interaction terms are given by 
\begin{align}
\label{eq;VF}   
-\Lcal_{F,\mathrm{scal}} =&\ 
 m_{\wt{L}}^2 \wt{L}^{\dag}\wt{L}+ m^2_{\wt{E}} \wt{E}^{\dag} \wt{E} 
        +\left(A\; \wt{L}^\dag H \wt{E} + \mathrm{h.c.} \right)  \\ \notag 
& \quad   + \la_{LH} \left(\wt{L}^{ \dag}H\right)\left(H^{\dag}  \wt{L}\right)  
  +\la^\prime_{LH} \left(\wt{L}^{\dag}H\right)\left(\wt{L}^\dag H\right).  
\end{align}
Since the quartic coupling in Eq.~\eqref{eq;VF} is irrelevant to DM physics,  
we simply neglect these terms.   
 The Yukawa couplings involving the DM are given by 
\begin{align}
\label{eq-yukF}
\Lcal_{F,\mathrm{Yukawa}}=&\ 
 \la_L^i \ol{\lp}_L^i \chi_R  \wt{L}+\la^{i}_{R}\wt{E}^\dagger  \ol{\chi}_L e^i_{R}  +\mathrm{h.c.} .
\end{align}
For simplicity, we use the common notation, $\la_L^i$ and $\la_R^i$, 
for the portal Yukawa couplings in both scalar and fermion DM models. 
As in the scalar DM models, 
we assume that one of the portal Yukawa couplings is sizable in the minimal setup. 
Models with the both couplings are discussed in Sec.~\ref{sec;DMLR}.

If both $\wt{L}$ and $\wt{E}$ exist in the model,  
the trilinear terms induce the mixing between these states. 
The mass matrix is given by 
\begin{align}
\label{eq-slepmat}
  \begin{pmatrix}
   \widetilde{E}^{\prime \dagger} & \widetilde{E}^\dagger
  \end{pmatrix}
\begin{pmatrix}
 m_{\wt{L}}^2 & A  v_H \\ A^* v_H & m_{\wt{E}}^2 
\end{pmatrix}
\begin{pmatrix}
   \widetilde{E}^{\prime} \\ \widetilde{E}
\end{pmatrix},
\end{align}
where $\wt{E}^\prime$ is the charged component of the doublet slepton $\wt{L}$. 
The mass eigenstates of the sleptons are defined as 
\begin{align}
\label{eq-slepmix}
 \begin{pmatrix}
  \widetilde{E}^\prime \\ \widetilde{E}
 \end{pmatrix}
= 
\begin{pmatrix}
 c_\theta & s_\theta \\
 -s_\theta & c_\theta
\end{pmatrix}
\begin{pmatrix}
 \tilde{E}_1 \\ \tilde{E}_2
\end{pmatrix},
\end{align} 
where $c_\theta$, $s_\theta$ satisfy $c^2_\theta+s^2_\theta=1$.
Their masses are denoted as $m_{\widetilde E_1}$ and $m_{\widetilde E_2}$. 
The Yukawa couplings involving the DM and sleptons are given by 
\begin{equation}
\Lcal_{F,\mathrm{Yukawa}} \supset \ol{e}_i \left[ \left( \la_L^i P_R c_\theta  - \la^{i*}_R P_L s_\theta \right ) \widetilde{E}_1+ \left( \la^i_L P_R s_\theta + \la^{i*}_R P_L c_\theta \right) \widetilde{E}_2 \right] \chi +h.c..
\end{equation}

The new physics contribution to $\Delta a_\mu$ in the fermion DM model 
is given by~\cite{Dermisek:2013gta,Jegerlehner:2009ry,Lynch:2001zs }  
\begin{align}
 \Delta a_\mu = - \frac{m_\mu}{16\pi^2 m_\chi^2}  
   \Bigl[  & 
     \left( c_\theta^2 |\la^\mu_L|^2 + s^2_\theta |\la^{\mu}_R|^2 \right) m_\mu F_s(y_1) 
                -c_\theta s_\theta \text{Re}\left(\la^\mu_L\la^\mu_R \right)  m_\chi G_s(y_1)   \\ \notag 
   &\quad    +  \left( s_\theta^2 |\la^\mu_L|^2 + c^2_\theta |\la^{\mu}_R|^2 \right) m_\mu  F_s(y_2) 
              +  c_\theta s_\theta \text{Re}\left(\la^\mu_L\la^\mu_R \right)  m_\chi  G_s(y_2)  
  \Bigr],  
\end{align}
where $y_i = m_\chi^2/m_{\tilde{E}_i}^2$.  
The loop functions are defined as 
\begin{align}
 F_s(x) &\ =  x \frac{1-6x+3x^2+2x^3-6x^2\log x}{6(1-x)^4},\quad
 G_s(x) = x \frac{1-x^2+2x\log x}{(1-x)^3}. 
\end{align}
As in the scalar DM models, 
both singlet and doublet sleptons are necessary to explain $\Delta a_\mu$ 
so that the chirality-flip effect proportional to $m_\chi$ is substantial. 

Hereafter, when we analytically study generic features of the models, we will not assume any specific structure of the portal Yukawa couplings. The couplings are supposed to have arbitrary values and, therefore, entirely flavor violating there. 
On the other hand, if DM has sizable couplings to two or more generations at the same time, 
such setups will be excluded in reality due to large LFV processes induced. 
Therefore, when we numerically analyze the models to calculate physical quantities, derive constraints and show some plots, we will identify a structure of the portal couplings such that DM couples exclusively to one of the three generations. 
Then, strong constraints from LFV processes are satisfied. 
In the analysis later, we will mainly focus on the namely muon-philic case, i.e. 
$\la_{L,R}^\mu \gg \la_{L,R}^{e}$, $\la_{L,R}^{\tau}$, motivated by the discrepancy in $\Delta a_\mu$.

\section{The LHC limits on extra lepton and slepton} 
\label{sec-LHC}
We discuss experimental limits from the LHC in this section.  
In the lepton portal models, 
the lightest mediator decays to a SM lepton and a DM particle. 
The pair production of the vectorlike leptons $E_1$ and sleptons $\wt{E}_1$ is 
\begin{align}
 p p \to E_1 E_1 \to e_i e_i + X X,
\quad \mathrm{and}\quad 
 p p \to \wt{E}_1 \wt{E}_1 \to e_i e_i + \chi\chi, 
\end{align}
respectively. 
Both processes give two SM leptons and a large missing energy 
which has been studied in slepton searches
at the LHC~\cite{Aad:2019byo,Aad:2019qnd,Aad:2019vnb,CMS:2017wox,Sirunyan:2018nwe,CMS:2019hos}.
We will study a case that the (s)lepton dominantly decays to a muon and a DM particle. 
The LHC limits in the electron-philic case will be similar to the ones in the muon-philic case, 
while the limits will be much weaker in the tau-philic case.

We estimate the experimental limits on sleptons and vectorlike leptons 
from the results in Ref.~\cite{Aad:2019vnb}. 
To illustrate our method, we study a limit on slepton first.  
The limit is estimated from the ATLAS analysis for a doublet smuon $\wt{\mu}_L$ pair production.  
We define an efficiency $\eps_\mathrm{SR}$ in each signal region (SR) as 
\begin{align}
\sig_\mathrm{prod}^\mathrm{ref}  \times \eps_\mathrm{SR} 
= \sig_\mathrm{SR}^\mathrm{exp},    
\end{align}
where $\sig_\mathrm{prod}^\mathrm{ref}$ is a cross section of $\wt{\mu}_L\wt{\mu}_L$ production 
and $\sig_\mathrm{SR}^\mathrm{exp}$ is an experimental upper bound on the effective cross section 
in the signal region. Here, $\wt{\mu}$ decays to a muon and a DM particle exclusively. 
The efficiency $\eps_\mathrm{SR}$ is a probability 
that pair-produced smuons pass the kinematic-cut of signal region.

We estimate the efficiency $\eps_\mathrm{SR}$ from the experimental limits   
as a function of mass difference between slepton and DM, 
$\Delta m \equiv m_{\wt{E}_1} - m_\chi $. 
On the experimental bound in ($m_{\wt{E}_1}$, $m_\chi$) plane, 
the efficiency is given by  
\begin{align}
\label{eq-efficiency}
\eps_\mathrm{SR} (\Delta m)  
\sim \frac{\sigma_\mathrm{SR}^\mathrm{exp}}{\sigma_\mathrm{prod}^\mathrm{ref}(m_{\wt{E}_1}(\Delta m))}.  
\end{align}
Here, the efficiency is assumed to be independent of slepton mass approximately.  
The production cross section is, on the other hand, determined by slepton mass. 
When a mass difference is larger than a limit on slepton mass with a massless DM,   
denoted by $m_{\wt{E}}^\mathrm{max}$,  
the efficiency is set at $\eps_\mathrm{SR}(m_{\wt{E}}^\mathrm{max})$. 
By using this efficiency function of $\Delta m$, a point ($m_{\wt{E}_1}$, $m_\chi$) is excluded 
if 
\begin{align}
\sigma_\mathrm{prod} \left(m_{\wt{E}_1}  \right) \times \eps_\mathrm{SR} (\Delta m) 
> \sigma_\mathrm{SR}^\mathrm{exp}   
\end{align}
in any signal region.  
In the analysis of Ref.~\cite{Aad:2019vnb}, there are two types of signal regions. 
One is designed for a mass spectrum with large mass difference
and requires no jets in an event. 
The other is designed for a mass spectrum with small mass difference
and requires a jet in an event. 
We refer to the signal regions without (with) jet for $\Delta m > 200$~($\le 200$)~GeV.

For vectorlike lepton, we estimate a limit by analogy with slepton search.
We estimate the efficiency in Eq.~\eqref{eq-efficiency} 
from the limit on the degenerate slepton scenario, 
where both left- and right-handed selectron and smuon have a common mass. 
Our estimations could be improved, 
using a limit which gives more similar shape as that for the vectorlike lepton.

\begin{figure}[t]
\centering 
\begin{minipage}[c]{0.48\hsize}
 \centering
\includegraphics[height=70mm]{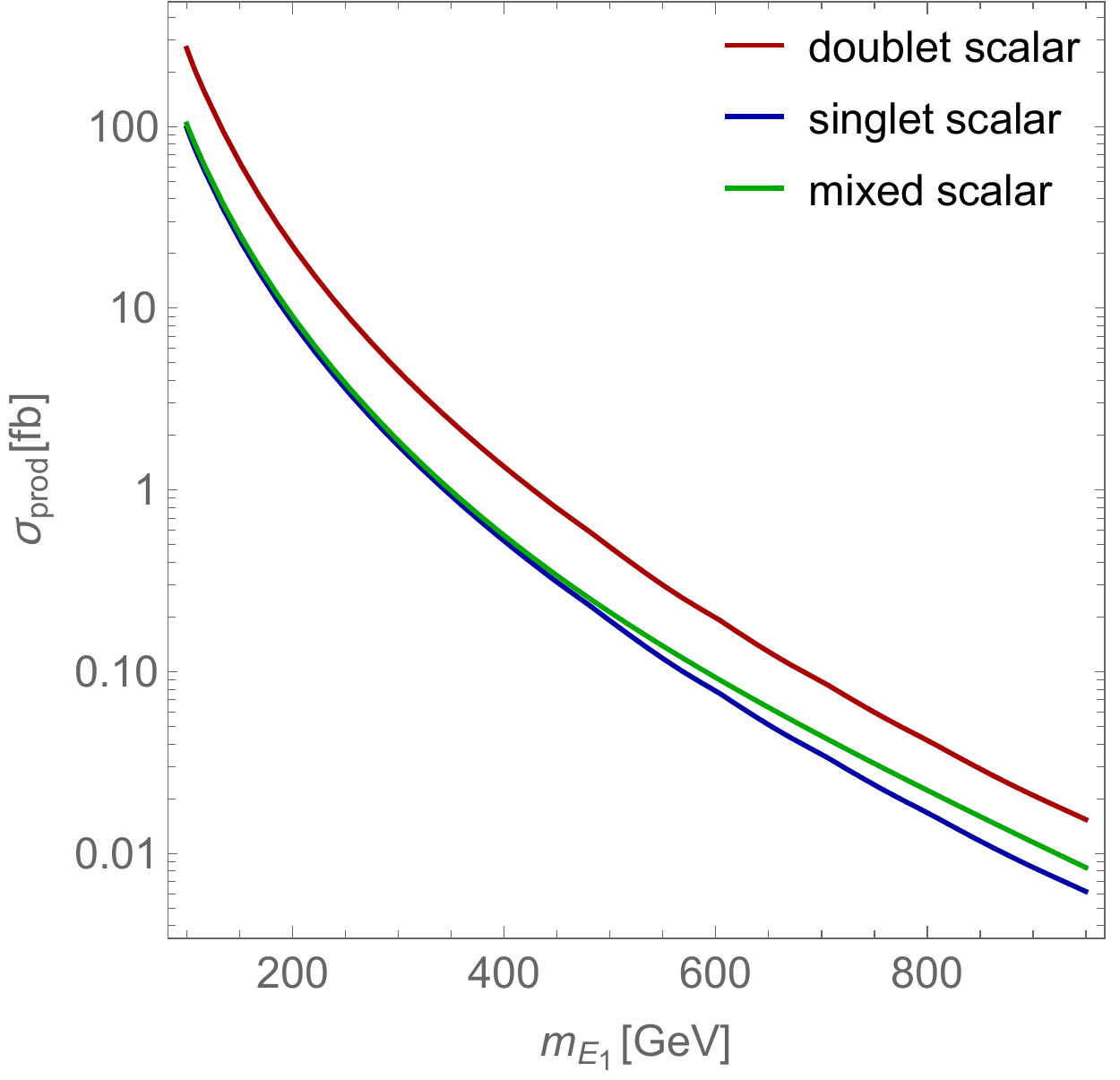} 
\end{minipage}
\begin{minipage}[c]{0.48\hsize}
 \centering
\includegraphics[height=70mm]{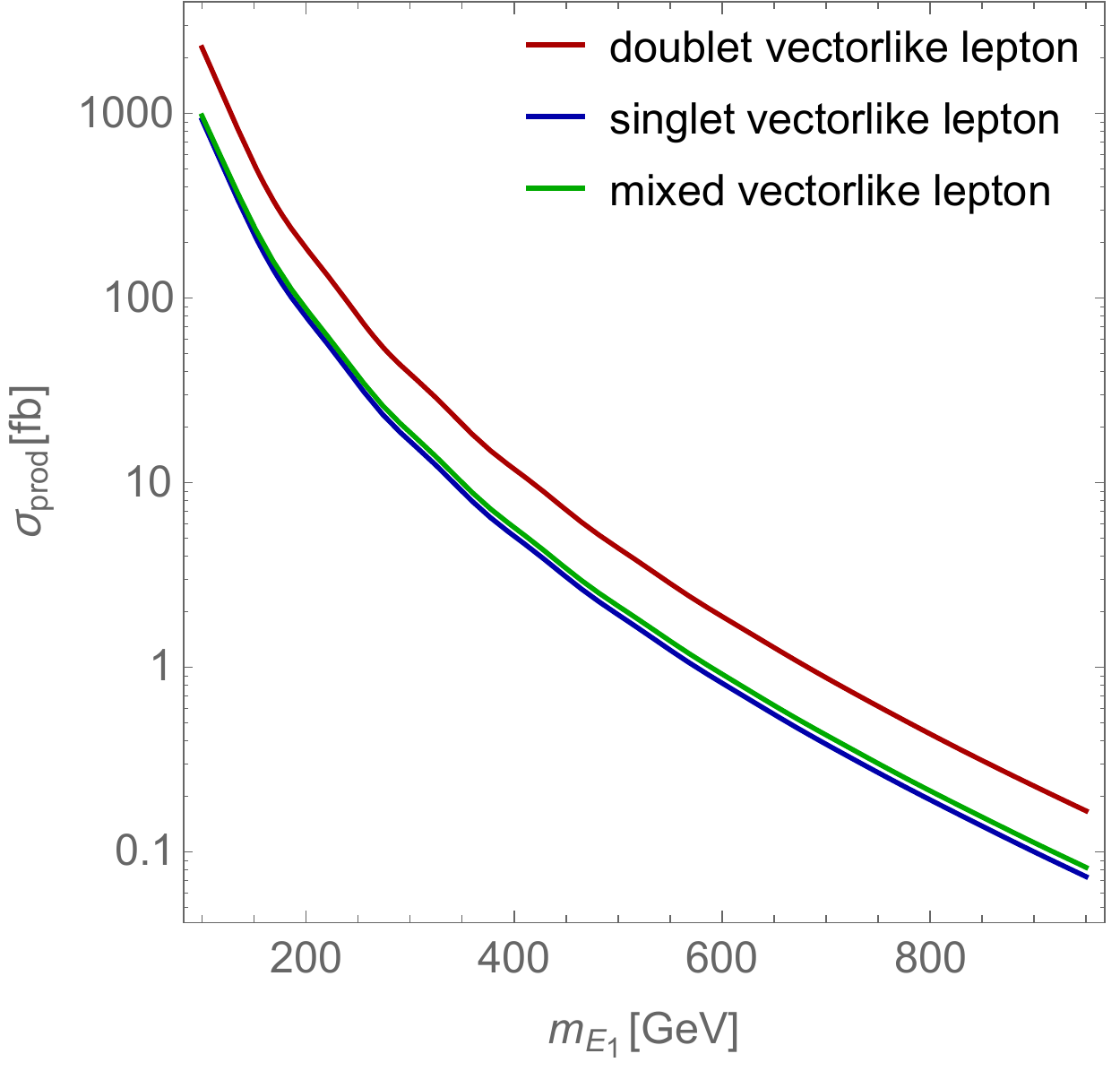} 
\end{minipage}
\caption{\label{fig-prodution}
Production cross sections of the lightest slepton (left) and vectorlike lepton (right) with $\sqrt{s}=13$ TeV. 
}
\end{figure}
\begin{figure}[t]
\centering 
\begin{minipage}[c]{0.48\hsize}
 \centering
\includegraphics[height=70mm]{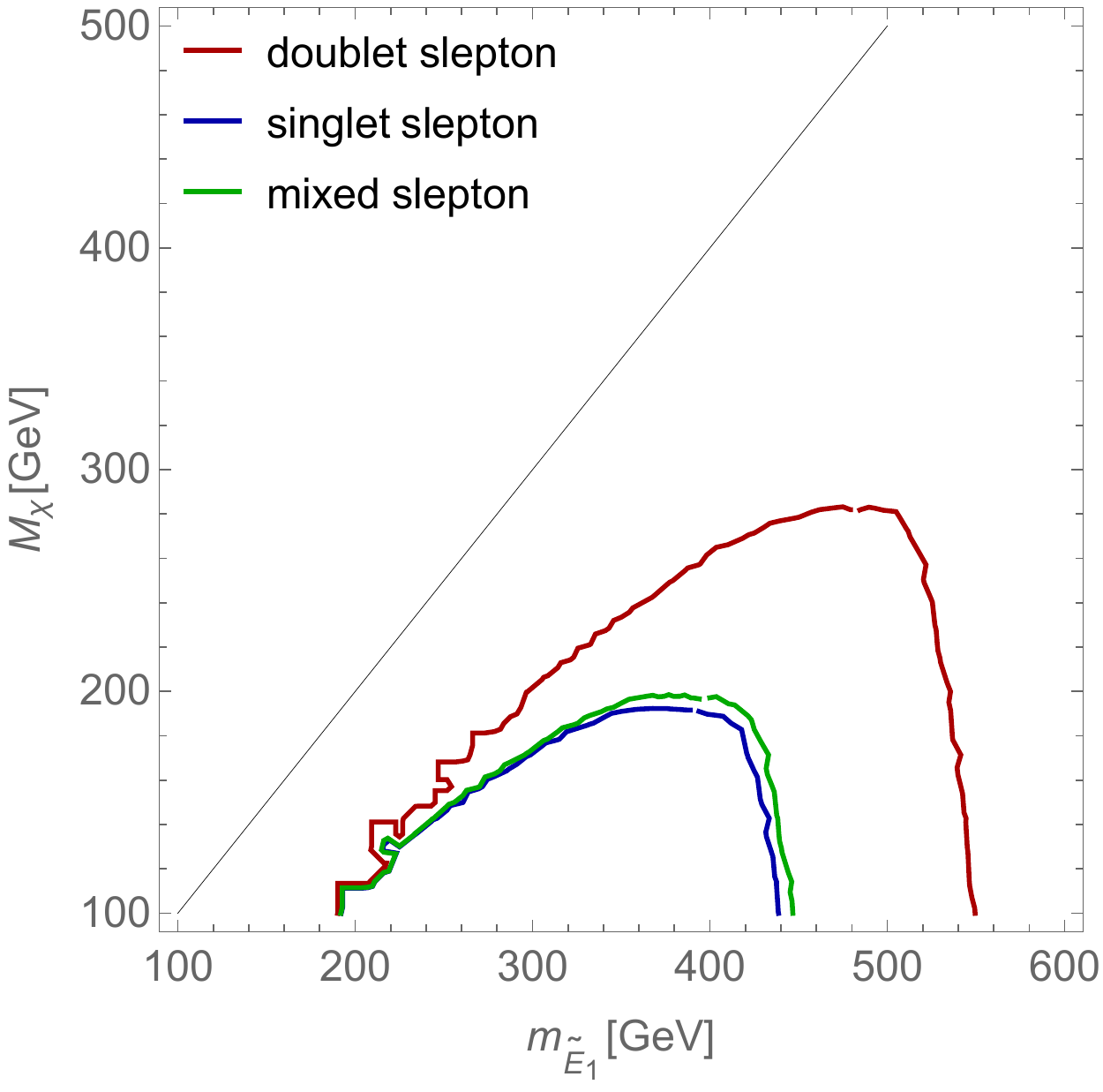} 
\end{minipage}
\begin{minipage}[c]{0.48\hsize}
 \centering
\includegraphics[height=70mm]{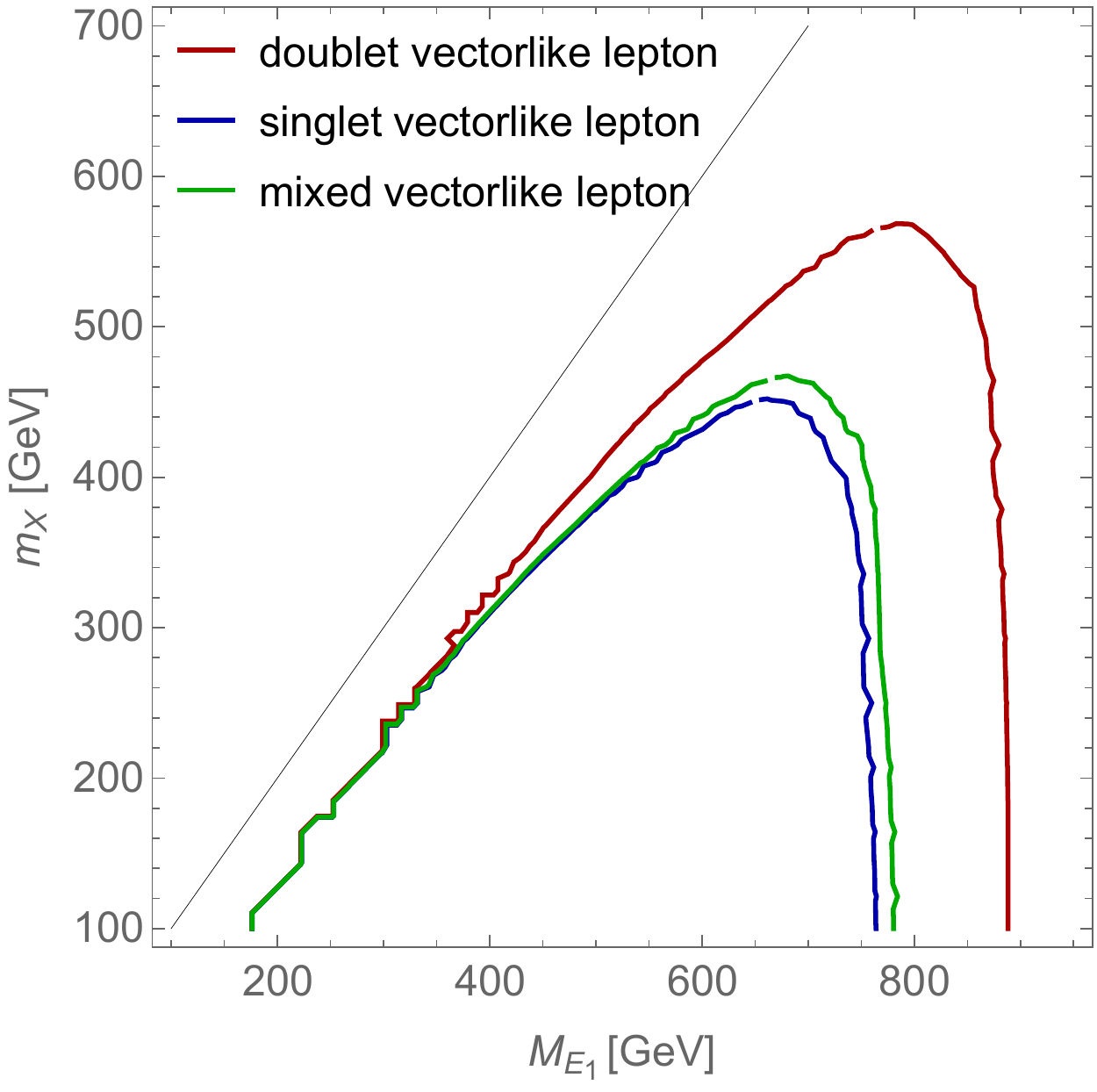} 
\end{minipage}
\caption{\label{fig-LHClim}
Experimental limits on the slepton (left) and vectorlike lepton (right) 
estimated from the ATLAS data~\cite{Aad:2019vnb}.  
}
\end{figure}
Figure~\ref{fig-prodution} shows production cross sections 
of a pair production of sleptons (left) and vectorlike leptons (right) with $\sqrt{s}=13$ TeV. 
For the production cross sections of sleptons,  
we refer to the result of LHC SUSY Cross Section Working Group~\cite{Bozzi:2007qr,Fuks:2013vua,Fuks:2013lya,Fiaschi:2018hgm,Beenakker:1999xh}.
The mixed slepton is defined as 
the lighter slepton when $s_L=s_R=s_\theta=1/\sqrt{2}$.  

Production cross sections for vectorlike leptons are calculated by 
\texttt{MadGraph5$\_$2$\_$6$\_$5}~\cite{Alwall:2014hca} 
based on an \texttt{UFO}~\cite{Degrande:2011ua} model file 
generated with \texttt{FeynRules$\_$2$\_$3$\_$32}~\cite{Christensen:2008py,Alwall:2014hca}.
As in the vectorlike lepton case, the mixed extra lepton is defined as  
the lighter one when $s_L=s_R=s_\theta={1}/{\sqrt 2}.$
 
Figure~\ref{fig-LHClim} shows the estimated $95\%$ C.L. upper limits on the mediator lepton masses and DM masses. 
The red, blue and green lines show the limit for the doublet, singlet and maximally mixed sleptons (vectorlike leptons), 
respectively.  
The doublet leptons are more constrained than the other cases 
because of the larger production cross sections. 
The limit on singlet slepton shows good agreement with the experimental limit in Ref.~\cite{Aad:2019vnb}.

The analysis in Ref.~\cite{Aad:2019vnb} excludes parameter space 
where the mass difference is larger than about 100 GeV. 
More degenerate region would be excluded by more dedicated searches 
exploiting an initial state radiation and soft leptons~\cite{ATLAS:2019lov,Aad:2019qnd}. 
The limits, however, exist only in restricted parameter space where the mass difference is about 10 GeV 
and the tightest limit is about 250 GeV for the degenerate four sleptons scenario.  
The limits from these searches are not shown in the following analyses, 
but we shall note that a shallow parameter space with $\Delta m \lesssim 10$ GeV 
would be constrained by these searches.

Let us comment on the heavier mediator leptons. 
The heavier states $E_2$ and $\wt{E}_2$ can decay to a SM lepton and DM particle 
as the lightest ones. 
In addition, these may decay to a lighter mediator lepton and a SM boson. 
For instance, $E_2 \to E_1 Z \to \mu Z X$ is possible 
and can give clean signals with three charged leptons and large missing energy per one vectorlike lepton $E_2$.  
The branching fractions to a SM boson can be comparable with or even dominate over 
that to a lepton and a DM particle depending on the couplings and mass spectrum. 
This is an interesting possibility to discover the lepton portal models  but this is beyond the scope of this paper. 
In the following, 
we only show the limits from a pair production of the lightest mediator leptons in Fig.~\ref{fig-LHClim}.


\section{Dark matter physics in the minimal models} 
\label{sec-minimalDM}
In this section, we study DM-related physics and 
discuss constraints on the models with the minimal matter contents,
where either $SU(2)_L$ doublet or singlet mediator field exists.
There are four types of DM (real/complex/Dirac/Majorana), and 
the mediator field is $(\rnum{1})$ an $SU(2)_L$ doublet~$L/\wt{L}$ or $(\rnum{2})$ an $SU(2)_L$ singlet~$E/\wt{E}$. The next-to-minimal models with both mediator fields will be discussed in the next section.

\subsection{Annihilation cross sections}
\label{sec:ann}
Pair annihilation is a basic property of particle DM. 
It governs the DM abundance based on the thermal freeze-out mechanism.  
If annihilation occurs in halo, it also contributes to cosmic ray flux 
which can be probed by indirect searches at telescopes.   
Since DM particles are non-relativistic at freeze-out and also in halo, 
it is useful to expand the pair annihilation cross section in terms of relative velocity $v$ of DM particles: 
\begin{equation}
\left(\sigma v\right)_{\fst} = a_{\fst} + b_{\fst} v^2 + c_{\fst} v^4 + \order{v^6}, 
\label{eq;cs}
\end{equation}
where $a_{\fst}$, $b_{\fst}$ and $c_{\fst}$ are dubbed 
as partial $s$-wave, $p$-wave and $d$-wave contributions, respectively. 
Here, the subscript $\fst$ represents a final state of the annihilation process.  
In the lepton portal models, 
the Yukawa couplings in Eqs.~\eqref{eq-yukS} and~\eqref{eq-yukF} 
induce various annihilation processes. 
In this section, we summarize the important features of the DM annihilation 
to $\lp_i\ol{\lp}_j$, $\lp_i\ol{\lp}_j V$ and $VV^\prime$, 
where $\lp_i$ is a SM charged lepton or neutrino  
and $V^{(\prime)}$ is a SM gauge boson.   
Here, $i=1,2,3$ is the flavor index. 
Note that the neutrinos have only left-handed component.    
The sample diagrams are shown in Fig.~\ref{fig;ann}. 
The full analytical formulas are shown in Appendix~\ref{app:analytics}. 

\begin{figure}[t]
\centering 
\includegraphics[viewport=170 600 440 770, clip=true, scale=0.5]{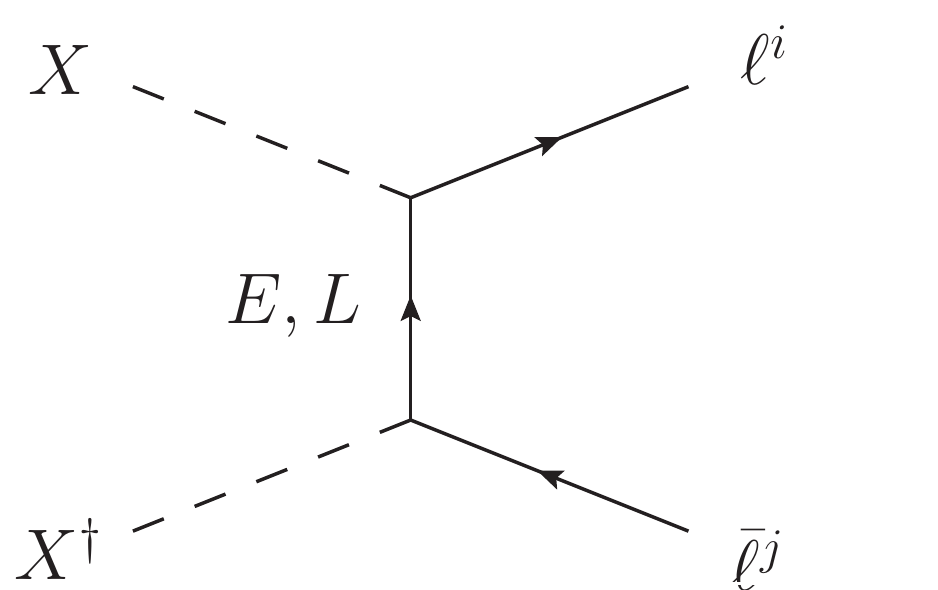}
\includegraphics[viewport=160 600 440 770, clip=true, scale=0.5]{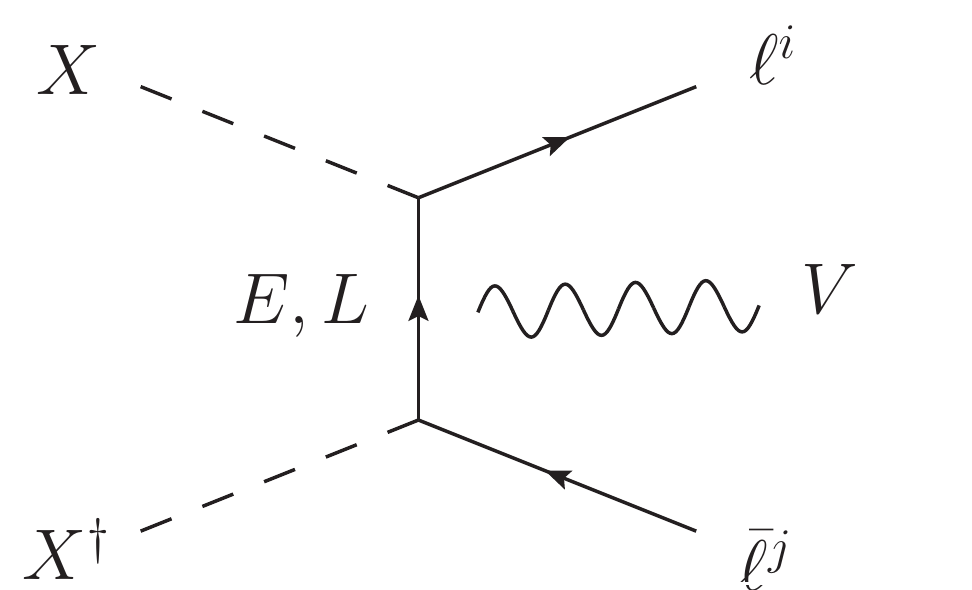}
\includegraphics[viewport=140 600 440 770, clip=true, scale=0.5]{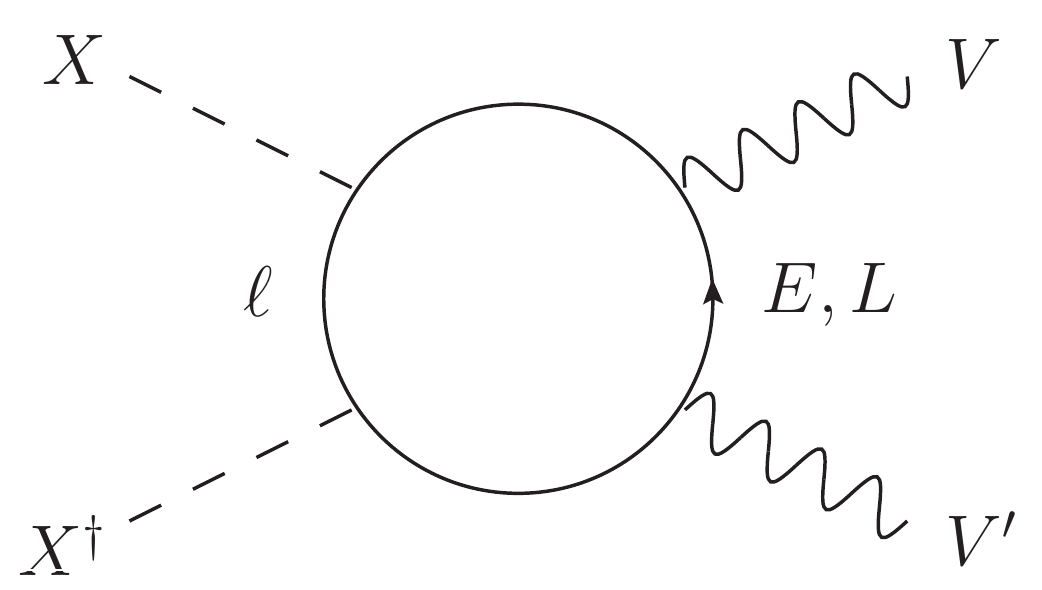}
\caption{\label{fig;ann}
Example diagrams of DM pair annihilation in the scalar DM model. 
In the fermion DM model, the dashed lines for $X$ and solid lines for $E,L$  
are replaced by the solid lines for $\chi$ and dashed lines for $\wt{E},\wt{L}$, respectively.
}
\end{figure}

\begin{figure}[!t]
\centering 
{\epsfig{figure=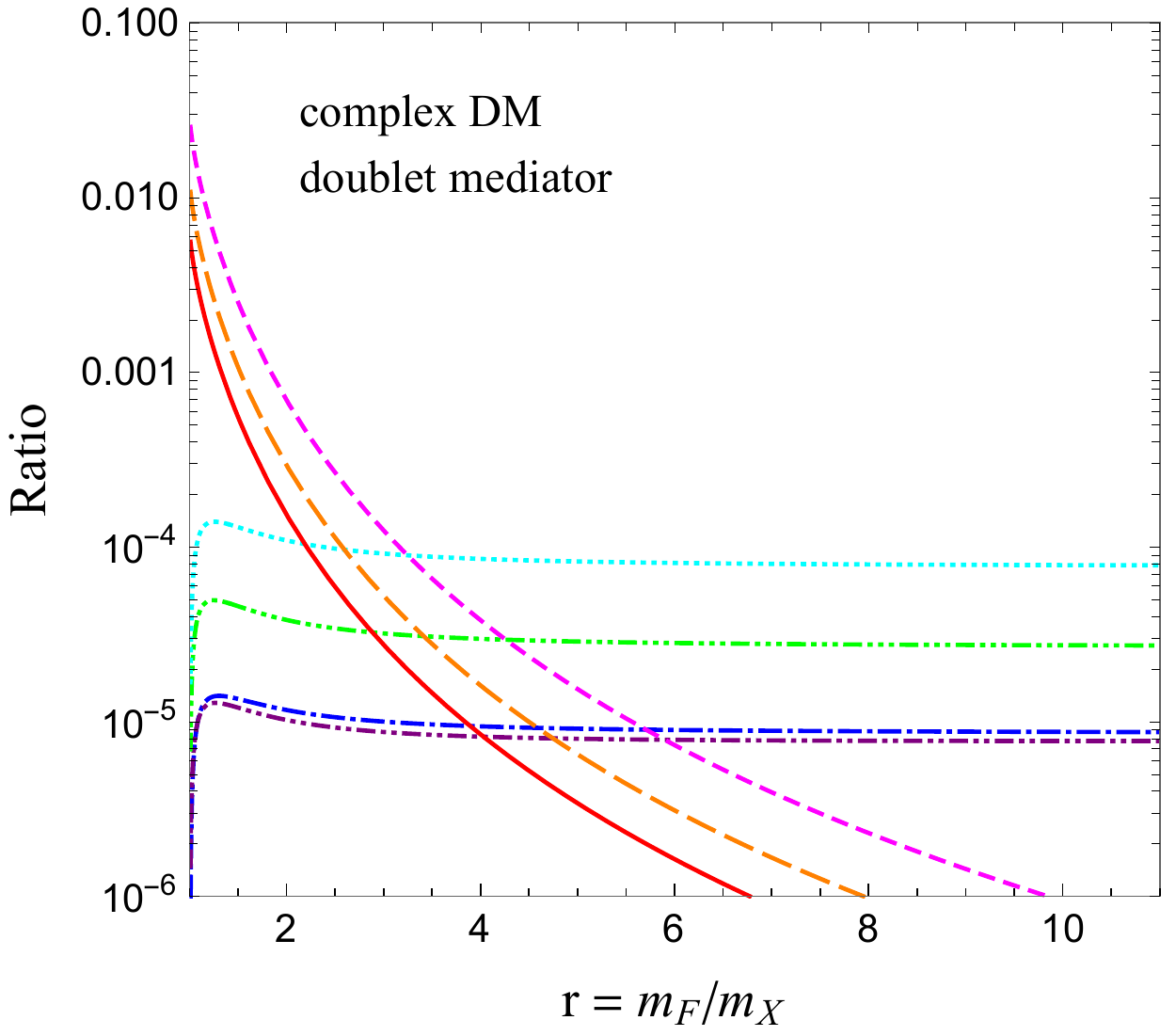,width=0.45\textwidth}}\hspace{.5cm}
{\epsfig{figure=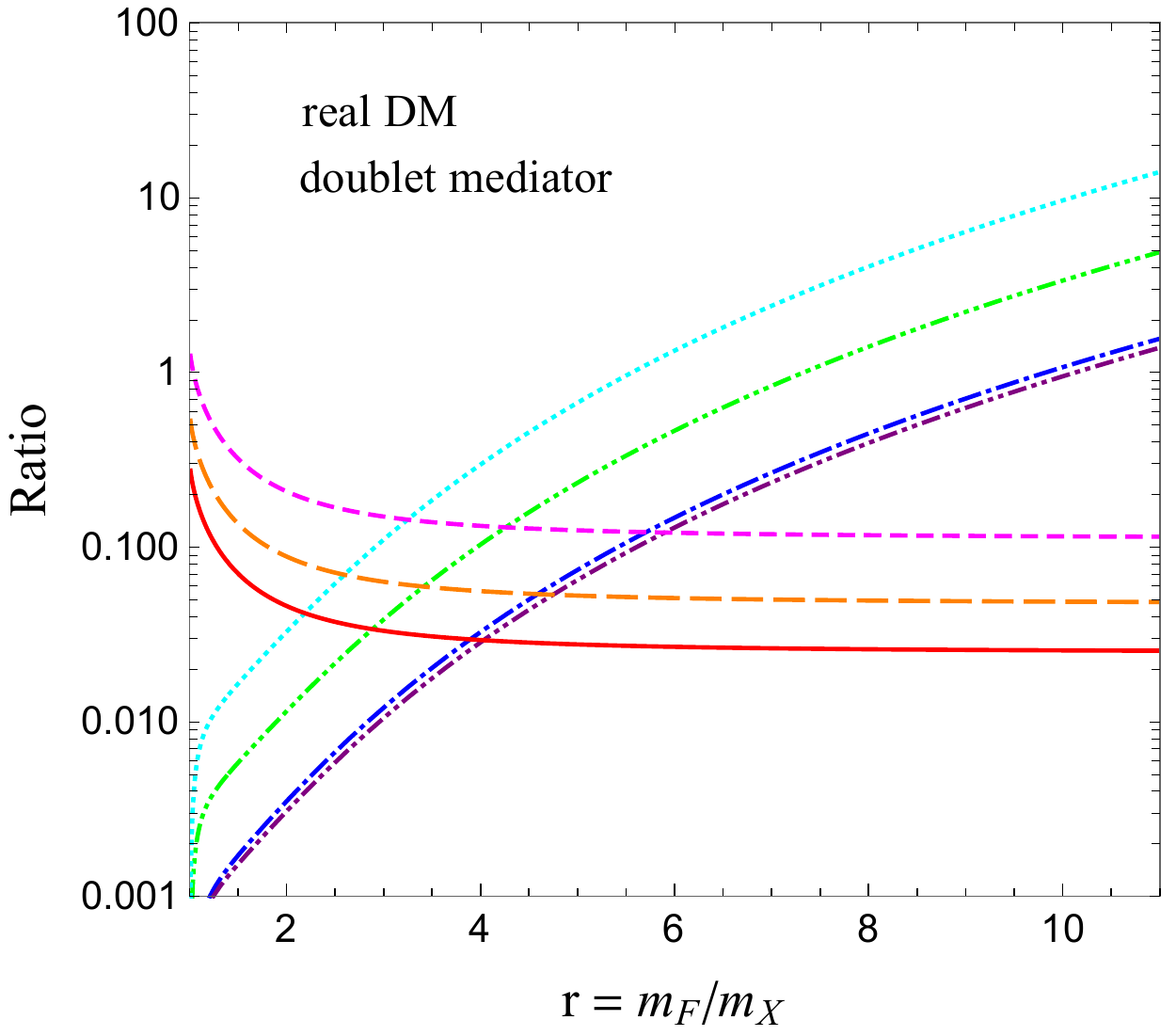,width=0.45\textwidth}}\vspace{.5cm}
{\epsfig{figure=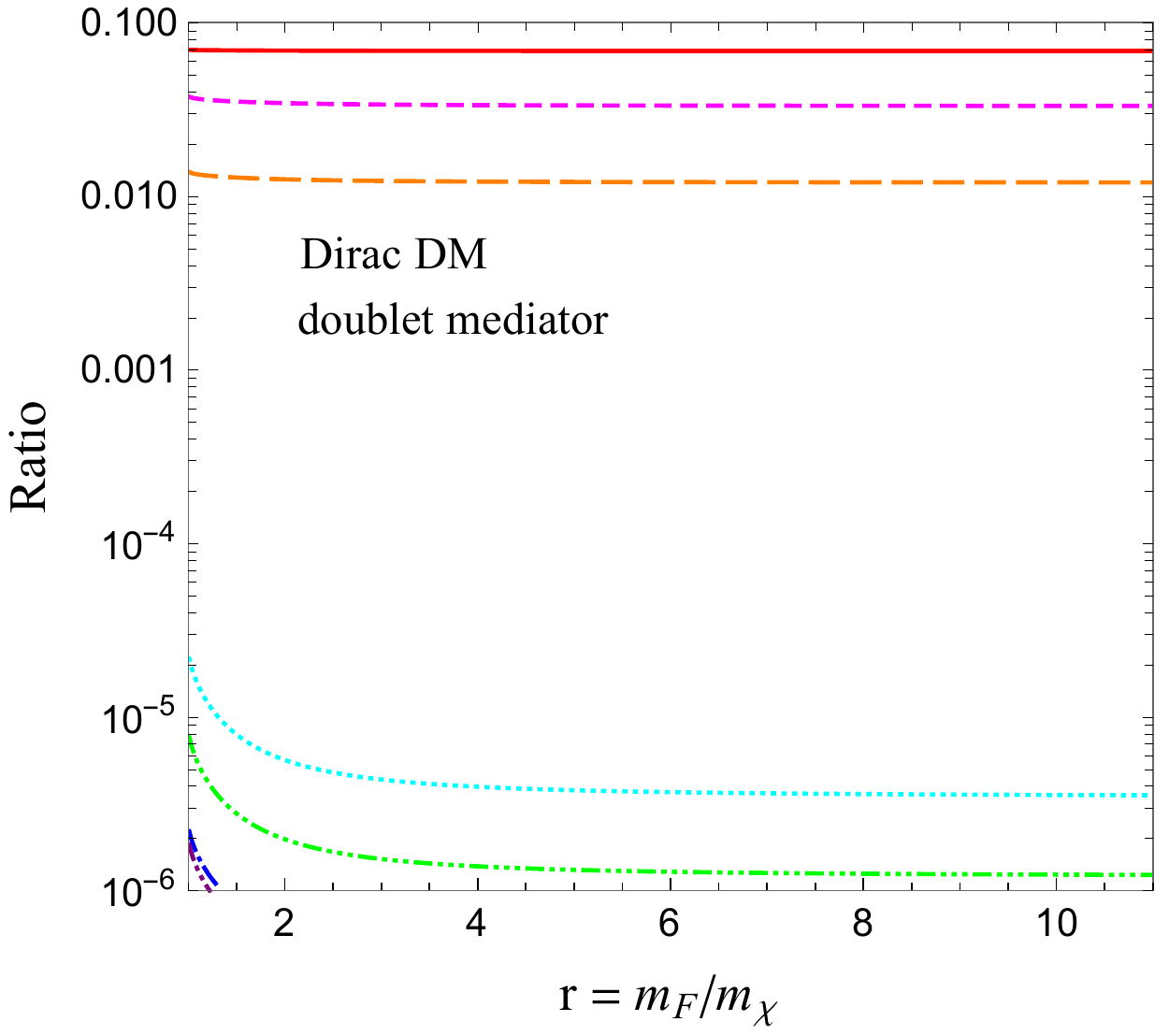,width=0.45\textwidth}}\hspace{.5cm}
{\epsfig{figure=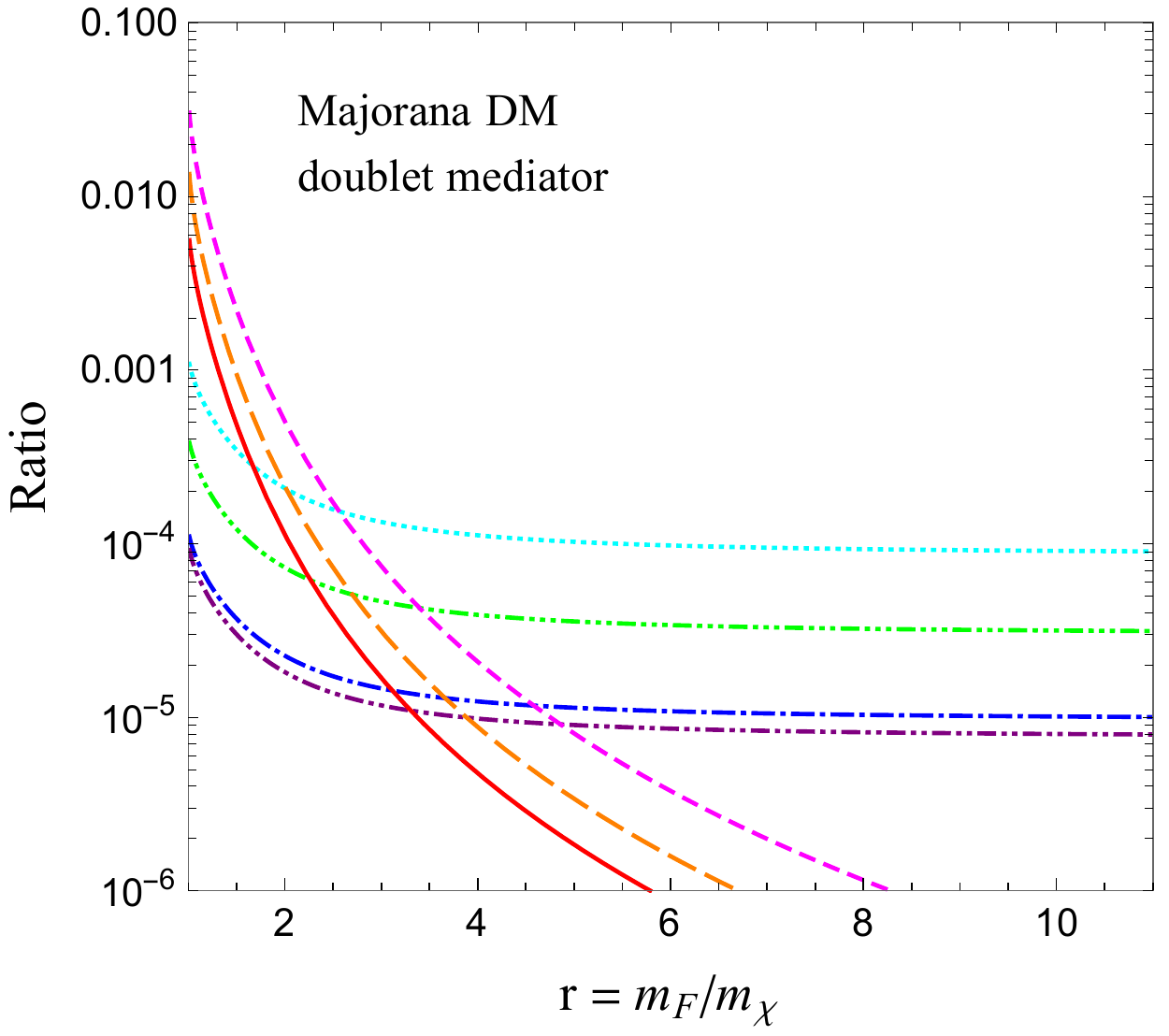,width=0.45\textwidth}}\vspace{.3cm}
{\epsfig{figure=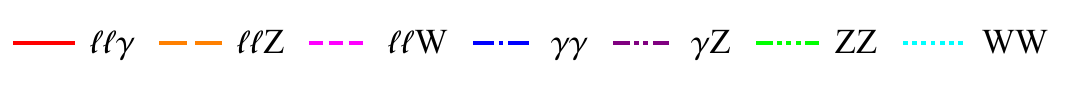,width=0.75\textwidth}}
\caption{\label{fig;AnnRatio}
Ratio of cross section $\VEV{\sigma v}_A / \VEV{\sigma v}_{\lp\bar{\lp}}$ 
in various annihilation channels as a function of $r$, 
in the complex (top-left), real (top-right), Dirac (bottom-left) and Majorana (bottom-right) DM models. 
We assume the $SU(2)_L$ doublet mediators and a muon-philic coupling. 
The DM mass is fixed at $500$\,GeV in all cases and 
DM velocity is set to thermal averaged values at freeze-out: 
$\VEV{v^2} \simeq 0.24$ and $\VEV{v^4}\simeq 0.1$.
The processes in associated with W-boson are absent in the models with the $SU(2)_L$ singlet mediator.}
\end{figure}

Figure~\ref{fig;AnnRatio} shows ratios of thermal averaged cross sections, 
$\VEV{\sigma v}_{\lp\lp V}$ and $\VEV{\sigma v}_{VV^\prime}$, 
to that of the tree-level two-body annihilations $\VEV{\sigma v}_{\lp\lp}$, 
with $m_{DM}=500$\,GeV, the $SU(2)_L$ doublet mediators and 
a muon-philic coupling ($\la_L^\mu\neq0$, $\la_L^e=\la_L^\tau=0$). 
Here, $\VEV{\sigma v}_{\lp\lp(V)} = 
\VEV{\sigma v}_{\mu\bar{\mu}(V)} + \VEV{\sigma v}_{\nu_\mu\bar{\nu}_\mu(V)}$
and 
$\VEV{\sigma v}_{\lp\bar{\lp}W} = 
\VEV{\sigma v}_{\mu\bar{\nu}_\mu W^+} + \VEV{\sigma v}_{\bar{\mu} \nu_\mu W^-}$. 
The velocity suppressed processes are evaluated at the freeze-out temperature: 
$\langle v^2 \rangle \simeq 0.24$ and $\langle v^4 \rangle \simeq 0.1$.
The plots are independent of the value of the coupling $\la_L^\mu$, 
since it cancels out among the numerator and denominator. 
In the real, complex and Majorana cases, these are also independent of the choice of the lepton flavor 
as far as we assume only one of the couplings to be non-vanishing. 
This is not the case for the Dirac DM, however. 
There is a slight flavor dependence in $\VEV{\sigma v}_{\lp\bar{\lp}\gamma}$ 
due to a collinear divergence in the limit $m_{\lp}\to0$. 
This will be discussed later more concretely. 
Phenomenologically important effects in these processes are commented in the following.

\subsubsection*{- DM pair annihilation into $\lp_i \bar{\lp}_j$}
Figure~\ref{fig;ann} (left) shows the DM pair annihilation into a pair of SM leptons 
$\lp_i \bar{\lp}_j$, where $\lp_i=e_i,\nu_i$.   
In the case ($\rnum{1}$)~(weak doublet mediator), the $s$-wave contributions are given by 
\begin{align}
 a_{\lp_i\lp_j}= \frac{|\la_L^i\la_L^j|^2}{32\pi m_{DM}^2(1+r^2)^2} \times 
\begin{cases}
 4(\eps_i+\eps_j) & \quad  \mathrm{real~scalar} \\
\eps_i+\eps_j  & \quad  \mathrm{complex~scalar}  \\
(\eps_i+\eps_j)/2    & \quad \mathrm{Majorana~fermion}  \\
1                             & \quad \mathrm{Dirac~fermion} \\
\end{cases}, 
\end{align}
where $m_{DM}$ denotes the DM mass: $m_{DM}$ is identical to $m_X$ in the scalar DM model and $m_{\chi}$ in the fermion DM model.
$r$ and $\eps_i$ are defined as $r \equiv m_L/m_{X}$ and $\eps_i=m_i^2/m_{X}^2$ in the scalar DM model, and $r \equiv m_{\wt{L}}/m_{\chi}$ and $\eps_i=m_i^2/m_{\chi}^2$ in the fermion DM model, with $m_i$ being mass of the lepton $\lp_i$. 
Here the sub-leading order in $\eps_i$ are neglected. 
In the models except for the Dirac DM model, the $s$-wave contribution is helicity suppressed ($\eps_i\ll1$). 
In the real DM, the $p$-wave contribution $ b_{\lp_i \lp_j}$ is also helicity suppressed 
and the leading contribution is the $d$-wave, so that 
the annihilation cross section is very suppressed 
and other processes discussed below are relatively important. 
The expressions for the case ($\rnum{2}$) is obtained 
by replacing $\la^i_L \to\la^{i*}_R$, $m_L \to m_E$ and $m_{\wt{L}}\to m_{\wt{E}}$.  
The full analytical expressions of the expansion coefficients are in Appendix \ref{app:analytics}.

\subsubsection*{- DM pair annihilation into $\lp_i \bar{\lp}_jV$} 
Figure~\ref{fig;ann} (middle) shows a diagram for the DM annihilation into a pair of leptons, 
accompanied with a gauge boson $V$, where $V=\gamma, Z, W$. 
In all types of DM, these processes have $s$-wave contributions without the helicity suppressions, 
while these are suppressed by a gauge coupling strength and three-body phase space 
by a factor $\sim\alpha/\pi$. 
Using the parametrization of Eq.~(\ref{eq;cs}), the relative importance at freeze-out is evaluated by 
\begin{equation}
\frac{\VEV{\sigma v}_{\lp\bar{\lp}V}}{\VEV{\sigma}_{\lp\bar{\lp}}} 
\sim \frac{\alpha/\pi}{\VEV{v^n}} ,
\end{equation}
where $n$ means the dominant partial wave of $\VEV{\sigma v}_{\lp\bar{\lp}}$ in each model. 
As seen in Fig.~\ref{fig;AnnRatio}, this ratio is ${\cal O}$(1--0.1) in the real DM model ($n=4$) 
depending on $r$, while no more than 0.1 in the other three models. 
We will therefore include these processes in calculating DM thermal abundance only in the real DM model. 
The concrete expressions of the cross sections as well as their squared amplitudes 
are listed in Appendix~\ref{app:analytics}. 

\begin{figure}[!t]
\centering 
{\epsfig{figure=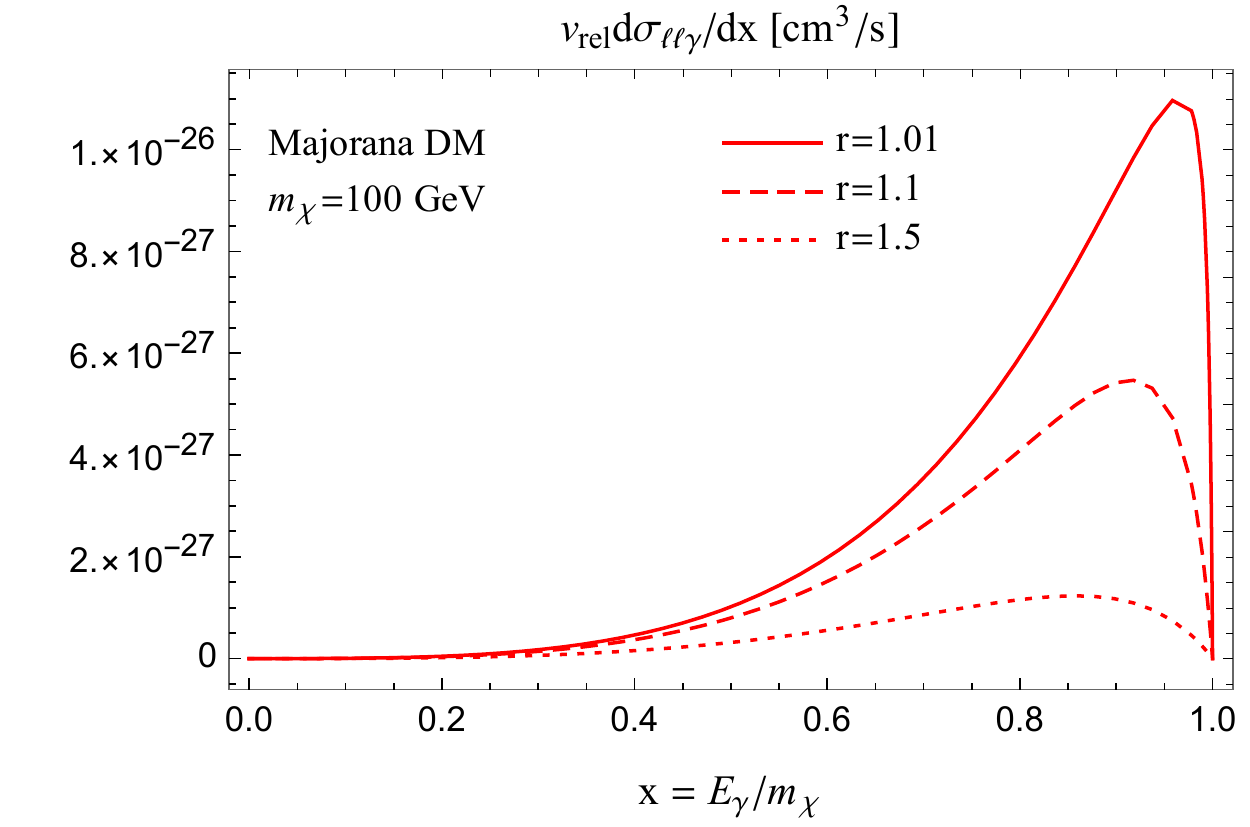 ,width=0.65\textwidth}}
\caption{Photon spectrum from $\chi\bar{\chi} \to \lp \bar{\lp}\gamma$ in the Majorana DM model. 
Here, the DM mass is $m_\chi = 100$\,GeV and $r = m_{\wt{L}}/m_\chi$ 
and the portal Yukawa coupling is unity. 
The spectrum for the scalar DM is completely the same.
}
\label{fig;spectrum}
\end{figure}

In general, the three-body processes are superposition of the final state radiation (FSR) from on-shell leptons, and an emission from the off-shell intermediate state, the namely virtual internal bremsstrahlung (VIB). 
The differential cross section of the FSR is related to the two-body cross section~\cite{Birkedal:2005ep}, 
\begin{equation}
\frac{d(\sigma v)_{\lp\ol{\lp}\gamma}^{\rm FSR}}{dx} \approx (\sigma v)_{\lp\ol{\lp}} \, \frac{Q_\lp^2 \alpha}{\pi} \frac{(1-x)^2+1}{x} \log\left(\frac{4m_{DM}^2(1-x)}{m_\lp^2}\right) ,
\label{eq;FSR}
\end{equation}
independently of the types of DM. 
Here, $x$ is defined as $x=2E_\gamma/\sqrt s$ with the photon energy, $E_\gamma$. 
If $(\sigma v)_{\lp\ol{\lp}}$ is helicity suppressed, the FSR contribution is also suppressed and negligible. 
The real, complex and Majorana DM models meet this condition. 
In these models, the three-body processes are dominated by the VIB, and exhibit a sharp spectrum 
of emitted vector bosons $V$ around $E_V = m_{X(\chi)}$
if the DM and mediator masses are nearly degenerate~\cite{Bringmann:2012vr,Garny:2013ama,Giacchino:2013bta,Toma:2013bka,Giacchino:2014moa,Ibarra:2014qma}. 
Figure~\ref{fig;spectrum} shows a photon spectrum 
from the three-body annihilation $\chi\chi\to\lp_i\ol{\lp}_j\gamma$ in the Majorana DM model. 
The spectrum in the scalar DM models is the same as this.  
An emitted photon in the VIB process fairly looks like 
a monochromatic spectrum within detector resolution. 
This sharp spectrum will be a distinctive signal of these types of DM models  
as will be discussed in Sec.~\ref{sec:indirect}. 

In the Dirac model, $(\sigma v)_{\lp\bar{\lp}}$ is not helicity suppressed, 
then the FSR entirely dominates the three-body process. 
It follows from Eq.~(\ref{eq;FSR}) that in the massless lepton limit, 
the FSR cross section is not well-defined, so that we have to keep the lepton mass finite to evaluate it. 
This leads to the slight flavor dependence of 
$\VEV{\sigma v}_{\lp\bar{\lp}\gamma}/\VEV{\sigma v}_{\lp\bar{\lp}}$, as mentioned above. 
Moreover, when we integrate over $x$, we encounter a divergence at the infrared edge $x=0$. 
This can be eliminated by taking radiative corrections to the two-body processes into account. 
In Fig.~\ref{fig;AnnRatio} (bottom-left), to evaluate the size of the three-body process, 
we just regularized the infrared divergence by integrating over the range $0.1<x<1$. 
This corresponds to introducing a sharp infrared cutoff at $E_\gamma = 0.1 m_{DM}$. 
We have confirmed that the numerical result obtained in that way is smaller than 
the so-called double logarithm approximation, 
$(\sigma v)_{\lp\bar{\lp}} \, (\alpha/\pi) \log(4m_{DM}^2/m_\lp^2) \log(m_{DM}^2/\Lambda_{IR}^2)$, 
by a factor $\sim1.5$.

\subsubsection*{- DM pair annihilation into $VV^\prime$}

The DM can pair-annihilate into two gauge bosons via loop diagrams 
shown in the Fig.~\ref{fig;ann} (right).   
The possible final states are $VV^\prime = \gamma\gamma,\,\gamma Z,\,ZZ,\,W^+W^-$. 
We do not consider the annihilation into $hZ,\,hh$, 
since these are further suppressed by the small Higgs Yukawa couplings of the charged leptons. 
These processes cannot be leading contributions at the freeze-out in any case 
due to the suppression via the gauge couplings and loop factor. 

Nevertheless, $XX \to \gamma\gamma,\gamma Z$ will be significant at indirect detection 
of the real scalar DM. 
For large $r = m_{E_1}/m_X$, the cross section is scaled as 
\begin{align}
(\sigma v)_{VV^\prime} \propto \frac{1}{r^4},   
\quad 
\mathrm{while}  
\quad 
(\sigma v)_{\lp\lp (V)} \propto \frac{1}{r^8}.   
\end{align}
Hence, the loop annihilation $XX \to \gamma\gamma,\gamma Z$ 
can be a sizable fraction of the total annihilation cross section for large $r$. 
For example, $(\sigma v)_{\gamma\gamma}/(\sigma v)_{\lp \ol{\lp}} \sim 0.01$ is realized for $r=3$.   
This small fraction is irrelevant to the DM abundance, 
but it can be crucial in gamma-ray searches 
since produced photons in $XX \to \gamma\gamma,\gamma Z$ are monotonic
and can be strongly constrained.

\subsection{Relic density}

In the thermal freeze-out scenario, it is assumed that DM abundance is produced in the thermal bath. 
The produced number density is determined by the Boltzmann equation, 
\begin{equation}
\frac{dn_{DM}}{dt} + 3 H n_{DM} = - \VEV{\sigma v}_{\rm eff} \left[ n_{DM}^2 - (n_{DM}^{\rm eq})^2 \right] ,
\end{equation}
with $H$ the Hubble rate and $n_{DM}^{\rm eq}$ the equilibrium density of DM. 
Here, $\VEV{\sigma v}_{\rm eff}$ is the effective annihilation cross section of DM, and is expressed in terms of DM pair annihilation and coannihilation. 
As a concrete example, in the real scalar DM model, it is given by 
\begin{equation}
\VEV{\sigma v}_{\rm eff} \simeq \VEV{\sigma v} + \left( \VEV{\sigma_{XL} v} +\VEV{\sigma_{X\ol{L}} v} \right) \, e^{-\frac{\Delta m}{T}} + \VEV{\sigma_{L\ol{L}} v} \, e^{-2\frac{\Delta m}{T}} ,
\label{eq;sigmaveff}
\end{equation}
where $\Delta m = m_L - m_X$. 
$\VEV{\sigma v} = \VEV{\sigma v}_{\lp\ol{\lp}} + \VEV{\sigma v}_{\lp\ol{\lp} V} + \VEV{\sigma v}_{VV^\prime}$ represents thermal average of DM pair annihilation cross section mentioned in the previous section, while $\langle\sigma_{XL(X\ol{L})} v\rangle$ and $\VEV{\sigma_{L\ol{L}} v}$ those of coannihilation cross section whose initial state is $XL$ ($X\ol{L}$) and $L\ol{L}$, respectively. 
Since the DM number density is frozen at $T_f\simeq m_{DM}/20$, the effective cross section $\VEV{\sigma v}_{\rm eff}$ at that temperature is crucial. It follows from Eq.(\ref{eq;sigmaveff}) that coannihilation processes are important when the exponential suppressions are not strong. 
Naive estimate suggests 
\begin{equation}
e^{-\frac{\Delta m}{T_f}} = e^{-\frac{m_X}{T} \frac{\Delta m}{m_X}} \simeq e^{- 20 \frac{\Delta m}{m_X}} , 
\end{equation}
is not too small, i.e. $\Delta m/m_X \lesssim {\cal O}(0.1)$ is a necessary condition that coannihilation is effective.  
Indeed, as we will see in Sec.~\ref{sec;resutls}, coannihilation processes help to deplete DM thermal abundance down to the observed value with a smaller Yukawa coupling in such a mass region. 
In our analysis, to take all coannihilation processes into account, we employ \texttt{micromegas$\_$4.3.5}~\cite{Belanger:2014vza} to numerically solve the Boltzmann equation. 
In the real scalar DM, we also consider the higher order processes, such as three-body process and loop processes, by including only their $s$-wave contributions. 
In the other types of DM, we ignore those contributions, since these are no more than 10\,\% of the leading process as seen in Fig.~\ref{fig;AnnRatio}.

\subsection{Indirect detection}
\label{sec:indirect}
Indirect dark matter searches look for cosmic ray fluxes, 
such as gamma ray, anti-proton, positron and neutrino, 
originated from DM annihilation on top of astrophysical backgrounds. 
These are good complementary tools of direct detections to probe DM, 
although these suffer from large systematic uncertainties of astrophysical contributions. 
In this paper, we will not perform any new data analysis, 
and will simply rescale the sensitivity curves derived in the literature. 
Here, we shall discuss the constraints from the indirect searches 
on the lepton portal models. 
As discussed below, 
the meaningful constraints are obtained from the DM annihilation 
into $\lp\ol{\lp}$ or $\lp\ol{\lp}\gamma$.  
In these processes,  
the constraint on the complex scalar DM is very similar to that on the Majorana DM, 
because the squared amplitudes are the same in these two types\footnote{See Appendix~\ref{app:analytics} for the explicit forms.}. 
Thus, we only mention real scalar, Majorana and Dirac fermion DM, 
to avoid repeating the same comments. 

The most prominent target in cosmic ray searches for DM annihilation 
is a spectral feature, such as gamma line or VIB photon. 
Such a sharp spectrum can be well disentangled from uncertain astrophysical backgrounds, 
since attributing it to one astrophysical process is difficult in general.   
The search for a spectral feature is often a unique way to observe DM signatures from the sky,  
particularly when the tree-level two-body annihilation is suppressed. 


The spectral features in the lepton portal models have been examined in the literature. 
It has been found that the sensitivity can be improved when we exploit the photon spectrum instead of the continuum photon flux. 
In the Majorana DM, for example, dedicated searches for the spectral features set orders of magnitude stronger upper bounds on $\VEV{\sigma v}_{\lp\bar{\lp}\gamma}$ 
than the bounds from the continuum gamma-ray observation 
of dwarf spheroidal (dSph) galaxies~\cite{Bringmann:2012vr}. 
The study of the spectral feature is then extended to the pair annihilation into $\gamma\gamma$, showing the upper limits on a combined annihilation cross section: 
$\VEV{\sigma v}_{\lp\bar{\lp}\gamma}+2\VEV{\sigma v}_{\gamma\gamma} \lesssim 10^{-26}$--$10^{-27}$ cm$^3$/s for DM mass ranging from 40 GeV to 10 TeV~\cite{Garny:2013ama}. 
Note that EW gauge invariance requires the existence of weak VIB processes, such as $\lp\bar{\lp}Z$, 
that exhibit a spectral feature of an anti-proton flux from decays and hadronization of the weak bosons,
similarly to that of photon flux.  
The impacts of the weak VIB emission on the PAMELLA anti-proton search are studied in Majorana DM in \cite{Garny:2011cj,Garny:2011ii}. 
Unfortunately, these cosmic-ray signatures of the Majorana DM cannot be observed 
even at future telescopes unless a boost factor is larger than $\order{10}$.

The search for the spectral features can impose good complementary bounds on the real scalar DM. 
As will be discussed in Sec.~\ref{sec:direct}, 
this type of DM is almost free from the direct detection bound. 
On the other hand, 
the fraction of annihilation into $\lp\bar{\lp}\gamma$ and $\gamma\gamma$ 
in the total annihilation is sizable because of the strong suppression in the two-body process $XX\to \lp\bar{\lp}$.  
This suggests that more fluxes are predicted than in the Majorana DM, 
opening up a possibility to discover DM signatures in gamma-ray spectrum. 
It is shown in Ref.~\cite{Ibarra:2014qma} that 
the future GAMMA-400~\cite{Galper:2012fp} and CTA~\cite{Consortium:2010bc} experiments 
can probe the real scalar DM where $m_X\geq100$~GeV and $m_{{E}_1}/m_X \geq 1.2$. 
The sensitivity prospect will be shown in Fig.~\ref{fig;plots1} later.

For the Dirac DM, the spectral feature will not be applicable, 
since the gamma-ray flux is dominated by the smooth FSR and a continuum secondary gamma-ray. 
The latter originates from the decay and fragmentation of the SM particles produced by DM annihilation. 
Gamma-ray observation of dwarf spheroidal (dSph) galaxies has therefore the best sensitivity.  
The strongest limit is set on the tau-philic DM. 
In that case, the lower limit on $\VEV{\sigma v}_{\tau\bar{\tau}}$ lies below the canonical thermal relic cross section $\sigma_0 \sim 3\times10^{-26}$~[cm$^3$/s] for $m_\chi\lesssim100$~GeV. 
In the muon-philic or electron-philic DM models with $\VEV{\sigma v}_{\mu\ol{\mu},e\ol{e}} = \sigma_0$, the lower bound on the DM mass is around 10~GeV~\cite{Ackermann:2015zua}. 
These bounds can be simply applied to the Dirac DM,  
since the unsuppressed two-body process $\chi\bar{\chi} \to \lp\bar{\lp}$ 
is mainly responsible for the thermal freeze-out. 
In the other DM types, 
the limit from the observation of dSph galaxies is much weaker than 
those from the gamma line limit~\cite{Garny:2013ama}.

We briefly comment on other possible constraints. 
Of particular importance is constraint from positron flux observations. 
In \cite{Bergstrom:2013jra,Cavasonza:2016qem}, 
upper limits on annihilation cross section into leptonic final states 
are derived based on the AMS-02 data of the positron fraction~\cite{Aguilar:2013qda}.  
The most stringent limit is posed on $e^+e^-$ channel, 
since the positron spectrum is so sharp even after propagation in galactic space 
that the spectral search is applicable. 
The resulting 95\% C.L. lower limit on DM mass is 100~GeV, 
when the thermal relic annihilation cross section into $e^+e^-$ is assumed~\cite{Cavasonza:2016qem}.
The constraint on $\mu^+\mu^-$ and $\tau^+\tau^-$ channels (or associated VIB processes) is much weaker, since the positron spectrum is broader and it becomes difficult to disentangle the DM signature from smooth astrophysical background.  
For further detail of the analysis, see Refs.~\cite{Bergstrom:2013jra,Cavasonza:2016qem}. 
In the lepton portal  models, the positron bound is relevant only to the Dirac DM 
and gives the limit on $m_X\lesssim100$\,GeV for $e^+e^-$ channel. 
Note that the other three types predict a sharp positron spectrum 
in $e\bar{e}\gamma$ process similarly to the VIB photon, 
nevertheless the cross section is too small to bring the positron constraints into play. 

The lepton portal DM models predict neutrino flux as well. 
If the leptonic mediator is a weak doublet, the cross section can be sizable 
since tree-level annihilation into $\nu\ol{\nu}$ is possible. 
The produced flux can be detected at neutrino telescopes. 
So far, the observations at neutrino telescopes have found no significant excess of neutrinos over the background. 
This is interpreted as an upper bound on annihilation cross section. 
For instance, the ANTARES neutrino telescope has searched for self-annihilation of DM in the center of the Milky Way, and reported bounds on the five representative annihilation channels~\cite{Albert:2016emp,ANTARES:2019svn}. Of these, $\tau^+ \tau^-$, $\mu^+ \mu^-$, and $\nu\ol{\nu}$ are relevant to us. 
Using the latest 11 years data, and assuming the NFW halo profile and 100~\% branching ratio, the upper limit on $\VEV{\sigma v}_{\nu\ol{\nu}}$ is $10^{-23}$--$10^{-24}$~[cm$^3$/s] for DM mass ranging from 50~GeV to 100~TeV~\cite{ANTARES:2019svn}. 
The limits on $\tau^+ \tau^-$ and $\mu^+ \mu^-$ channels are weaker than the former under the same assumption. 
The searches at IceCube neutrino observatory set similar upper bounds on them~\cite{Aartsen:2017ulx}. 
In our models, $\VEV{\sigma v}_{\nu\ol{\nu}}$ is less than $10^{-26}$~[cm$^3$/s] in every setup, so that the constraint from the neutrino flux has no impact on the models.

\subsection{Direct detection}
\label{sec:direct}
\begin{figure}[t]
\centering 
\includegraphics[viewport=180 530 400 760, clip=true, scale=0.55]{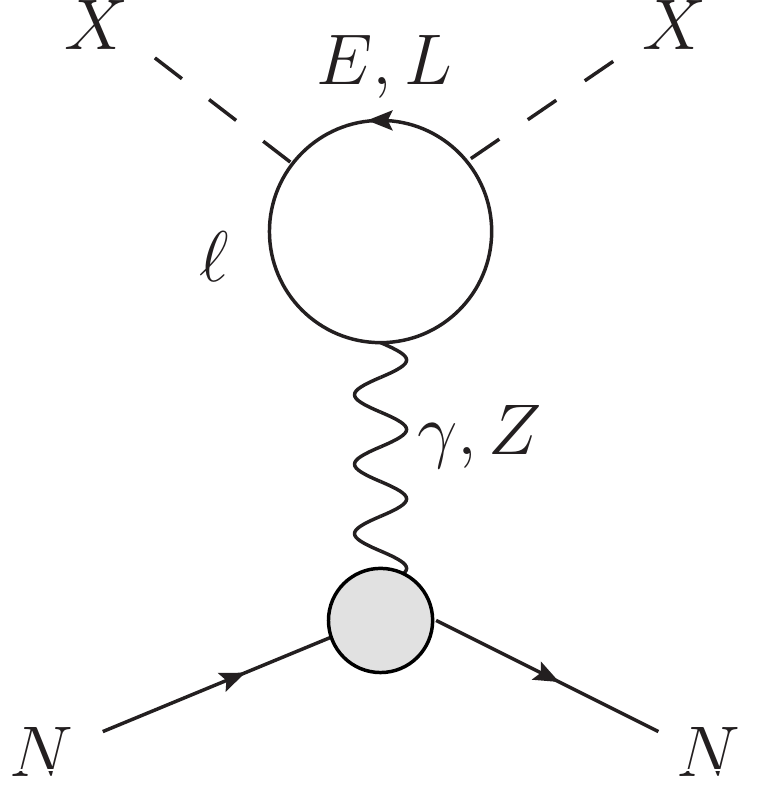}
\hspace{1cm}
\includegraphics[viewport=160 530 410 760, clip=true, scale=0.55]{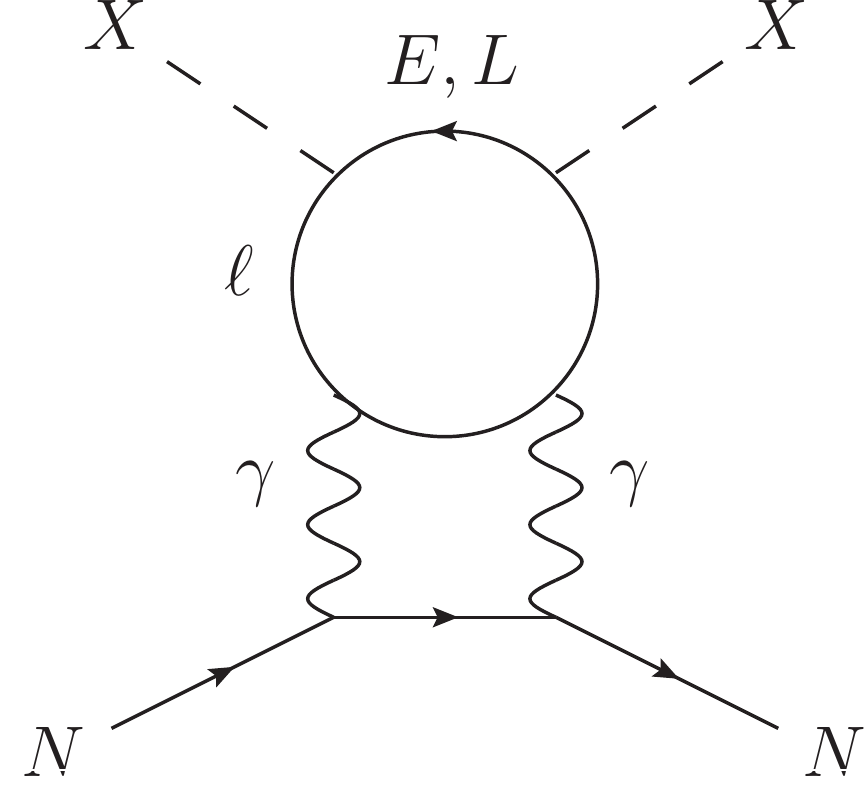}
\caption{\label{fig;penguin}
Example diagrams for DM-nucleon scattering at direct detection: (left) photon and $Z$ penguin contributions for the complex DM; (right) leading 2-loop contribution for the real DM. There are similar penguin contributions in the fermion DM.}
\end{figure}

In the lepton portal DM models, 
there is no tree-level scattering between the DM and a nucleus, 
so that the leading contribution arises at loop levels. 
As is well known, 
primary contribution is photon exchanging in all types of the DM models. 
The diagrams are shown in Fig.~\ref{fig;penguin}.
The relevant DM-photon effective interactions are dependent on the DM types and masses. 
There are also similar $Z$ exchanging contributions in which photons in the leading diagrams 
are replaced by $Z$ bosons. 
These are, however, suppressed by lepton masses and therefore sub-leading. 
In the following, we will summarize typical features of the leading process in each type of DM.  
See Appendix~\ref{app:analytics}
for the full analytical expressions for the photon and $Z$ contributions. 

In the complex scalar DM model,  
the photon penguin diagram shown in Fig.~\ref{fig;penguin} (left) is the leading contribution.
The induced DM-nucleon effective operator is given by 
\begin{equation}
{\cal L}_{\rm eff}^S \supset C_{V,N} (iX^\dagger \overleftrightarrow{\partial_\mu} X)(\ov{N} \gamma^\mu N) ,
\end{equation}
where $N=p,n$. 
Here, we define $\phi_2 \overleftrightarrow{\partial_\mu} 
\phi_1 \equiv \phi_2 \partial_\mu \phi_1 - \partial_\mu \phi_2 \cdot \phi_1$. 
In the limit $m_{L} \gg m_i, m_X$,  the coefficient $C_{V,N}$ is dominantly given by the photon-penguin diagram, $C^\gamma_{V,N}$,
\begin{align}
C^{\gamma}_{V,N} & \simeq \frac{ \alpha Q_f Q_N }{ 12 \pi m_L^2 } 
\sum_i |\la_L^i|^2 \left( \frac{3}{2} + \log \frac{m_i^2}{m_L^2} \right),  
\label{eq;photon}
\end{align}
in the case ($\rnum{1}$). 
The expression for the case ($\rnum{2}$) is obtained by formally replacing 
$\la_L^i \to \la_R^{i*}$ and $m_L \to m_E $. 
When the vectorlike lepton is heavy, 
the cross section can be considerably enhanced by the logarithmic term, 
leading to strong limits from direct detection experiments.

In the real scalar DM model, there is no penguin-type contribution.   
The leading contribution arises at two-loop level via two photon-exchanging~\cite{Kopp:2009et}.    
The diagram is shown in Fig.~\ref{fig;penguin} (right). 
The DM-nucleon scattering cross section is so suppressed that  
there are no significant constraints from the direct detection. 
When DM couples to the electron, a DM-electron scattering is induced at tree level. 
 This scattering is limited in light mass region typically $m_X \lesssim1$~GeV. In this paper, however, we focus only on $m_X\gtrsim100$~GeV, since such a light DM scenario would be strictly constrained by, e.g. the LEP experiment in our model.
We note that the recent study of limits on DM-electron scattering includes Refs.~\cite{Essig:2017kqs,Pandey:2018esq,Baxter:2019pnz,Agnes:2018oej}.

In the fermion DM models, there are multi-pole interactions, 
\begin{equation}
{\cal L}_{\rm multipole}^F = b_\chi \ol{\chi} \gamma^\mu \chi \partial^\nu F_{\mu\nu} 
+ \frac{\mu_\chi}{2} \ol{\chi} \sigma^{\mu\nu} \chi F_{\mu\nu} 
+ a_\chi \ol{\chi} \gamma^\mu \gamma^5 \chi \partial^\nu F_{\mu\nu} 
+ i \frac{d_\chi}{2} \ol{\chi} \sigma^{\mu\nu} \gamma^5 \chi F_{\mu\nu} ,
\label{eq:multipole}
\end{equation}
in addition to contact-type DM-nucleon effective interaction,
\begin{equation}
{\cal L}_{\rm eff}^F = \sum_{N=p,n} \left( C_{S,N} \bar{\chi}\chi\ol{N}N + C_{V,N} \ol{\chi}\gamma^\mu\chi\ol{N}\gamma_\mu N \right) .
\end{equation}
The differential cross section for elastic DM scattering off a nucleus is expressed in terms of these Wilson coefficients:
\begin{align}
\frac{d\sigma}{dE_R} =&\ \frac{\alpha Z^2}{v^2} 
\left[ \mu_\chi^2 \left( \frac{v^2}{E_R} - \frac{m_N}{2\mu_\mathrm{red}^2} \right) 
       +d_\chi^2 \left( \frac{1}{E_R}-\frac{1}{m_\chi} \right) \right] \abs{F(E_R)}^2 
      + \frac{m_N f_A^2}{2\pi v^2}  \abs{F(E_R)}^2 \notag \\ 
&+\frac{\alpha Z^2}{v^2} a_\chi^2 \left[ 2 m_N v^2-\frac{(m_N+m_\chi)^2}{m_\chi^2} E_R\right] \abs{F(E_R)}^2 
 \notag \\
&+\frac{m_N \mu_A^2}{2\pi v^2} \left(2\mu_\chi^2+d_\chi^2v^2+4m_NE_Ra_\chi^2\right) 
\frac{J_A+1}{3J_A} \abs{F_{\rm spin}(E_R)}^2 , 
\label{eq;dsigma}
\end{align}
where we assumed the Dirac DM. 
Here, $\mu_\mathrm{red}=m_\chi m_N/(m_\chi+m_N)$ is the reduced mass of DM and a nucleus, 
$f_A=Z \left( C_{S,p} + C_{V,p} - eb_\chi - e\mu_\chi/(2m_\chi) \right) 
+ (A-Z) \left( C_{S,n} + C_{V,n} \right)$ 
with $Z$ an atomic number of a nucleus, and 
$m_N$, $J_A$ and $\mu_A$ nuclear mass, spin and magnetic moment, respectively. 
In our models, $C_{S,N}$ and $C_{V,N}$ are induced by $Z$ and Higgs penguin diagrams. 
We use the Helm form factor normalized with $F(0)=1$ for $F(E_R)$, 
and use a spin form factor with thin-shell approximation 
derived in \cite{Lewin:1995rx} for $F_\mathrm{spin}(E_R)$. 
The first two lines are the spin-independent contributions
and the third line is the spin-dependent one. 
For the spin-independent one, 
there are non-contact contributions which appear with $1/E_R$ and lead to infrared enhancement, 
while these are absent in the spin-dependent part.   
It should be noted that due to the non-contact contributions, 
we cannot simply refer to the exclusion curve reported in the experimental papers, 
in which the contact type interaction is assumed.  
The different dependencies on $E_R$ and $v$ from the contact ones should be taken into account, 
if the rate is affected by the non-contact contributions. 
The method to translate the null results at direct detection experiments into limits on the parameter space in our model is explained in Appendix~\ref{dipole}. 

In the Dirac DM model, interactions via the charge radius $b_\chi$, the magnetic dipole $\mu_\chi$, and electric dipole $d_\chi$ are particularly important, since these are not suppressed by the DM velocity $v\sim10^{-3}$ and nuclear recoil energy $E_R={\cal O}$(10~keV). 
The anapole $a_\chi$ is suppressed by $v$ or $E_R$, thus it has a negligible effect on the Dirac case. 
In the minimal setup, the electric dipole $d_\chi$ is also vanishing. 
It is CP-violating, while the photon penguin contribution in Fig.~\ref{fig;penguin} (left) is always proportional to $|\la_L^i|^2$ in the case (i), or $|\la_R^i|^2$ in the case (ii). 
Thus, CP-violating contribution is not generated in the minimal setup. 
If there are both double and singlet mediators, a non-vanishing $d_\chi$ proportional to $\mathrm{Im}(\la_L^i \la_R^i)$ appears. 
The asymptotic behaviors of $b_\chi$ and $\mu_\chi$, as $m_{\wt{L}} \to \infty$, are given by 
\begin{align}
b_\chi & \simeq \frac{eQ_f}{96\pi^2 m_{\wt{L}}^2} 
\sum_i |\la_L^i|^2 \left( \frac{3}{2} + \log\frac{m_i^2}{m_{\wt{L}}^2} \right) ,\\
\mu_{\chi} & \simeq - \frac{eQ_f m_\chi}{64\pi^2 m_{\wt{L}}^2} 
\sum_i |\la_L^i|^2,  
\end{align}
in the case ($\rnum{1}$). 
The expression of the case ($\rnum{2}$) is obtained by $\la_L^i \to \la_R^{i*}$ and $m_{\wt{L}}\to m_{\wt{E}}$. 
There is a logarithmic enhancement in $b_\chi$ as in the complex DM, 
while such an enhancement is absent in $\mu_\chi$. 
As pointed out in Ref.~\cite{Ibarra:2015fqa}, 
the charge radius operator gives dominant contribution to the cross section for $m_\chi\lesssim1$ TeV, 
while the magnetic dipole operator becomes dominant one for larger masses. 
Since the former is a dimension six operator and the latter is a dimension five, 
their asymptotic behaviors in large $m_\chi$ are scaled as $1/m_\chi^2$ and $1/m_\chi$, respectively. 
As a consequence, 
the magnetic dipole interaction remains more relevant than the charge radius operator, as $m_\chi\to\infty$.  

For the Majorana DM, only the anapole moment $a_\chi$ is non-vanishing in Eq.~(\ref{eq:multipole}) due to the Majorana condition, $\chi^c=\chi$. 
The differential cross section is obtained by 
setting $b_\chi=\mu_\chi=d_\chi=0$ and replacing $a_\chi \to 2a_\chi$ in Eq.~(\ref{eq;dsigma}).
The reason for the latter replacement is that 
there are twice as many possible contributing diagrams as the Dirac DM. 
The $Z$ and Higgs penguin contributions are also non-vanishing, 
but these are suppressed by lepton masses and thus negligible.  
The direct detection bounds on the Majorana case can be obtained 
by the same method as in the Dirac DM, 
although these are essentially very weak due to the suppressions 
by the DM velocity $v$ and nuclear recoil energy $E_R$.

\subsection{Current status}
\label{sec;resutls}

In this section, we summarize current limits on the minimal models. 
Free parameters in the models 
are the DM mass, the mediator mass and the portal Yukawa couplings $\la_{L(R)}^i$.  
In this paper, we assume that the Yukawa coupling to one generation of SM leptons is sizable  
and those to the other generations are negligible, to suppress the LFV processes. 
In this section, we show plots in the muon-philic case, $\la_{L(R)}^\mu \gg \la_{L(R)}^{e,\tau}$. 
The qualitative behavior will be similar to the ones in the electron- and tau-philic cases.
For simplicity, the Yukawa couplings to the muon $\la^\mu_{L,R}$ will be simply denoted by $\la_{L,R}$ in the following. 

\begin{figure}[ph]
\centering 
{\epsfig{figure=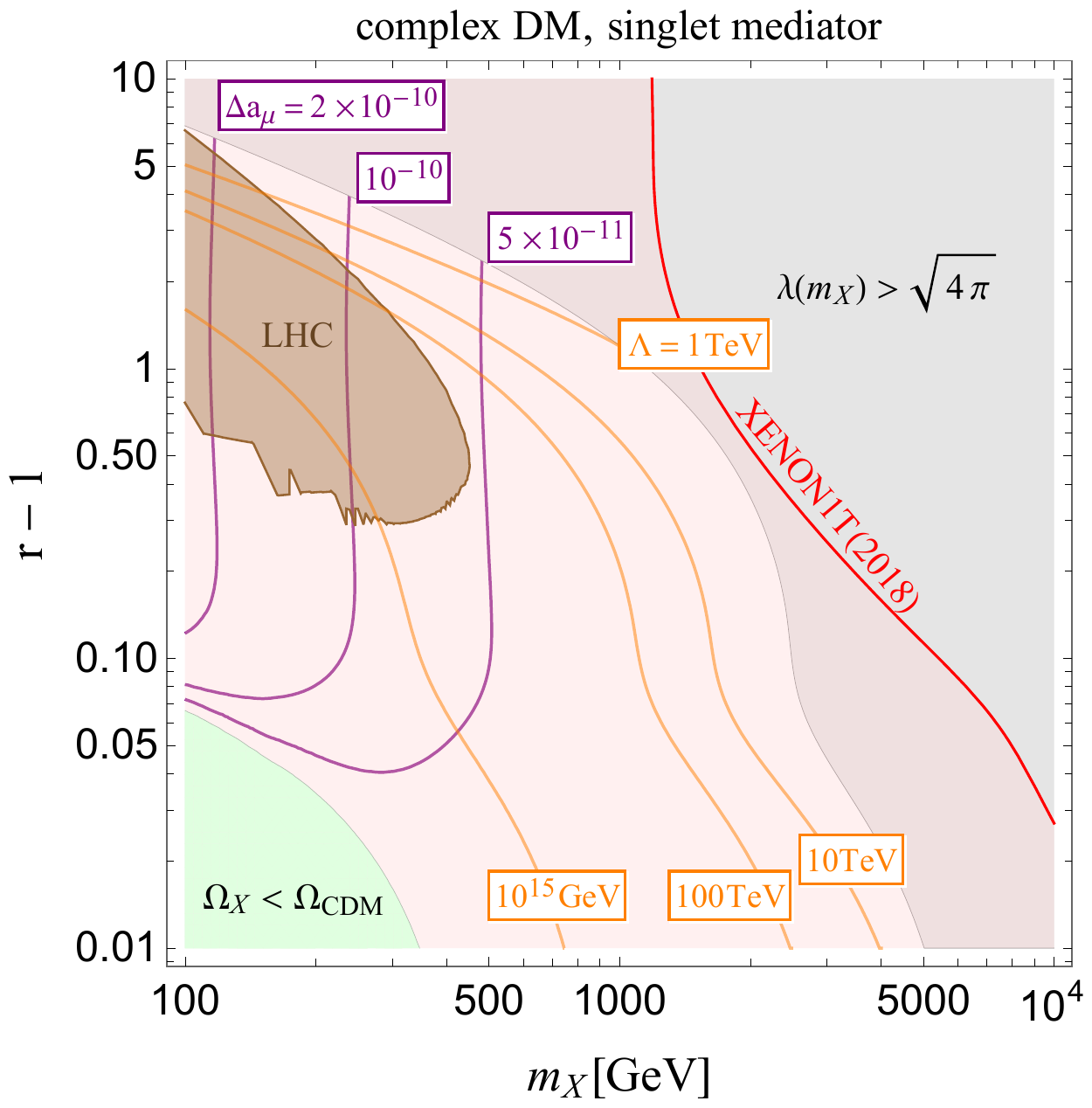,width=0.48\textwidth}}\hspace{.3cm}
{\epsfig{figure=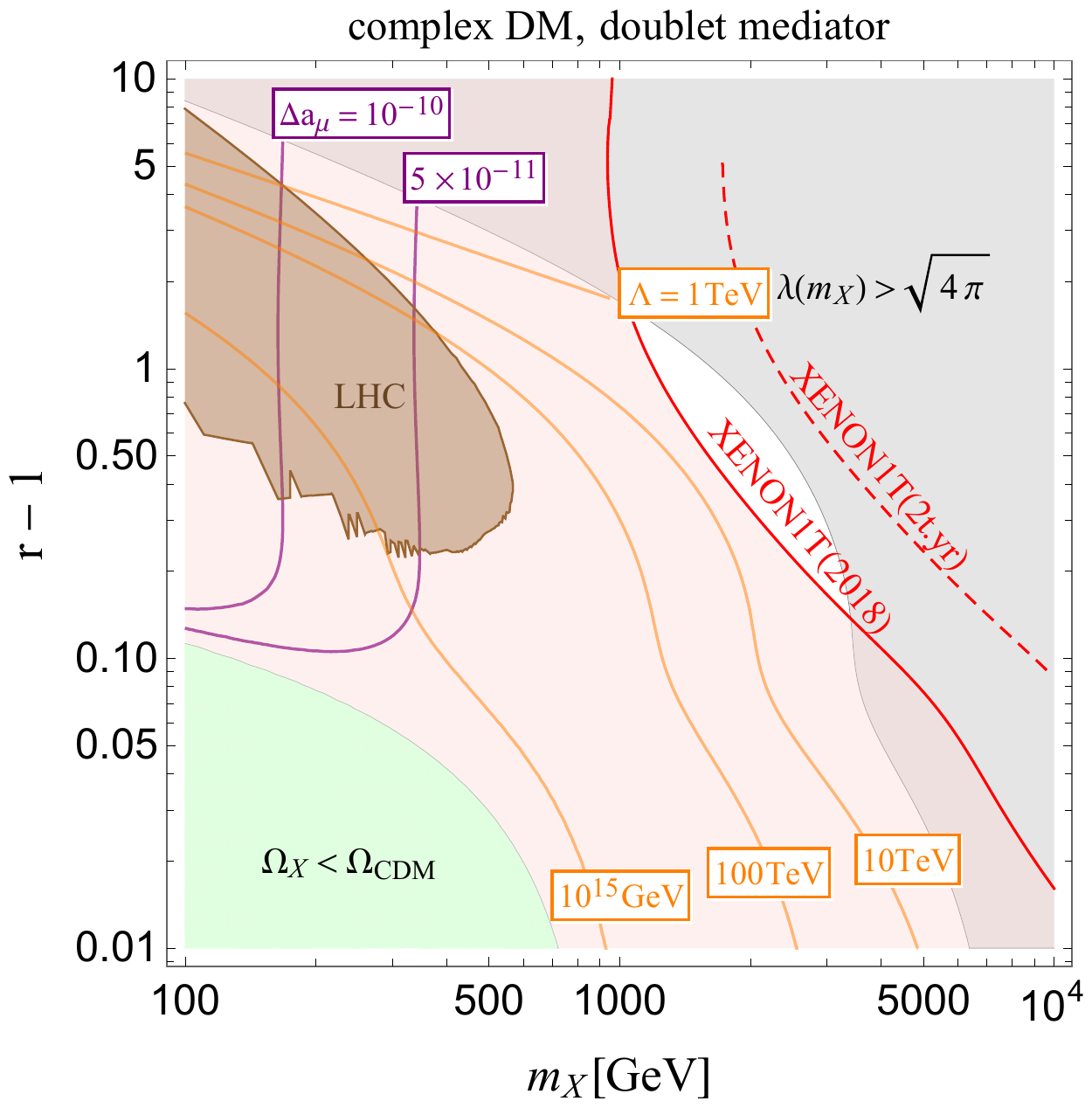,width=0.48\textwidth}}\vspace{0.5cm}
{\epsfig{figure=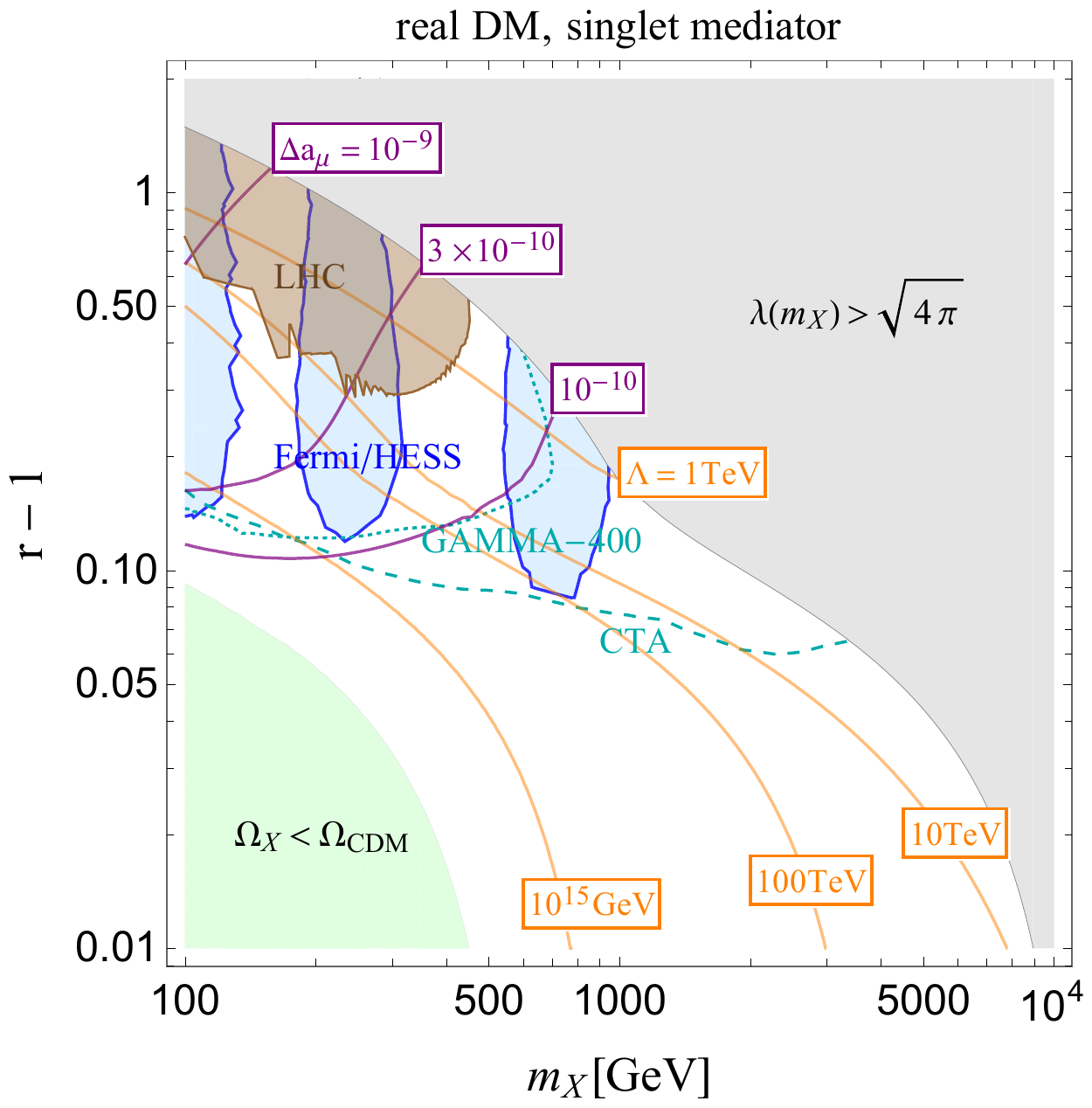,width=0.48\textwidth}}\hspace{.3cm}
{\epsfig{figure=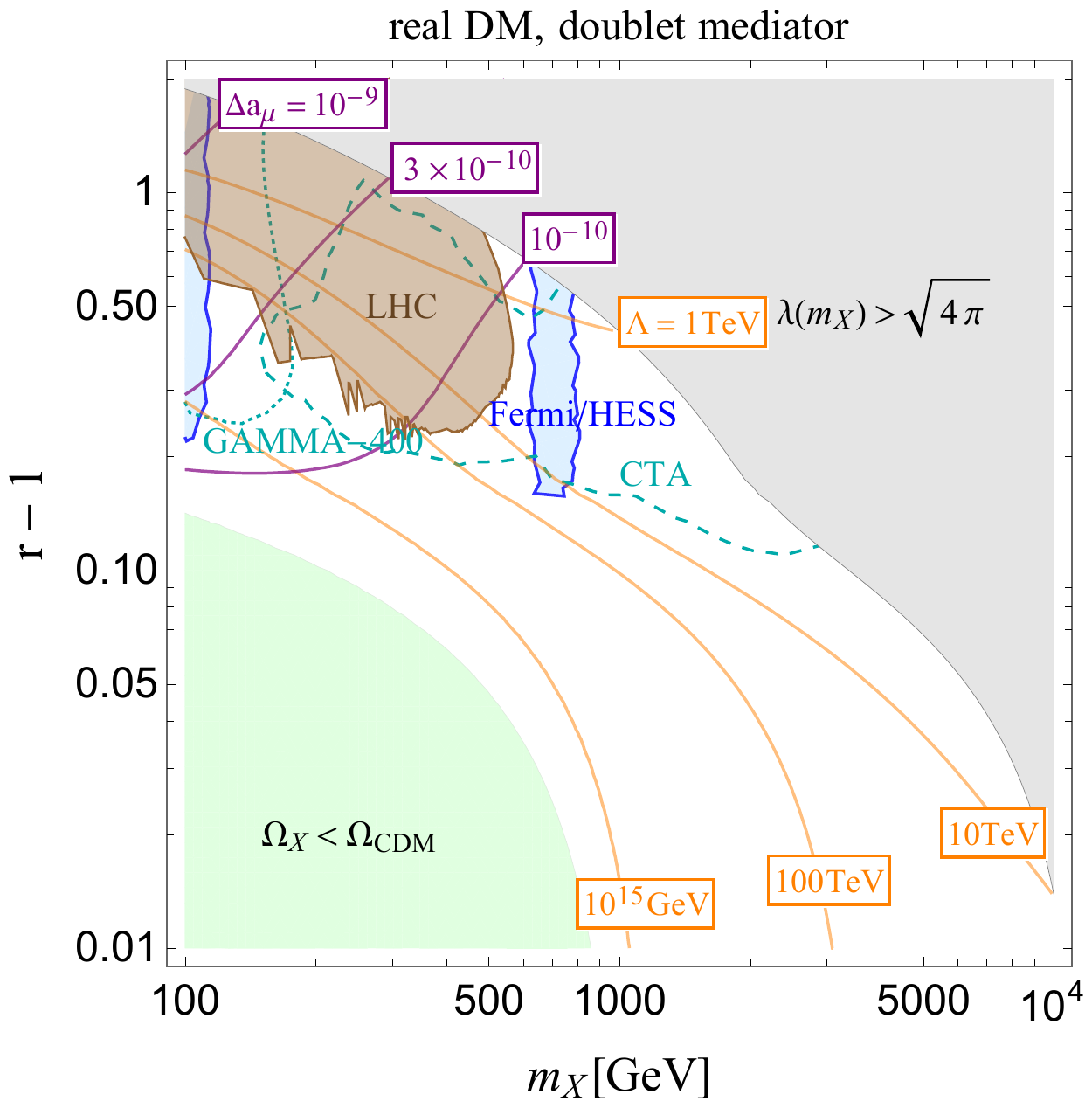,width=0.48\textwidth}}
\caption{\label{fig;plots1}
Plots for the complex (top panels) and real (bottom panels) scalar DM.  
The vectorlike lepton is a weak singlet (left panels) or doublet (right panels). 
Direct detections of the DM at the XENON1T~\cite{Aprile:2018dbl} excludes the red region. 
In the gray region on the upper corner, 
the DM thermal abundance requires a non-perturbatively large coupling 
$\la_{L(R)}>\sqrt{4\pi}$ below TeV scale. 
In the green region, 
coannihilation is too efficient to explain the thermal abundance observed at the Planck~\cite{Tanabashi:2018oca}. 
The orange contours stand for the Landau pole scales and the purple contours stand for $\Delta a_\mu$. 
The brown regions are excluded by the slepton searches at the LHC, 
where the limits are projected from Fig.~\ref{fig-LHClim}. 
The shaded blue regions in the bottom panels   
are excluded by the current gamma line observations 
at the Fermi-LAT~\cite{Ackermann:2013uma} and the HESS~\cite{Abramowski:2013ax}. 
The cyan lines show the future sensitivity 
at the GAMAM-400 (dotted)~\cite{Galper:2012fp} and the CTA (dashed)~\cite{Consortium:2010bc}.}
\end{figure}

Figures \ref{fig;plots1} and \ref{fig;plots2} show the limits from the direct and indirect detection 
as well as new physics contribution to the muon anomalous magnetic moment $\Delta a_\mu$ 
and a scale of Landau pole in the four types of DM models.   
The coupling $\la_L$ (or $\la_R$) is fixed to explain the central value of the observed DM abundance: $\Omega_{\rm CDM} h^2 = 0.1186\pm0.0020$~\cite{Tanabashi:2018oca}. 
We define a scale of Landau pole $\Lambda$ where $\la_{L(R)}(\Lambda) = \sqrt{4\pi}$.
The beta functions for the Yukawa coupling constants are listed in Appendix~\ref{sec-beta}. 
Similar study can be found in Ref.~\cite{Ibarra:2014qma} for the real DM, 
in Ref.~\cite{Kawamura:2018kut} for the complex DM, 
in Refs.~\cite{Kopp:2014tsa,Garny:2014waa} for the Majorana DM 
and in Ref.~\cite{Ibarra:2015fqa} for the Dirac DM.

In these figures, 
the latest result of the XENON1T experiment~\cite{Aprile:2018dbl} excludes the red region. 
In the gray region on the upper corner, 
the DM thermal abundance requires a non-perturbatively large coupling 
$\la_{L(R)}>\sqrt{4\pi}$ below TeV scale. 
In the green region, 
coannihilation is too efficient to explain the thermal abundance observed by the Planck collaboration~\cite{Tanabashi:2018oca}. 
The orange contours stand for the Landau pole scales and the purple contours stand for $\Delta a_\mu$. 
The brown regions are excluded by the slepton searches at the LHC, 
where the limits are projected from Fig.~\ref{fig-LHClim}. 

The shaded blue regions in the bottom panels   
are excluded by the current gamma line observations 
at the Fermi-LAT~\cite{Ackermann:2013uma} and the HESS~\cite{Abramowski:2013ax}. 
The cyan lines show the future sensitivity 
at the GAMAM-400~\cite{Galper:2012fp} and the CTA~\cite{Consortium:2010bc}. 
To obtain these limits, we refer to the 95\% C.L. upper limit on the combined cross section 
$\VEV{\sigma v}_{\lp\bar{\lp}\gamma}+2\VEV{\sigma v}_{\gamma\gamma}$ shown in Fig.~5 of \cite{Garny:2013ama}. 
The referred limit is obtained in the Majorana DM model with $r=1.1$, so that strictly speaking, 
it cannot simply be applied to the real DM case.  
As shown in Appendix~\ref{app:analytics}, however, 
the photon spectrum of $XX\to \lp\ol{\lp}\gamma$ in the real DM model 
is the same as that of $\chi\chi \to \lp\ol{\lp}\gamma$ in the Majorana DM.  
In addition, 
the three-body processes dominate over the two-body annihilation to $\gamma\gamma$ 
for $r\lesssim 3$ as shown in Fig.~\ref{fig;AnnRatio}. 
Thus, we apply the upper limit on the Majorana DM to the real DM, assuming that 
the difference between two limits is marginal as far as the models are perturbative.
In the following, we discuss more details of the current limits type-by-type.

\subsubsection*{ - Complex scalar DM}

Two top panels in Fig.~\ref{fig;plots1} show the results of the complex DM. 
The two-body annihilation $XX^\dagger \to \lp\bar{\lp}$ is the leading one at freeze-out. 
${\cal O}(1)$ Yukawa coupling is required to explain the DM abundance,   
since $s$-wave contribution is helicity suppressed and the $p$-wave is dominant. 
The other processes, such as the VIB process $XX^\dagger \to \lp\bar{\lp}V$, 
are sub-leading and less than 10\% of the total rate, as mentioned above. 
The large Yukawa coupling lowers the Landau pole scale.  
For instance, the DM mass should be smaller than 1~TeV 
so that the Yukawa couplings are perturbative up to the GUT scale around $10^{16}$ GeV.    
In compressed regions with $r\sim1$, 
the coannihilation further reduces the thermal abundance  
and the smaller Yukawa coupling $\la_{L,R}$ is enough to explain the observed value. 

The direct detection at the XENON1T~\cite{Aprile:2018dbl} has already excluded wide parameter space. 
In particular, it fully covers the theoretically allowed parameter space in the minimal setup with the singlet vectorlike lepton. 
It should be noted that the compressed mass region ($m_F \approx m_X$) looks 
already excluded by the XENON experiment in the figure. 
As pointed out in Ref.~\cite{Kawamura:2018kut}, allowing ${\cal O}(1\%)$ fine-tuning 
between $m_X$ and $m_F$, one should be able to find a narrow allowed region in this mass regime. 
In this paper, we do not focus on such a fine-tuned case. 
For readers who are interested in the fine-tuned case, see e.g. Ref.~\cite{Kawamura:2018kut}. 
In the $SU(2)_L$ doublet vectorlike lepton case, we can find the allowed region 
where $1\,{\rm TeV} \lesssim m_X \lesssim 3\,{\rm TeV}$. 
It will be covered by the future XENON1T experiment, whose projected limit is shown by a dashed red line, assuming that the sensitivity is 4.5 times better than the current one. 
The direct detection limit on the tau-philic case will be slightly weaker than that shown in Fig.~\ref{fig;plots1}, 
because of the smaller logarithmic factor $\ln m_\tau/m_{F}$ than in the muon-philic case. 
The indirect detection is not sensitive to this type of DM as discussed in Sec.~\ref{sec:indirect}.  

The discrepancy of the muon anomalous magnetic moment is hardly explained in this case.   
As mentioned in Sec.~\ref{sec-modelscal}, 
the sizable $\Delta a_\mu$ can be obtained only if 
it is enhanced by the vectorlike lepton mass due to the chirality flip. 
This is a common feature in the minimal lepton portal models, irrespectively of DM type. 
This fact motivates us to consider models with both singlet and doublet mediators.

\subsubsection*{- Real scalar DM} 

The bottom panels in Fig.~\ref{fig;plots1} show the results of the real DM model. 
Both the $s$- and $p$-wave contributions in the two-body annihilation $XX \to \lp\bar{\lp}$ 
are helicity suppressed and the $d$-wave is the dominant. 
The VIB processes $XX\to\lp\ol{\lp}V$ and the loop processes $VV^\prime$ 
have also too small cross sections to be dominant. 
As a result, the annihilation rates at freeze-out temperature are so small that 
the observed DM abundance can only be explained when $m_{L(E)} \leq 3m_X$ 
while keeping the perturbative coupling. 
In wide parameter space, the DM abundance is correctly produced with help of coannihilation.

At the DM direct detection, there is no contribution from the photon penguin diagram, unlike the complex DM. 
The leading DM-nucleon scattering is induced at 2-loop via diphoton exchange~\cite{Kopp:2009et}. 
As a result, the direct detection gives no constraints. 
On the other hand, 
the real DM can be probed by using cosmic ray fluxes at the indirect detection experiments. 
As discussed in Sec.~\ref{sec:indirect}, 
the sharp spectral feature of gamma ray is the promising signal. 
The blue regions where $r \gtrsim 1.1$ are excluded 
by the Fermi-LAT~\cite{Ackermann:2013uma} and HESS~\cite{Abramowski:2013ax}. 
A combination of future observations at GAMMA-400~\cite{Galper:2012fp} and CTA~\cite{Bernlohr:2012we}
would test the $SU(2)_L$ singlet vectorlike lepton case when $r \gtrsim1.1$. 
The sensitivity to the $SU(2)_L$ doublet vectorlike lepton case is slightly weaker, 
because $\VEV{\sigma v}_{\lp\ol{\lp}\gamma}+2\VEV{\sigma v}_{\gamma\gamma}$ is smaller than the singlet case.

\begin{figure}[ph]
\centering 
{\epsfig{figure=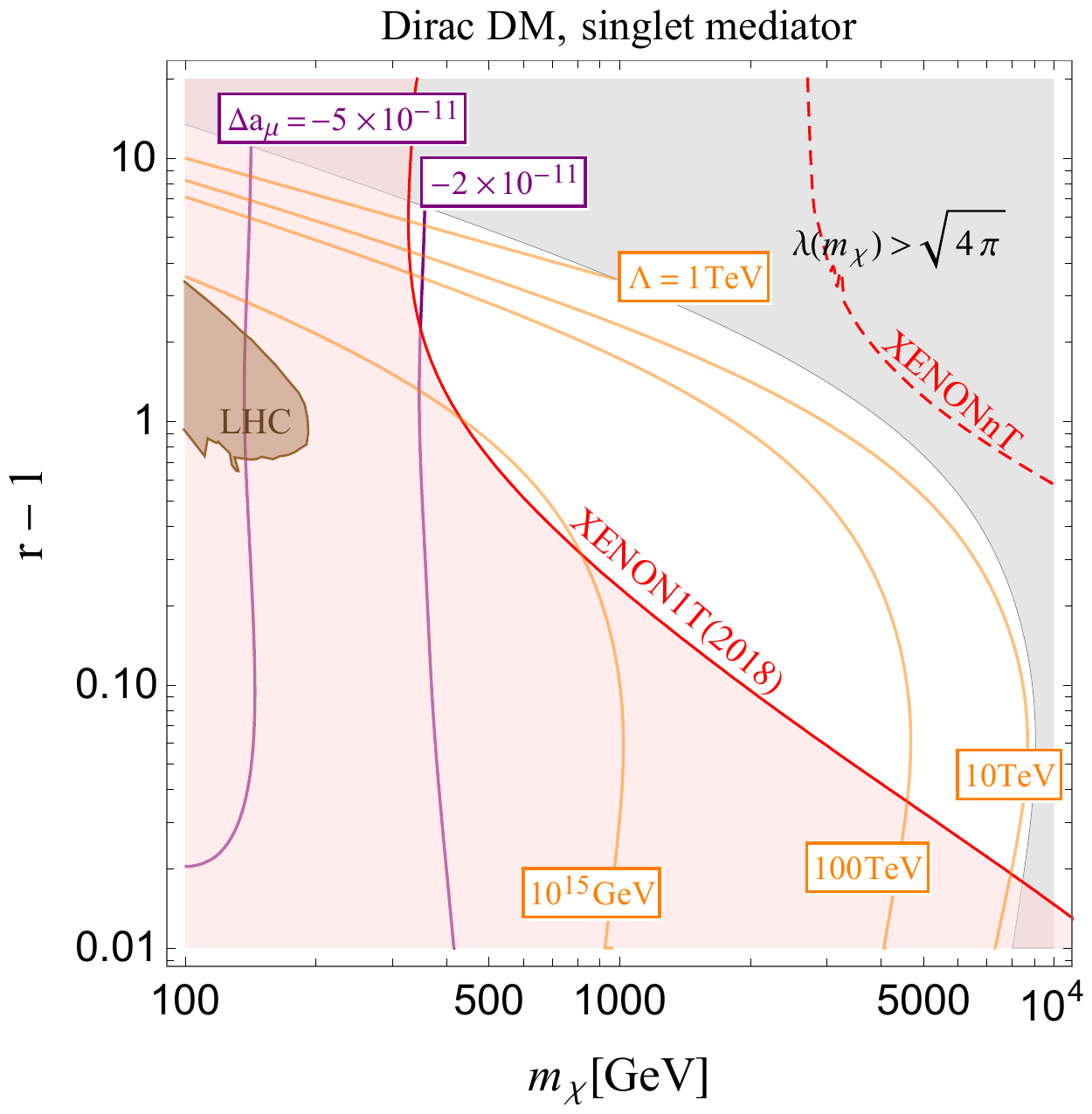,width=0.48\textwidth}}\hspace{.5cm}
{\epsfig{figure=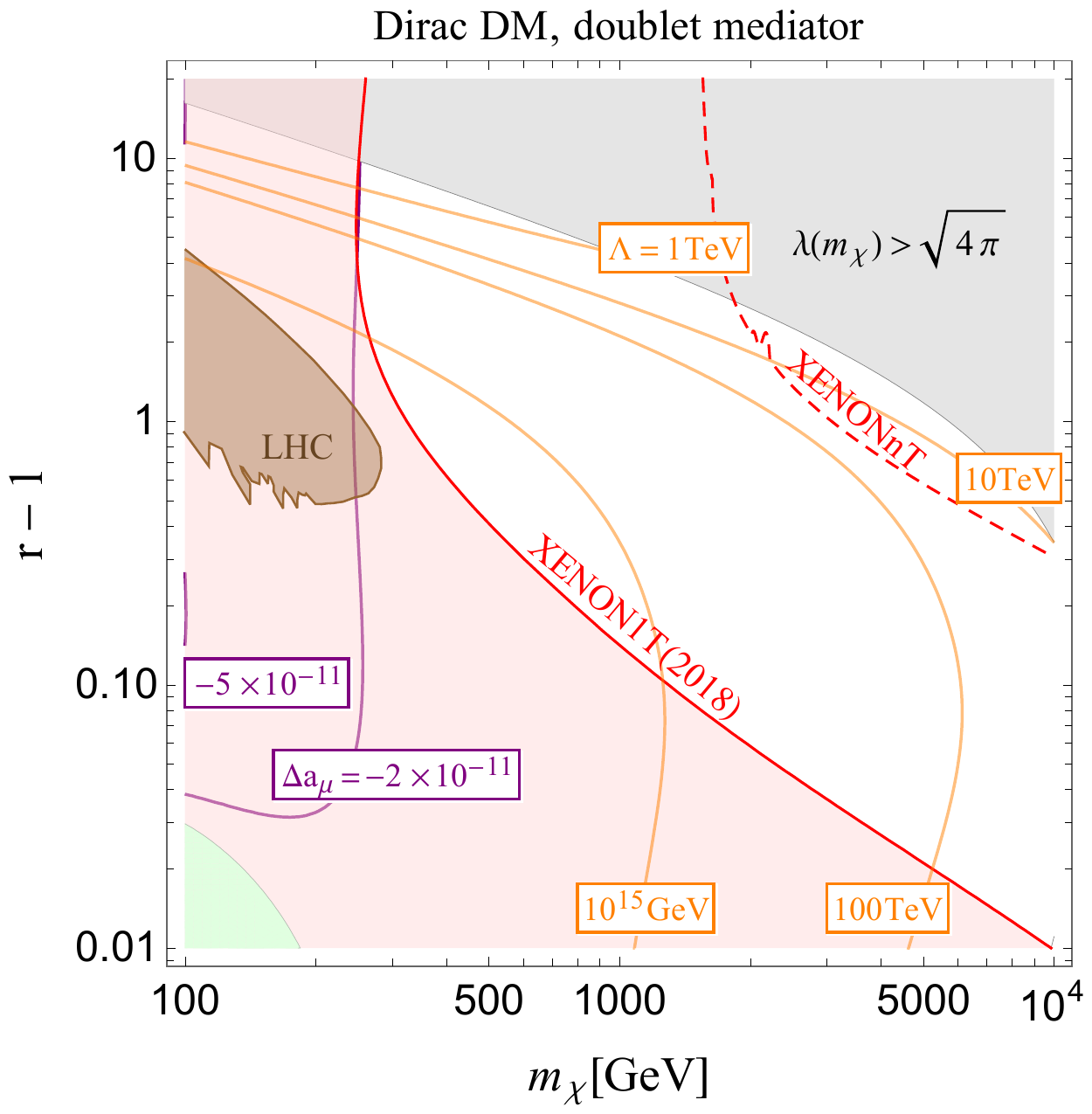,width=0.48\textwidth}}\vspace{0.5cm}
{\epsfig{figure=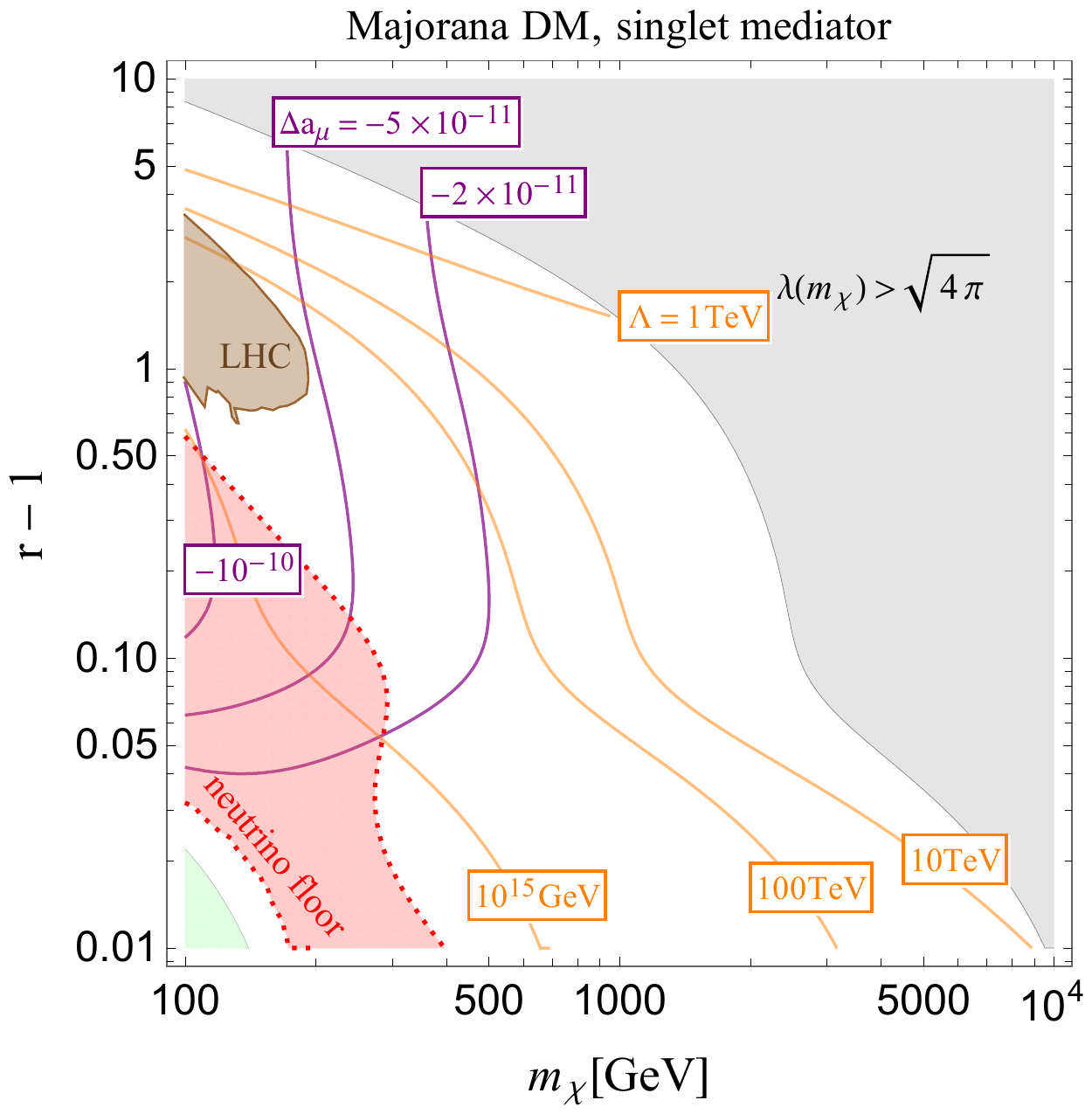,width=0.48\textwidth}}\hspace{.5cm}
{\epsfig{figure=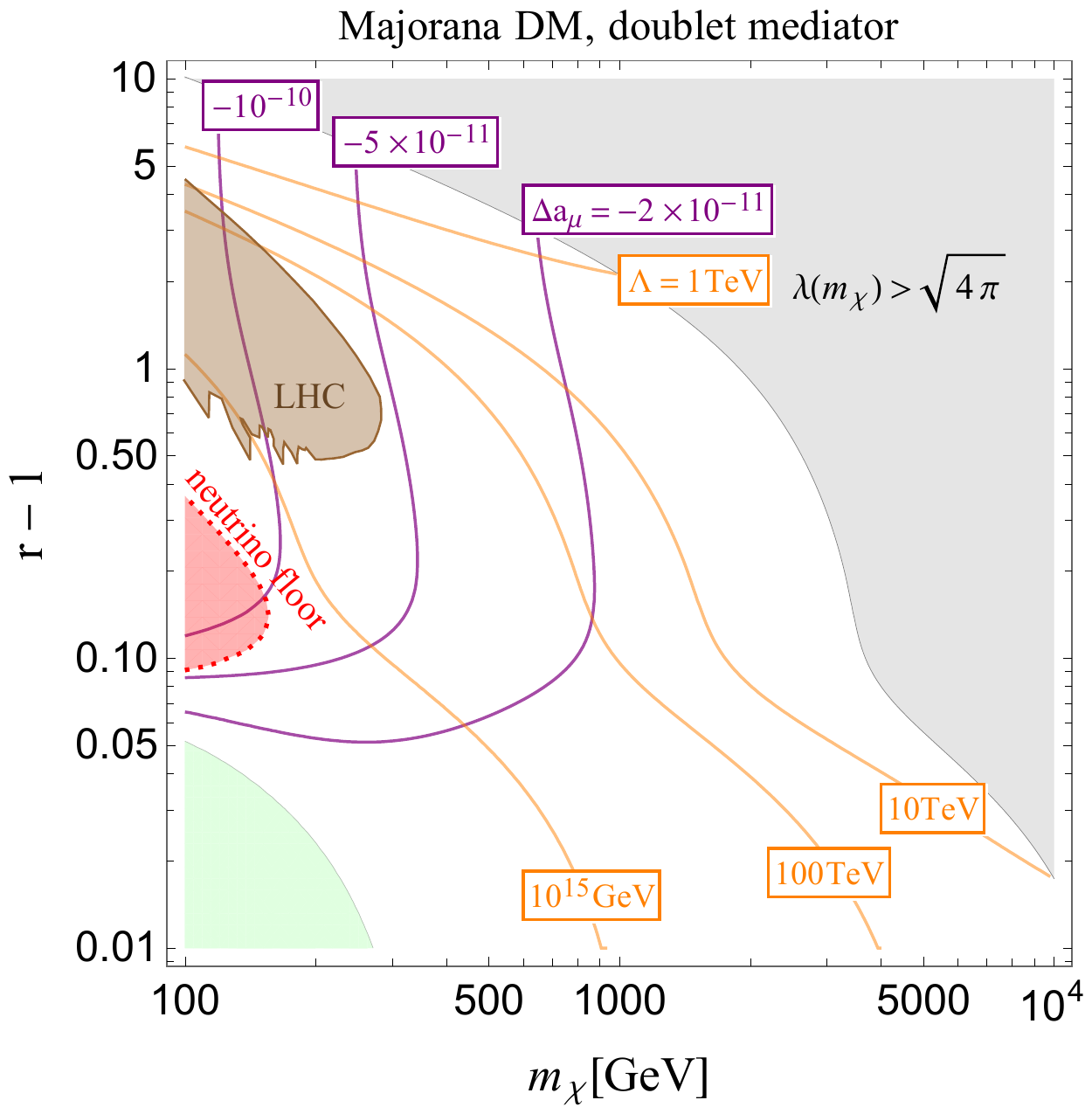,width=0.48\textwidth}}
\caption{\label{fig;plots2}
Plots for the fermion DM models with the same constraints as in Fig.~\ref{fig;plots1}.
The top (bottom) panels are for the Dirac (Majorana) fermion DM  
and 
the left (right) panels are for the singlet (doublet) slepton.} 
\end{figure}

\subsubsection*{- Dirac fermion DM}

The results of the Dirac fermion DM case are shown on the top panels in Fig.~\ref{fig;plots2}. 
The prime difference from the other types is that 
the partial $s$-wave in $\chi\bar{\chi}\to \lp\bar{\lp}$ annihilation is not helicity suppressed.  
The smaller Yukawa coupling is predicted to account for the DM abundance, 
and thus the Landau pole scale is higher than the other cases.

The constraints from the DM direct detection are much weaker than in the complex DM 
because of the smaller Yukawa coupling.
The current limits on the DM mass from the XENON1T 
are 300 (220)~GeV for the singlet (doublet) slepton case.  
The XENONnT experiment~\cite{Aprile:2015uzo} will probe the region below the red dashed lines 
and will cover most of the parameter space with perturbative Yukawa coupling. 
To refer to the XENONnT sensitivity, we assumed it is 50 times better than the current limit. 

There are no limits from the indirect searches in the muon-philic case as shown in Fig.~\ref{fig;plots2}. 
The limit from the observations of the dSph galaxies at Fermi-LAT~\cite{Ackermann:2015zua} 
is about 10 GeV in the muon-philic case, while it reaches about 100 GeV in the tau-philic case. 
In the electron-philic case, the stringent limit about 100 GeV will be set 
by the positron flux searching for annihilation into $e^+e^-$.

\subsubsection*{- Majorana fermion DM}
The pair annihilation of the Majorana DM is similar to that of the complex DM.  
The $p$-wave in $\chi\chi \to \lp\bar{\lp}$ dominates the freeze-out processes.  
The VIB and loop annihilations are subdominant and are no more than 10 \% contributions.
Hence, sizable Yukawa couplings are required to achieve correct DM density.  
This causes lower Landau-pole scales in analogy with the complex DM.

The constraint from the DM direct detection is, on the other hand, so weak that there are no exclusion lines in the figures. 
At 1-loop level, the non-vanishing contributions to DM-nuclei scattering are 
via the anapole interaction, $Z$-penguin and Higgs exchanging.  
The anapole interaction induces the spin-independent scattering.  
This gives the leading contribution although it is suppressed by DM velocity.  
The $Z$-penguin induces contact-type interactions, 
($\bar{\chi}\gamma_\mu\gamma_5\chi) (\bar{N}\gamma^\mu N)$
and 
($\bar{\chi}\gamma_\mu\gamma_5\chi) (\bar{N}\gamma^\mu\gamma_5 N)$. 
The former contributes to the spin-independent scattering and the latter to the spin-dependent one. 
These interactions, however, give only sub-leading effects due to the suppression by the lepton mass. 
The former contribution is further suppressed by the DM velocity. 
The Higgs exchanging contribution is also suppressed by the leptons mass. 
Altogether, the current sensitivity of direct detection cannot probe the Majorana DM. 
For a reference, we show the red dotted lines that correspond to the neutrino floor, which is assumed to have 10 times better sensitivity than the XENONnT reach. 
The regions between the dotted lines are within the neutrino floor sensitivity. 

Moreover, the indirect detection hardly constrains the parameter space  
since the $\chi\chi \to \lp\bar{\lp}$ process is suppressed by small DM velocity $v\sim10^{-3}$ in our Galaxy. 
It might be possible that 
the photon sharp spectral features from the VIB and loop processes are probed at the gamma ray telescopes
if a boost factor is $\order{10}$ or larger~\cite{Garny:2013ama}.  
We conclude that the Majorana DM is most invisible from the DM searches 
among the minimal lepton portal models.

\section{Lepton portal models for $\Delta a_\mu$}   
\label{sec;DMLR}
We have seen that it is hard to explain the discrepancy of the muon $g-2$ in the minimal lepton portal models that have either $SU(2)_L$ doublet or singlet mediator.  
In this section, we study the extended model with both the doublet and singlet mediators, 
and examine correlation between DM physics and $\Delta a_\mu$. 
We focus on the real scalar and Majorana fermion DM, 
since otherwise most of parameter space will be excluded by the DM direct detection.

\subsection{Real scalar DM} 
The presence of the vectorlike lepton mixing influences the DM annihilation. 
In particular, the $s$-wave contribution to $XX\to e_i\bar{e}_j$ appears as 
\begin{align}
(\sigma v)_{e_i\ol{e}_j} 
&= \frac{|\la_{L}^i\la_{R}^j|^2+|\la_{L}^j\la_{R}^i|^2}{2\pi} 
        \left( \frac{c_Rs_L m_{E_1}}{m_X^2+m_{E_1}^2} - \frac{c_Ls_R m_{E_2} }{m_X^2+m_{E_2}^2} \right)^2 + \order{v^2, \eps_i}. 
\label{eq;2Xto2l}
\end{align}
This is not suppressed by the light SM lepton masses 
and dominates the annihilation cross section. 
Note that the annihilation into a neutrino pair is not changed 
because of the absence of singlet vectorlike neutrino in our model.   

Let us discuss a correlation between $\Delta a_\mu$ and the DM abundance. 
First of all, we assume for simplicity that the DM abundance is determined solely by Eq.(\ref{eq;2Xto2l}), 
i.e. the $s$-wave contribution of $XX\to\mu\bar{\mu}$, and then estimate the induced $\Delta a_\mu$. 
This assumption implies that two portal couplings are not hierarchical ($\la_L\simeq\la_R$), 
the doublet-singlet mixing is not tiny, and any coannihilation process is not effective. 
Hence, we may regard $r=m_{E_1}/m_X$ as large in the estimate here. 
In this case, based on Eq.~\eqref{eq;2Xto2l}, the total annihilation cross section is approximately evaluated as 
\begin{equation}
\sigma_0 \sim  (\sigma v)_{\mu\bar{\mu}} 
\sim \frac{(\la_L^\mu\la_R^\mu)^2 ( c_R s_L m_{E_2} - c_L s_R m_{E_1})^2}{2\pi m_{E_1}^2 m_{E_2}^2} , 
\label{eq;2Xto2mu}
\end{equation}
where $\sigma_0\simeq 3\times10^{-26}$ [cm$^3$/s] 
is the canonical value to explain the observed DM density. 
Inserting this relation into Eq.~\eqref{eq-DelamuS}, we obtain
\begin{align}
\label{eq-DAMreal}
\Delta a_\mu \sim \frac{m_\mu}{16\pi^2} \frac{\la_L^\mu \la_R^\mu( c_R s_L m_{E_2} - c_L s_R m_{E_1})}{m_{E_1} m_{E_2}} 
\sim \frac{m_\mu}{16\pi^2} \sqrt{2\pi \sigma_0}.  
\end{align} 
Thus, $\Delta a_\mu \sim 5.0 \times 10^{-8}$ is predicted, which is clearly too large. 
This conclusion is independent of DM and mediator masses, the portal Yukawa couplings and mixing angle, as far as the partial $s$-wave of $XX\to\mu\bar{\mu}$ is responsible for DM production. 
Of course, the above estimate is not exact and there is a numerical error in fact, since our treatment is too rough. 
In Eqs.~(\ref{eq;2Xto2mu}) and (\ref{eq-DAMreal}), we ignored the finite $r$ correction
and did not take into account a loop function of $\Delta a_\mu$ etc.. 
Those corrections are, however, not so large, so that 
the consequence that the prediction of $\Delta a_\mu$ is too large is quite robust. 
Therefore, to generate the favorable value of $\Delta a_\mu$, we have to relax some of the above assumptions and prevent the $s$-wave contribution Eq.(\ref{eq;2Xto2l}) from dominating the DM annihilation. 
One simple way to do that is to employ large coannihilation. 
It requires a considerable fine-tuning between $m_X$ and $m_E$, 
while it allows for small pair-annihilation contributions in return, 
so that the Yukawa couplings become small, leading to a suppression of $\Delta a_\mu$.

Another way is that we consider a hierarchical coupling or small vectorlike lepton mixings. 
It is obvious from Eq.(\ref{eq;2Xto2l}) that the $s$-wave contribution is proportional to $(\la_L \la_R)^2$. 
If we take $\la_L=\veps \la_R$ ($\veps\ll1$), the contribution is suppressed by a factor of $\veps^2$. 
Therefore, as $\veps$ decreases, the relative importance of the partial $d$-wave grows, since it has the unsuppressed contribution $\propto \la_R^4$. 
The relative growth of the $d$-wave contribution breaks the correlation in Eq.(\ref{eq-DAMreal}), and enables us to explain the $\Delta a_\mu$ discrepancy. 
This is the case with small vectorlike lepton mixings. 
The $s$-wave part is also suppressed for the small mixings, while the $d$-wave is not. 
Then, we easily reach the same conclusion through a parallel discussion. 
In this paper, we will focus on the case with a hierarchical coupling, assuming an unsuppressed vectorlike mixing.

\begin{figure}[t]
\centering 
{\epsfig{figure=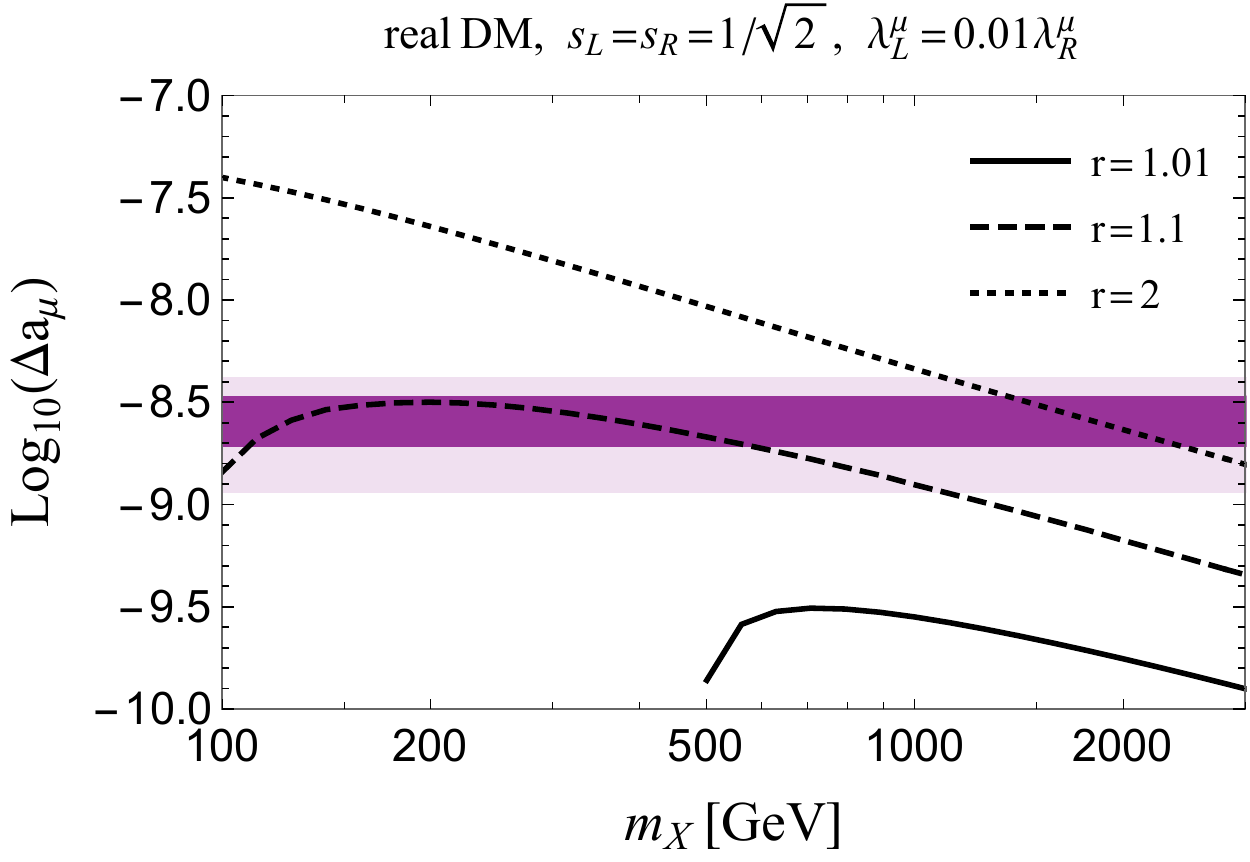,width=0.48\textwidth}}\hspace{.5cm}
{\epsfig{figure=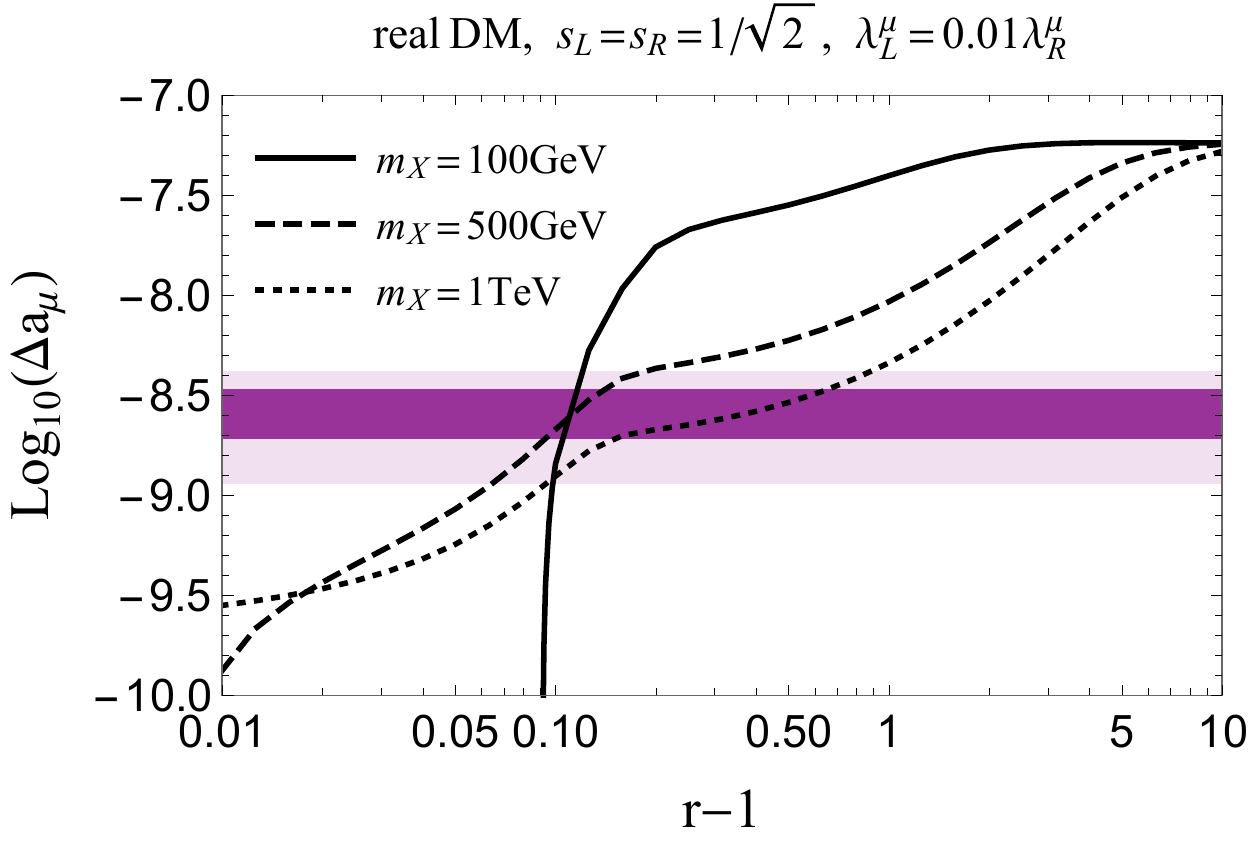,width=0.48\textwidth}}
\caption{\label{fig;amuS}
$\Delta a_\mu$ in the real scalar DM model as functions of  the DM mass $m_X$ 
and degeneracy $r-1$, where $r\equiv m_{E_1}/m_X$.  
The solid, dashed and dotted lines are $r=1.01$, $1.1$ and $2$ on the left panel 
and $m_X = 100$, $500$ GeV and $1$ TeV on the right panel. 
The Yukawa coupling $\la_L=0.01\la_R$ is fixed via the thermal relic abundance. 
$\Delta a_\mu$ is explained within 1$\sigma$ (2$\sigma$) uncertainties on the (light) purple band.
We assume the maximal mixing, $s_L = s_R = 1/\sqrt{2}$ 
and $\la_L^\mu=0.01\la_R^\mu$. 
}
\end{figure}

To see quantitative details, we show $\Delta a_\mu$ in Fig.~\ref{fig;amuS} as a function of the DM mass $m_X$ (left) and the mass ratio $r=m_{E_1}/m_X$ (right).  
In the analysis, 
we assume for simplicity that $m_L=m_E$ and $\kappa=\tilde{\kappa}$ 
which lead to the maximal mixings $s_L=s_R= 1/\sqrt{2}$. 
The resulting condition $s_L=s_R$ is also in favor of constraints from the EW precision observables (EWPOs).  
Furthermore, the mass difference between the vectorlike leptons are 
set to $m_{E_2}-m_{E_1}=2 \kappa v_H=100$~GeV.
The relative size of the Yukawa couplings is fixed to $\la_L=0.01\la_R$, 
and the absolute size of the couplings is determined via the observed DM abundance. 
$\Delta a_\mu$ is explained within 1$\sigma$ (2$\sigma$) uncertainties on the (light) purple band.
We see that $\Delta a_\mu$ is successfully explained together with the DM density.

The behavior of $\Delta a_\mu$ is understood as follows.  
With the relation $\la_L=0.01\la_R$, the DM pair annihilation is essentially dominated by the $d$-wave, 
which is scaled as $(\sigma v)_{\mu\bar{\mu}} \propto \la_R^4/(m_X^2 r^8)$. 
Then, as far as coannihilation is irrelevant to the DM production, we find the scaling of $\Delta a_\mu$, 
\begin{equation}
\Delta a_\mu \sim \frac{m_\mu \veps \la_R^2 c_R s_L (m_{E_2}-m_{E_1})}{16\pi^2 m_X^2 r^2} 
\propto \frac{\veps r^2}{m_X} ,
\label{eq;DAMreal2}
\end{equation}
where we assumed that $\la_R$ is determined via the DM abundance. 
It follows from the equation that $\Delta a_\mu$ decrease (increase) as 
the DM mass $m_X$ increase (decrease) in this regime. 
In Fig.~\ref{fig;amuS} (left), we can observe such a behavior in fact. 
Once coannihilation operates and becomes superior to the pair annihilation, 
smaller Yukawa couplings are predicted to explain the DM abundance 
and thus $\Delta a_\mu$ becomes small. 
Since coannihilation is significant as $r\to1$, 
$\Delta a_\mu$ is smaller as $r$ is closer to unity. 
We see this effect via the coannihilation  
in the solid ($r=1.01$) and dashed ($r=1.1$) lines in Fig.~\ref{fig;amuS} (left).  
As the DM mass decreases, 
$\Delta a_\mu$ increases until $m_X = 700$ ($200$) GeV for $r=1.01$ ($1.1$), 
and then $\Delta a_\mu$ starts decreasing due to the coannihilation dominance. 
This behavior cannot be found when $r=2$, because the coannihilation is not effective when $r\gtrsim1.2$. 

We see the effects of the coannihilation more explicitly in Fig.~\ref{fig;amuS} (right). 
With DM mass fixed, a small $\Delta a_\mu$ is induced when $r\approx1$, 
while $\Delta a_\mu$ increases monotonically as $r$ goes away from unity. 
Remarkably, as $r \to \infty$, $\Delta a_\mu$ looks approaching the asymptotic value 
Eq.(\ref{eq-DAMreal}), independently of DM mass. 
To understand that, we shall take a closer look at the DM pair annihilation. 
Keeping only the relevant contributions, we find the cross section, in the limit of $r\gg1$ and $\veps\ll1$, 
\begin{equation}
(\sigma v)_{\mu\bar{\mu}} \simeq \frac{\la_R^4}{8\pi m_X^2 r^4} 
\left( \frac{\veps^2(m_{E_2}-m_{E_1})^2}{m_X^2} + \frac{2v^4}{15r^4} \right).
\end{equation}
Despite of the small $\veps$, the $s$-wave contribution can overcome the $d$-wave one, 
if the following condition is satisfied: 
\begin{equation}
\frac{\veps^2(m_{E_2}-m_{E_1})^2}{m_X^2} \gtrsim \frac{2v^4}{15r^4} .
\end{equation}
Since we fix the vectorlike mass splitting to $m_{E_2}-m_{E_1}=100$\,GeV and 
$v^4 \sim 0.1$ at freeze-out, 
the right-hand side is about $10^{-4}$ when $m_X=100$\,GeV and $r=3$. 
Thus, in the case with $\la_L=0.01\la_R$ in Fig.~\ref{fig;amuS} (right), 
the above condition can be satisfied when $r\gtrsim3$. 
Once the $s$-wave contribution dominates the DM production, 
we can repeat the previous discussion to derive Eq.(\ref{eq-DAMreal}), 
and then arrive at the asymptotic value $\Delta a_\mu \sim 5.0 \times 10^{-8}$ for large $r$.

\begin{figure}[ph]
\centering 
{\epsfig{figure=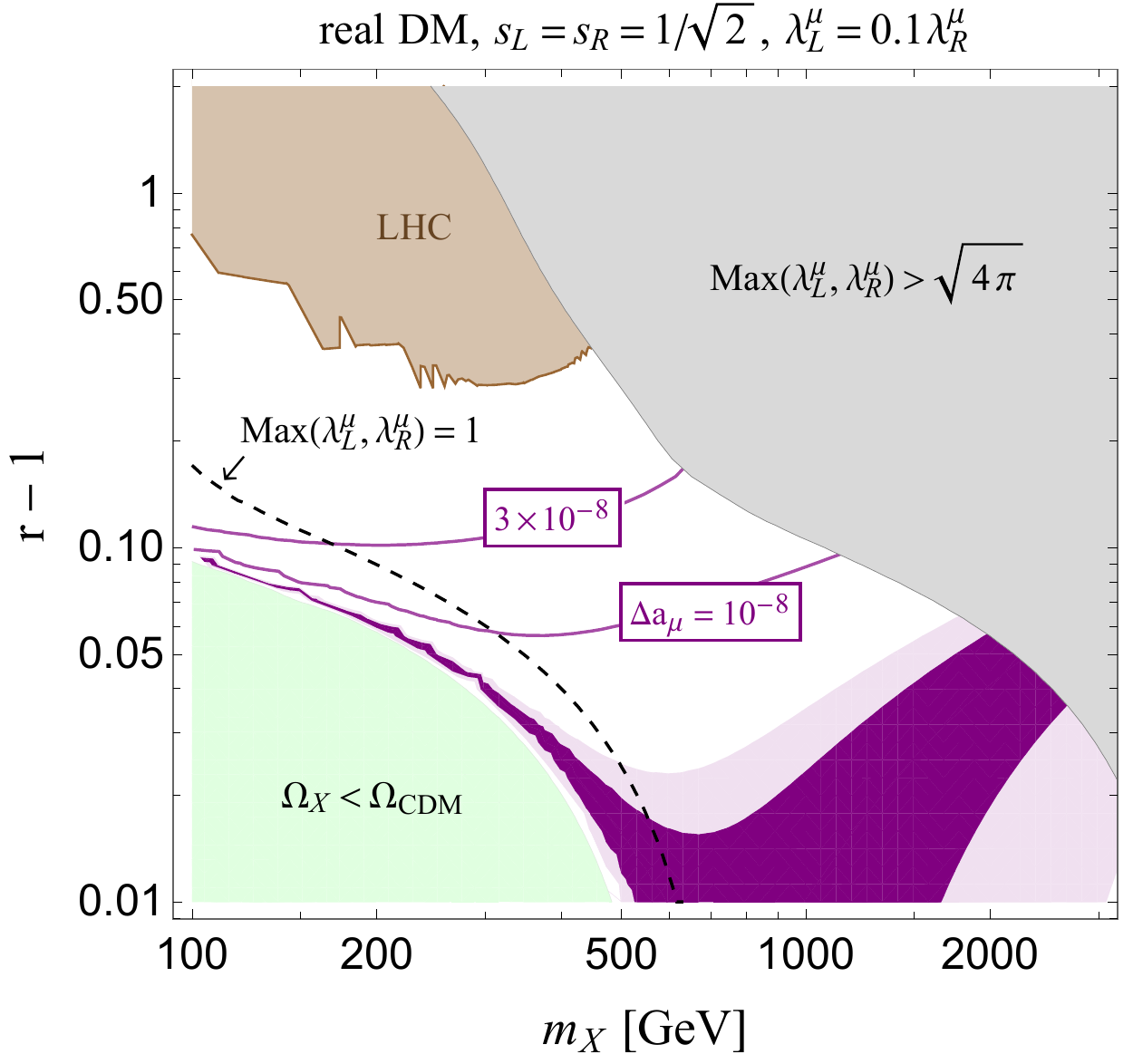,width=0.48\textwidth}}\hspace{.3cm}
{\epsfig{figure=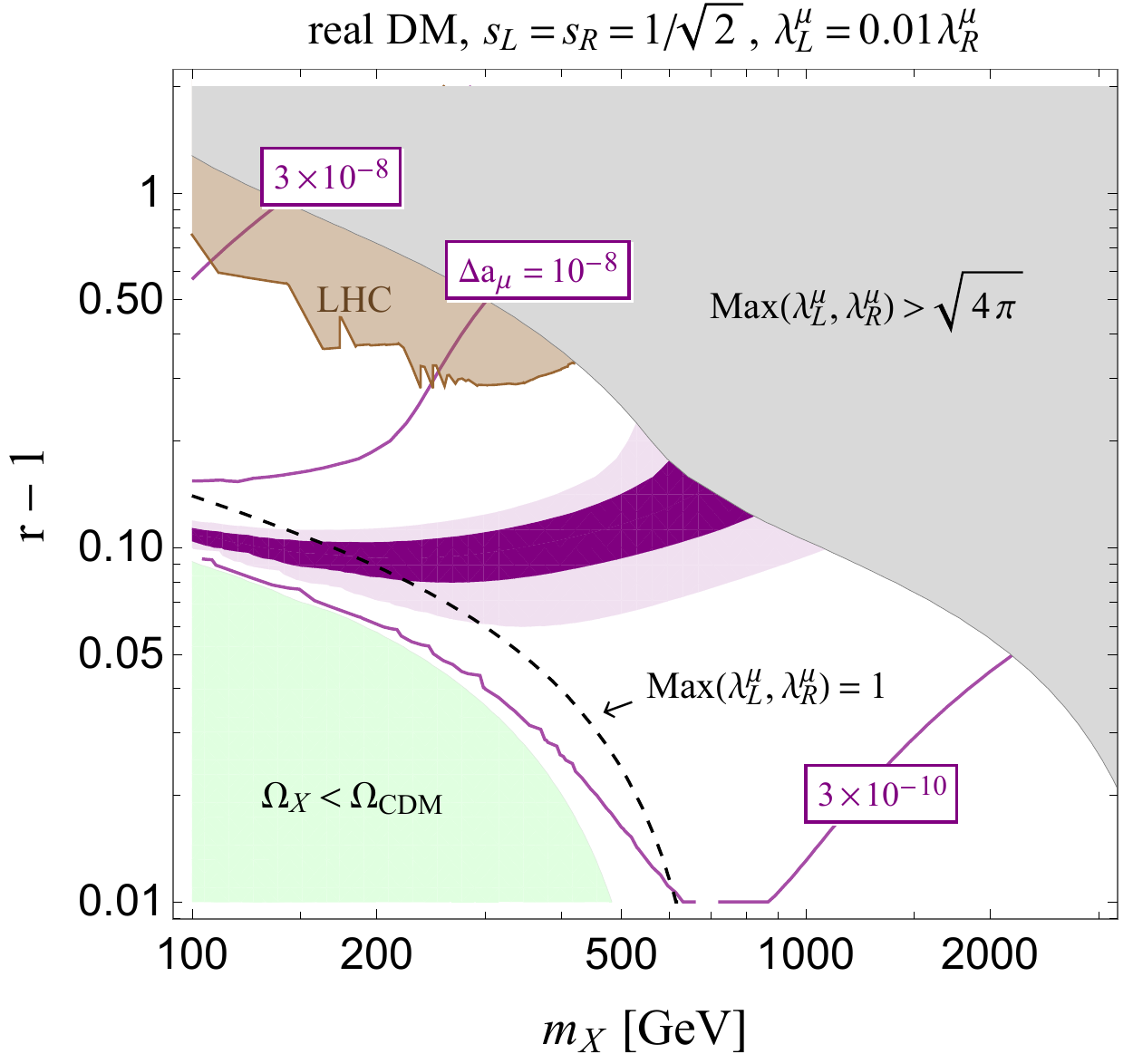,width=0.48\textwidth}}\vspace{.5cm}
{\epsfig{figure=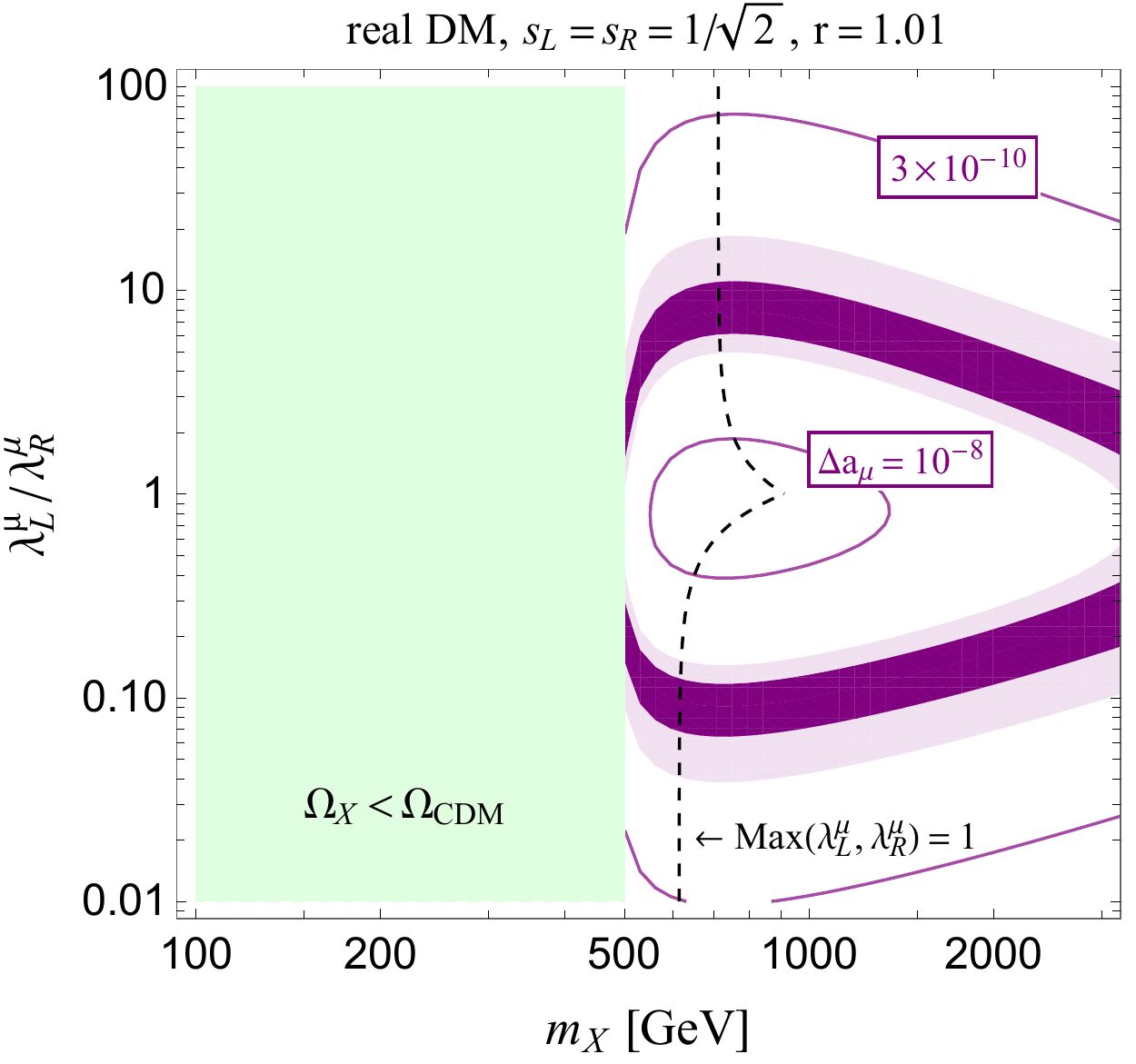,width=0.48\textwidth}}\hspace{.3cm}
{\epsfig{figure=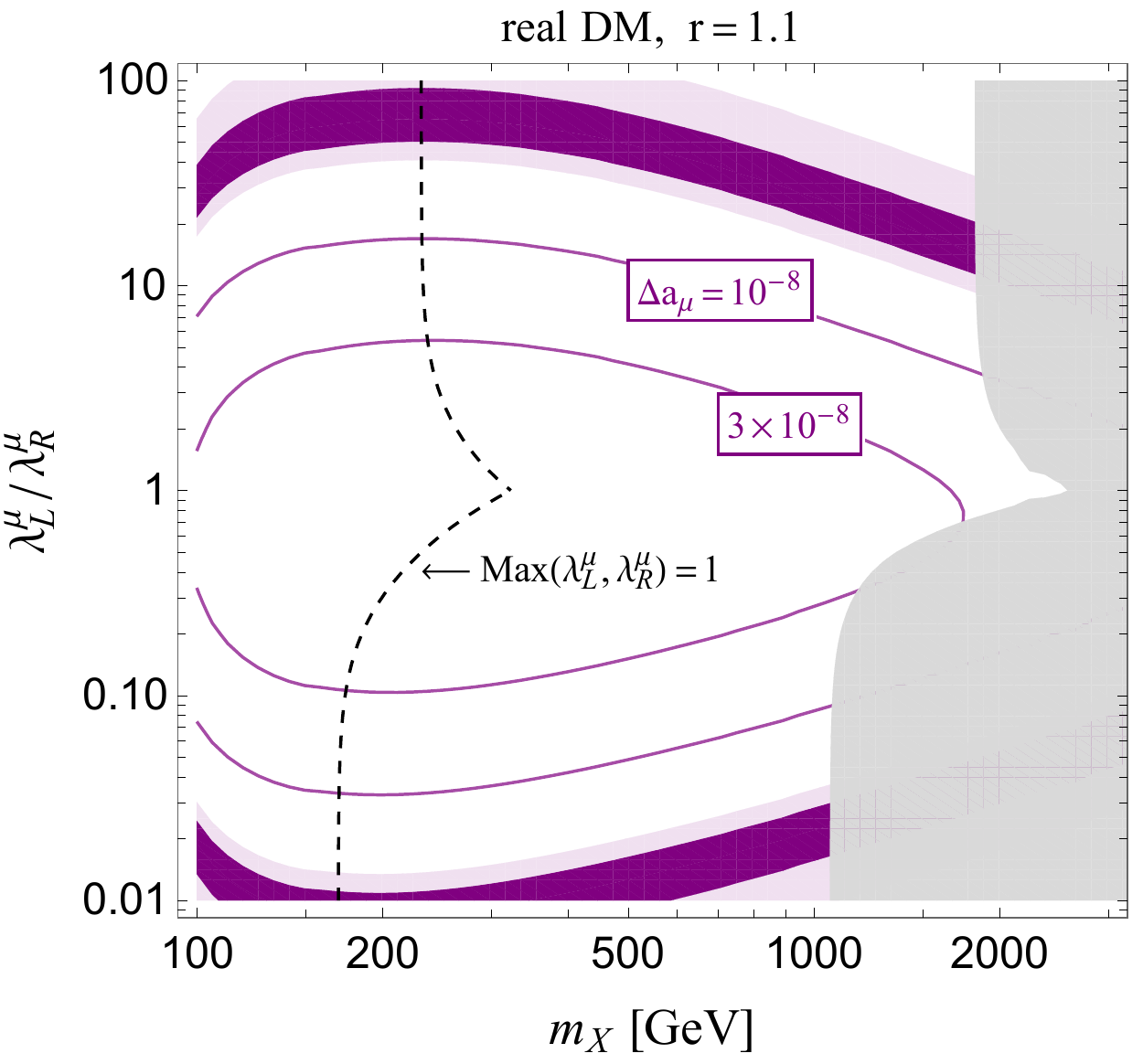,width=0.48\textwidth}}
\caption{\label{fig;Real} 
Parameter space consistent with the DM observations in the real DM model with the maximally mixed vectorlike leptons $s_L=s_R=1/\sqrt{2}$. 
We show the induced values of $\Delta a_\mu$ with the purple lines. 
$\Delta a_\mu$ is explained in the (light) purple region within 1$\sigma$ (2$\sigma$).}
\end{figure}

Figure \ref{fig;Real} shows the parameter space, where $\Delta a_\mu$ is explained consistently with the DM physics, on the ($m_X$, $r-1$) plane (top) and the ($m_X$, $\la^\mu_L/\la^\mu_R$) plane (bottom).  
The size of the Yukawa couplings is fixed to explain the DM abundance. 
The meanings of green and gray regions are the same in Figs.~\ref{fig;plots1} and~\ref{fig;plots2}. 
The purple lines show the induced values of $\Delta a_\mu$ in unit of $10^{-9}$. 
We also highlight the regions with the (light) purple band, where $\Delta a_\mu$ is explained within $1\sigma$ ($2\sigma$). 
Since $\Delta a_\mu$ can easily be reduced by relaxing our assumption for the maximal mixing, 
we conclude that the discrepancy $\Delta a_\mu$ can be resolved in the real DM model. 
Note, however, $\order{10\%}$ mild tunings may be required to realize the DM abundance.\\

We comment on the constraint from the EW precision observables (EWPOs).
In the scalar DM case, ${\cal O}(1)$ Yukawa coupling is required to achieve the correct
relict density since the annihilation cross section is strongly suppressed. 
We study the EWPOs concerned with the $T$ parameter and the partial $Z$ boson decay 
width to two $\mu$. We estimate the bound on the EWPOs based on the study in Refs. \cite{Tanabashi:2018oca,Kowalska:2017iqv}.
We conclude that the most stringent constraint comes from the partial decay width of $Z$ boson
to 2 $\mu$. The deviation of the decay width of $Z \to \mu \overline{\mu}$
is much less than ${\cal O}(0.1)$\% that is below the 
limit \cite{Tanabashi:2018oca}, in the case with $s_L=s_R={1}/{\sqrt{2}}$.
If $s_L$ is not the same as $s_R$, the deviation is enhanced and may be
tested by the future experiment \cite{Kowalska:2017iqv}.
In the same analogy with the scalar DM model, 
the constraint from the EWPOs in the Majorana DM case is not so strong.

\subsection{Majorana Fermion DM}
As in the real DM model, 
the annihilation $\chi\chi\to e_i\ol{e}_j$ has the $s$-wave contribution, 
\begin{align}
(\sigma v)_{e_i \ol{e}_j}
&= \frac{|\la_{L}^i\la_{R}^j|^2+|\la_{L}^j\la_{R}^i|^2}{16\pi} 
 \left( \frac{c_\theta s_\theta m_\chi}{m_\chi^2+m_{\wt{E}_1}^2} 
       -\frac{c_\theta s_\theta m_\chi}{m_\chi^2+m_{\wt{E}_2}^2} \right)^2 +\order{v^2, \eps_i} ,
\end{align}
that is not suppressed by the lepton masses. 

Let us study a correlation between DM and $\Delta a_\mu$ in the Majorana DM model. 
In this case, the DM pair annihilation $\chi\chi\to \mu\ol{\mu}$ will dominantly contribute to the DM production as far as the DM is enough lighter than the sleptons and the coannihilation is negligible.
We neglect these effects for simplicity.
In this case, the annihilation cross section is given by 
\begin{equation}
(\sigma v)_{\mu\ol{\mu}} \sim 
\frac{(c_\theta s_\theta \la_L^\mu\la_R^\mu)^2}{16\pi} \frac{(m_{\widetilde{E}_2}^2-m_{\widetilde{E}_1}^2)^2}{m_\chi^6r^8} + \frac{(\la_L^\mu)^4+(\la_R^\mu)^4}{48\pi m_\chi^2 r^4} v^2,  
\end{equation}
with $r \equiv m_{\wt{E}_1}/m_\chi$. 
Here, the second term is the $p$-wave contribution from the chirality-conserving interaction. 
If the $s$-wave part is dominant, $\Delta a_\mu$ is estimated as 
\begin{equation}
\label{eq-DAMmaj}
\Delta a_\mu 
\sim \frac{m_\mu}{16\pi^2} \frac{c_\theta s_\theta \la_L \la_R(m_{\wt{E}_2}^2-m_{\wt{E}_1}^2)}{m_\chi^3r^4} 
\sim \frac{m_\mu}{16\pi^2} \sqrt{16\pi\sigma_0} ,
\end{equation} 
where we used the fact that the cross section is approximately equal to $\sigma_0$. 
Thus, if the above assumption is valid, $\Delta a_\mu \sim 10^{-7}$ is predicted as in the real DM model. 

However, the Majorana case is not so simple as the real scalar case. 
Even if $\la_L=\la_R$ and the maximal mixing $s_\theta=1/\sqrt{2}$, 
the $p$-wave can have comparable contribution with the $s$-wave, 
in spite of a mild velocity suppression $v^2\sim0.24$. 
The cross section in this case is expressed by 
\begin{equation}
(\sigma v)_{\mu\ol{\mu}} \sim 
\frac{(\la_R^\mu)^4}{64\pi m_\chi^2 r^4} \left( \frac{(m_{\widetilde{E}_2}^2-m_{\widetilde{E}_1}^2)^2}{m_\chi^4 r^4} + \frac{8v^2}{3} \right) .
\end{equation}
As $r$ increases, the $s$-wave contribution is decaying more rapidly than the $p$-wave one.  
This indicates that the latter can still be leading for large $r$.  
For example, when we consider 100\,GeV DM and the slepton mass difference $m_{\widetilde{E}_2}^2-m_{\widetilde{E}_1}^2 = (100\,{\rm GeV})^2$, two contributions have the similar size at $r\simeq1.1$. 
With such a parameter set, the $p$-wave will therefore be leading contribution where $r\gtrsim1.1$. 
Note that $r=1.1$ is not large, so that coannihilation is operative to some extent in this regime. 
It may be illuminating to derive an asymptotic behavior in the Majorana DM case as well as the real scalar one. 
In general, both the $s$- and $p$-wave contributions are comparably important, 
but in the limit of $r\to\infty$ or $m_\chi\to\infty$, we find that the $p$-wave one is dominant. 
In this limit, $\Delta a_\mu$ is evaluated by 
\begin{align}
\label{eq-DAMmaj}
\Delta a_\mu 
\sim&\  
\frac{m_\mu c_\theta s_\theta (m_{\widetilde{E}_2}^2-m_{\widetilde{E}_1}^2)}{16\pi^2 m_\chi^2 r^2} \sqrt{\frac{(\la_L/\la_R)^2}{1+(\la_L/\la_R)^4}} \sqrt{ \frac{48\pi\sigma_0}{v^2}}.
\end{align} 
The expression in the Majorana DM case is more complicated than the one in the real DM case. 
$\Delta a_\mu$ depends on $m_\chi$, $r$, $\la_L/\la_R$, 
the slepton mixing and the slepton mass difference.

\begin{figure}[t]
\centering
{\epsfig{figure=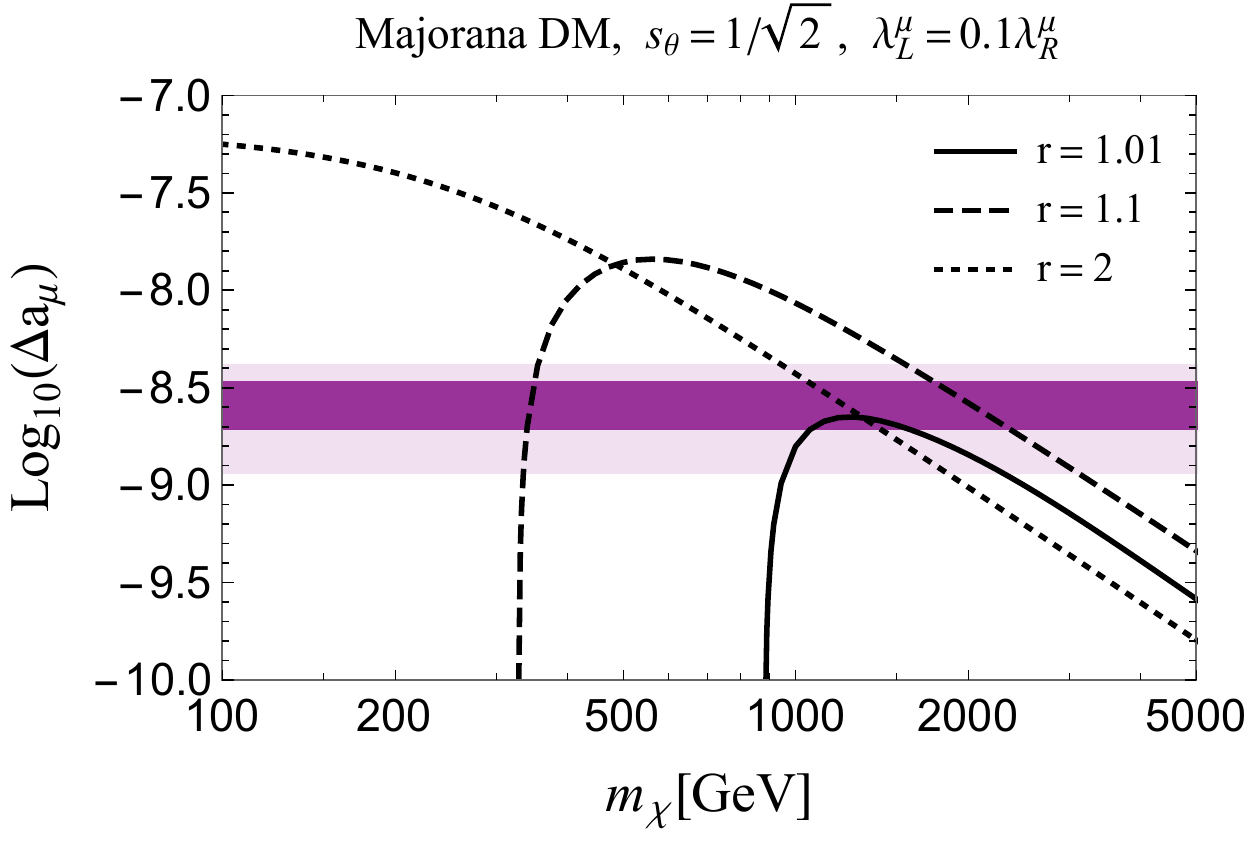,width=0.48\textwidth}}\hspace{.5cm}
{\epsfig{figure=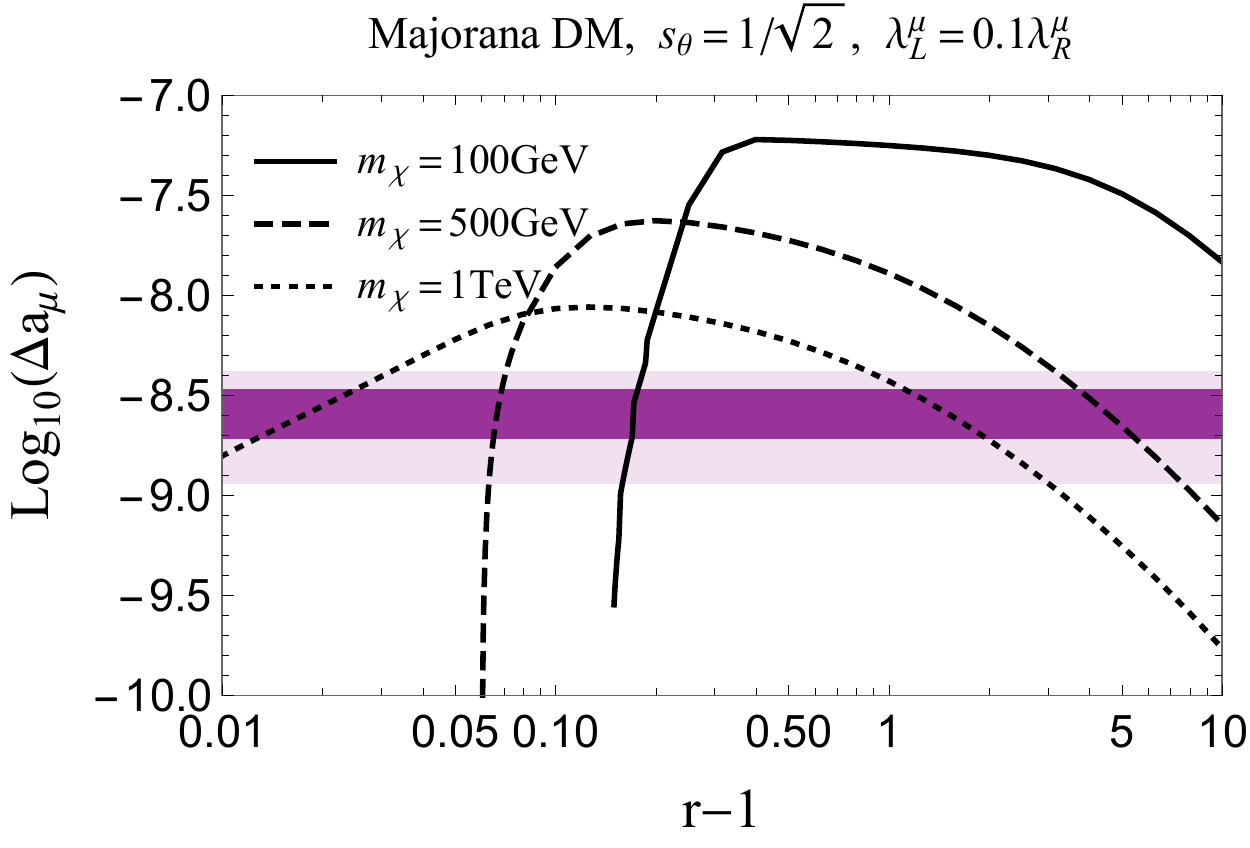,width=0.48\textwidth}}\vspace{.5cm}
\caption{\label{fig;MajoranaDAM}
The same plots for the Majorana DM as Fig.~\ref{fig;amuS}.
}
\end{figure}
Figure \ref{fig;MajoranaDAM} shows $\Delta a_\mu$ 
as a function of the DM mass $m_\chi$ and the ratio $r\equiv m_{\wt{E}_1}/m_\chi$. 
As in the real DM case, 
we consider $m_{\wt{L}}^2=m_{\wt{E}}^2$, so that the slepton mixing is maximized, $s_\theta=1/\sqrt{2}$.  
The trilinear coupling $A$ in Eq.~\eqref{eq-slepmat} is set to  
$A v_H=(500\,\mathrm{GeV})^2=(m_{\wt{E}_2}^2-m_{\tilde{E}_1}^2)/2$. 
The ratio of the two portal couplings is $\la_L = 0.1\la_R$, and 
the absolute size is determined to explain the DM density. 
We see that $\Delta a_\mu > \order{10^{-9}}$ can be realized where the DM is lighter than a few TeV.  

\begin{figure}[ph]
{\epsfig{figure=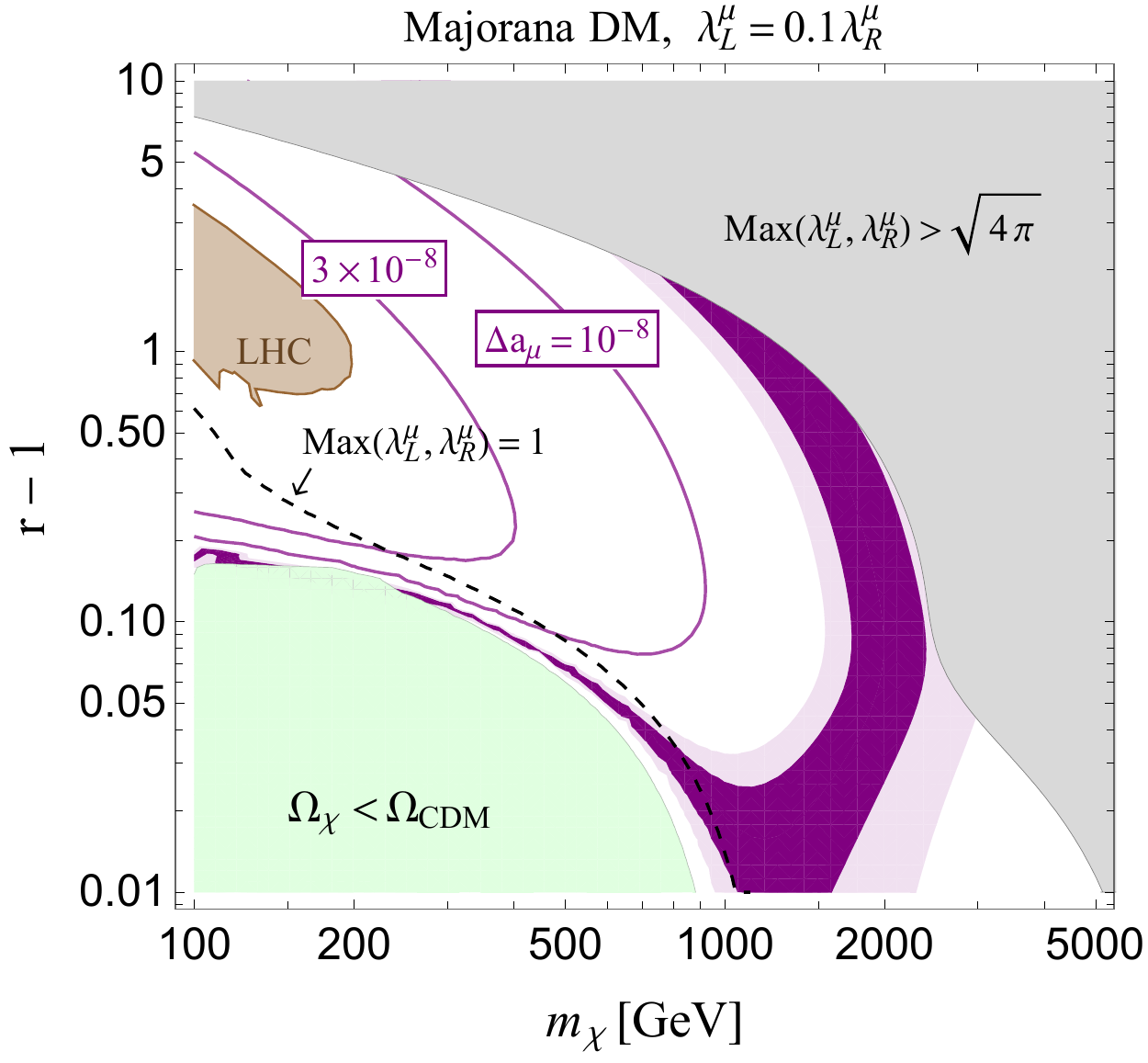,width=0.48\textwidth}}\hspace{.5cm}
{\epsfig{figure=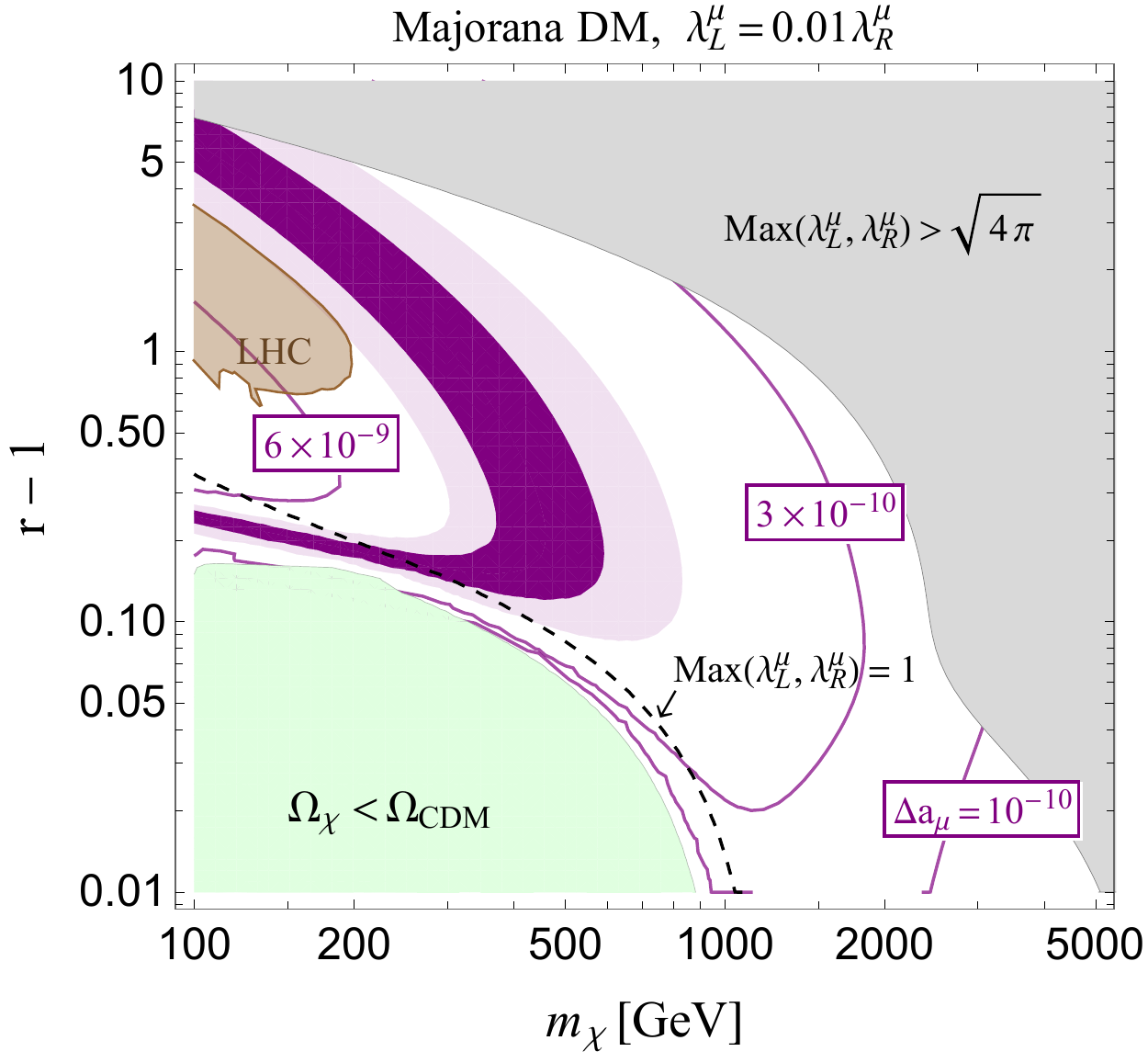,width=0.48\textwidth}}\vspace{.5cm}
{\epsfig{figure=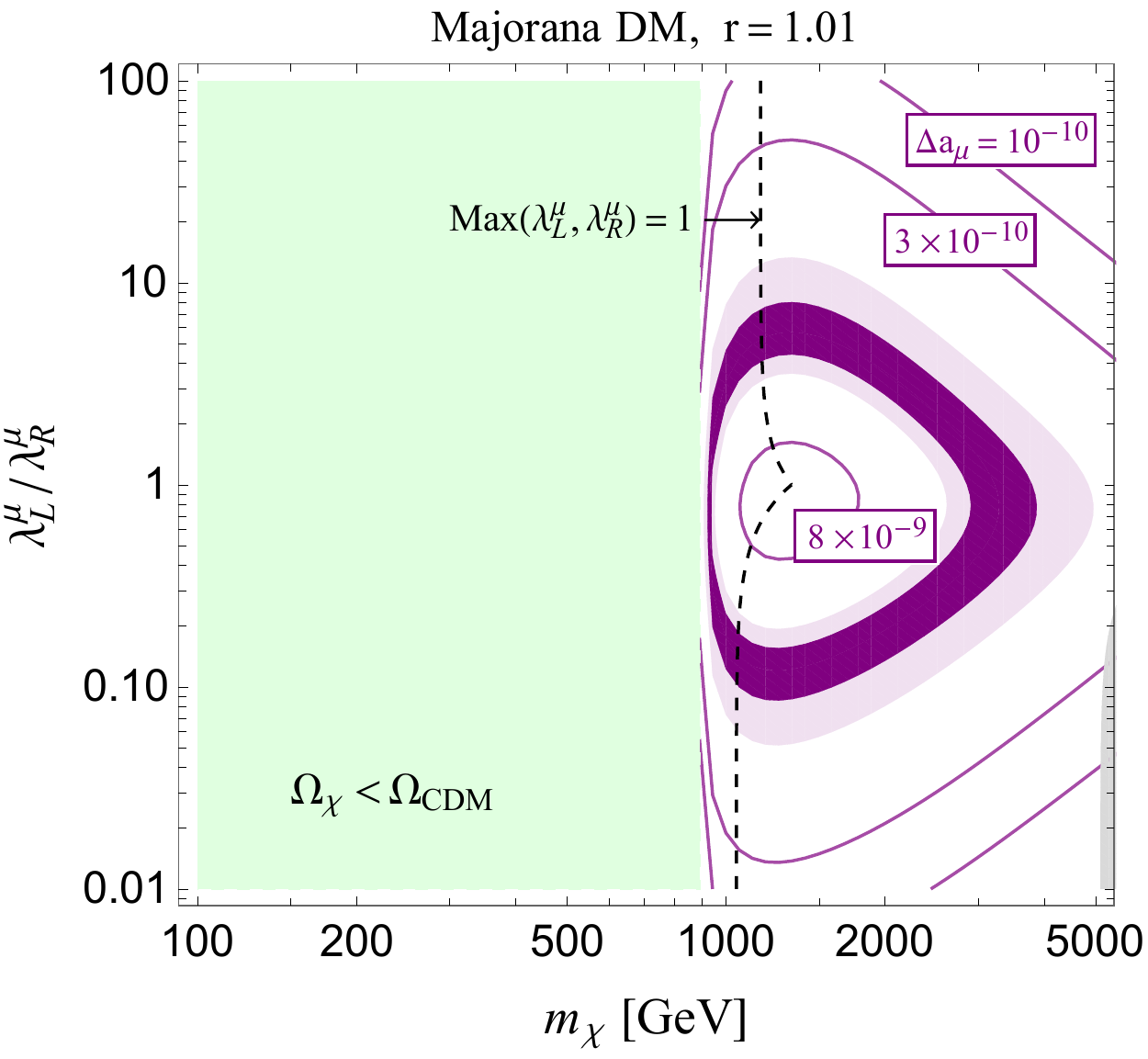,width=0.48\textwidth}}\hspace{.5cm}
{\epsfig{figure=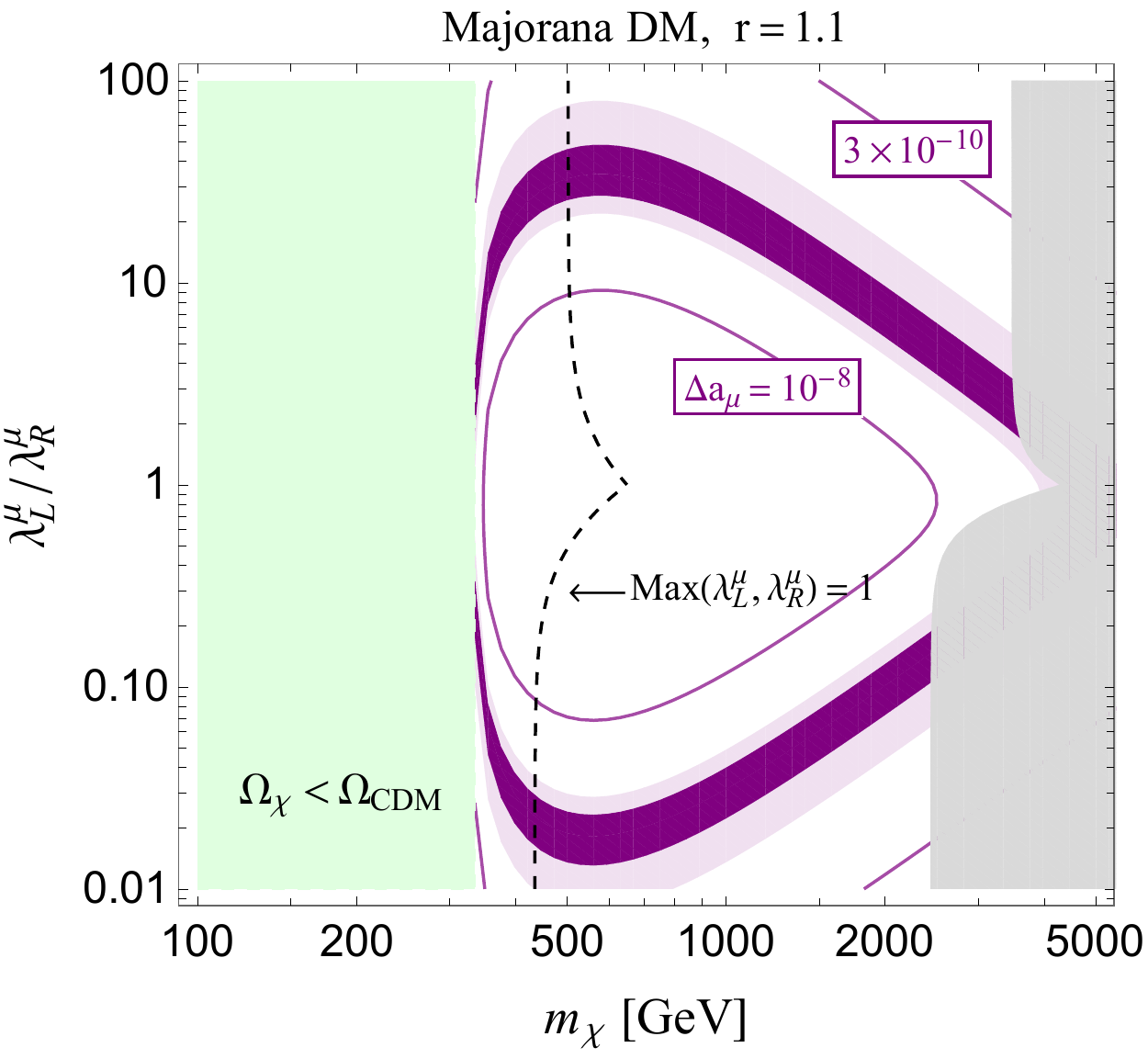,width=0.48\textwidth}}
\caption{\label{fig;Majorana}
The same plots for the Majorana DM as Fig.~\ref{fig;Real}. 
}
\end{figure}
The dependence on the DM mass and the ratio $r$ is slightly different from the one in the real DM case. 
Concerning the DM mass dependence, as $m_\chi$ increases, 
$\Delta a_\mu$ decreases more rapidly than in the real DM.   
This is understood by comparing Eqs.~(\ref{eq-DAMmaj}) and (\ref{eq;DAMreal2}). 
$\Delta a_\mu \propto 1/m_\chi^2$ in the Majorana case, 
while $\Delta a_\mu \propto 1/m_\chi$ in the real case. 
$\Delta a_\mu$ increases with decreasing $m_\chi$, 
but when coannihilation becomes active, $\Delta a_\mu$ starts decreasing. 
This behavior is similar to the real DM case. 
Further, we can see in Fig.\ref{fig;MajoranaDAM} (right) that the dependence on $r$ is also different. 
Since $\Delta a_\mu$ is scaling as $1/r^2$ for large $r$, 
$\Delta a_\mu$ decreases with increasing $r$ in the Majorana case. 
That is in contrast to the real case
where $\Delta a_\mu$ is nearly independent of $r$ in the large $r$ limit 
as shown in Eq.~\eqref{eq-DAMreal}. 
With decreasing $r$, $\Delta a_\mu$ is increasing, but when $r$ gets close to unity,  
coannihilation becomes effective, which makes the portal Yukawa couplings small. 
Then, Eq.~\eqref{eq-DAMmaj} is no longer valid. 
Altogether, $\Delta a_\mu$ is maximized at $r-1=$0.1--1 as shown in Fig.~\ref{fig;MajoranaDAM}. 

Figure \ref{fig;Majorana} is plotted in the same manner as Fig.~\ref{fig;Real}. We choose the same parameter set as used in Fig.~\ref{fig;MajoranaDAM}. 
Incompatibility of mass fine-tuning with coupling hierarchy is maintained in the Majorana case as well, but the restriction on the mass degeneracy is a bit loose. More concretely, when we take $\la_L/\la_R=0.1$ (the top-left of Fig.~\ref{fig;Majorana}), a parameter space, where $(m_{\widetilde{E}_1}-m_\chi)/m_\chi={\cal O}(1)$, is still allowed. As a consequence, we can accommodate both DM and $(g-2)_\mu$ explanation without terrible fine-tuning in mass and hierarchical coupling, but this time we have to cost a very large Yukawa coupling close to $\sqrt{4\pi}$, which in turn causes a low Landau pole scale.

\section{Summary}
\label{sec;summary}

In this paper, we have examined the lepton portal DM models, in which 
scalar or fermion DM couples to SM leptons via Yukawa interaction involving DM and leptons. 
New EW charged fields, namely vectorlike leptons or sleptons, are introduced, 
so that the renormalizable Yukawa interactions are allowed by the EW symmetry. 
Depending on the spin of DM, the new EW charged fields are different. 
We have classified our DM models as shown in Table \ref{tab-DMtypes} and
discussed the phenomenology in each model. 
The DM relic abundance can be achieved thermally in all DM models. 
One important issue is how we can evade strong bounds from DM direct detection experiments, without spoiling the thermal production of DM. 
We have found that the complex DM model has been almost excluded by the latest XENON1T result. 
In the Dirac DM model, $s$-wave contribution is dominant in the annihilation,
so the bound from the DM direct detection is relatively weak, and several hundreds GeV DM is still allowed. The future sensitivity of the XENONnT/LZ experiments will probe the remaining parameter space in the Dirac model. 
The real DM and the Majorana DM, on the other hand, could evade the direct detection bound,
although small mass difference between the DM and the extra EW charged particle is required
to evade too low Landau pole scale. 
One interesting point is that we can test the real DM model at the indirect detection. 

We have also investigated a possibility that the lepton portal DM models can explain the discrepancy 
in the muon anomalous magnetic moment, together with the DM thermal production. 
We have found that large enough $\Delta a_\mu$ cannot be induced in the minimal models, 
where either singlet or doublet vectorlike lepton (or slepton) exists. 
This consequence is common to the four DM types. 
However, when we consider the extended model with both singlet and doublet, 
$\Delta a_\mu$ can easily be accommodated. 
In Sec.~\ref{sec;DMLR}, we have demonstrated this possibility in both real and Majorana DM models, and clarified explicit correlations between the annihilation cross section of DM and $\Delta a_\mu$, as shown in Eqs.\,\eqref{eq-DAMreal} and \eqref{eq-DAMmaj}.
If there is DM behind the discrepancy, there are both $SU(2)_L$ doublet and singlet fields in addition to DM, and their couplings with DM are large. 
As far as detectability is concerned, 
it may be difficult to prove the explanations in both cases, depending on the parameter space.
If DM is lighter than a few TeV and the mass difference is larger than 10\,\%, 
the indirect searches for DM probably conclude the real DM scenario.
The LHC experiments possibly prove both real and Majorana DM scenarios,
if DM mass is several hundreds of GeV.
In other parameter region, we may be able to test our scenarios in flavor physics,
turning on the other couplings between DM and other leptons.

In our study, we have focused on heavy DM region, $m_{DM}\geq100$\,GeV. 
There would be allowed regions, even if DM is lighter than EW gauge bosons, 
since the direct detection experiments become insensitive to light mass DM. 
In such a light DM case, however, it may be difficult to achieve the relic abundance of DM thermally, 
due in part to lower limits on masses of the EW charged new particles from the collider experiments. 
The resulting large mass splitting between DM and the new charged particles gives rise to the inactive coannihilation mechanism. 
Without the coannihilation, the DM production relies on the DM pair annihilation, 
whose cross section scales as $\VEV{\sigma v} \propto \la^4 m_{DM}^2/m_E^4$. 
It is easy to see that DM cannot be so light when $m_E \gtrsim100$\,GeV while keeping perturbativity. 
This suggests that thermal production of DM will impose a lower limit on DM mass. 
In addition, as DM gets lighter, cosmological and astrophysical searches, 
such as CMB observations and indirect detection, will grow in importance. 
The bounds depend on detail of DM models and need a further dedicated analysis, 
that is beyond the scope of this paper.

\section*{Acknowledgment} 
This work is supported in part by the Grant-in-Aid for Scientific Research from the
Ministry of Education, Science, Sports and Culture (MEXT), 
Japan No.\ 18K13534 (J.K.),  No. 19H04614, No. 19H05101 and No. 19K03867 (Y.O.). 
The work of J.K. is supported in part by the Department of Energy (DOE) under Award No.\ DE-SC0011726. 
The work of S.O. is supported in part by NSERC, Canada, and JSPS Overseas Challenge Program for Young Researchers.


\appendix
\section{Analytic expressions}
\label{app:analytics}

In this Appendix, we summarize analytic expressions and intermediate results, which have been omitted in the main text. 
We focus on the minimal models, and consider only the case that DM couples to the left-handed leptons $\lp^i$, introducing the vectorlike doublet $L$ ($\wt{L}$), whose neutral and charge component are assumed to be degenerate $m_E=m_N$ ($m_{\wt{E}}=m_{\wt{N}}$). 
However, the results can easily be translated into the case of right-handed leptons $e_R^i$, by replacing $\la_L^i \to \la_R^{*i}$ and taking the weak charges of leptons into account. 
For abbreviation, we use $\mu \equiv m_L^2/m_X^2$ and $\eps_i \equiv m_i^2/m_X^2$. 
In a part of calculation of 1-loop annihilations into $VV^\prime$, we exploit \texttt{Package-X$\_$2.0} \cite{Patel:2016fam}.

\subsection{Complex scalar DM}
\label{app:Sc}

For scalar DM, the relevant interaction is given by 
\begin{equation}
{\cal L}_S = - \la_L^{*i} \, X^\dagger \ol{L}_R \, \lp_L^i + h.c. ,
\end{equation}
where $\lp^i=e^i,\nu_L^i$ denote charged leptons or neutrinos.

\subsubsection{Annihilation}

\begin{center}
{1. $XX^\dagger \to \lp^i \bar{\lp}^j$}
\end{center}
For the velocity expansion,
\begin{equation}
(\sigma v)_{XX^\dagger \to \lp^i \bar{\lp}^j} = a_{ij} + b_{ij} v^2 ,
\end{equation}
we find 
\begin{equation}
\begin{split}
a_{ij} & = \frac{(\eps_i + \eps_j)}{32\pi m_X^2 (1+\mu)^2} |\la_L^i \la_L^j|^2 ,\\
b_{ij} & = \frac{|\la_L^i \la_L^j|^2}{48\pi m_X^2 (1 + \mu)^2} ,
\end{split}
\end{equation}
where we only keep the leading order terms in $\eps_i$ and $\eps_j$.
\\
\begin{center}
{2. $XX^\dagger \to \lp^i \bar{\lp}^j V$}
\end{center}

The differential cross section is expressed by 
\begin{equation}
v d \sigma_{\lp^i \bar{\lp}^j V} 
= \frac{ \left| {\cal M}_V \right|^2 }{ 128 \pi^3 } dx dy ,
\end{equation}
where $ x \equiv 2 E_V / \sqrt{s}$ and $y \equiv 2 E_f / \sqrt{s}$. 
The cross section is obtained by performing $x$ and $y$ integrals over $\sqrt{\xi_V} \leq x \leq 1 + \xi_V/4$ and $y_-  \leq y \leq y_+$, 
where $\xi_V \equiv m_V^2 / m_X^2$ ($V=Z,W,h$) and 
\begin{equation}
y_{\pm} = \frac{1}{2} \left( 2 - x \pm \sqrt{ x^2 - \xi_V } \right) .
\end{equation}
The calculation has been done in the $s$-wave limit and we neglect lepton masses. 

\paragraph{\it Squared amplitudes:}
\begin{equation}
\begin{split}
\left| {\cal M}_\gamma^{S_c} \right|^2 
   &= \frac{ 32\pi \alpha Q_f^2 | \la_L^i \la_L^j |^2 }{ m_X^2 } \, f_\gamma^{S_c} (x,y) \\
\left| {\cal M}_Z^{S_c} \right|^2 
   & = \frac{ 32\pi \alpha | \la_L^i \la_L^j |^2 }{ m_X^2 } \frac{((T_3)_{\lp_L}-Q_\lp s_W^2)^2}{c_W^2 s_W^2} \, f_Z^{S_c} (x,y) \\
\left| {\cal M}_W^{S_c} \right|^2 
   & = \frac{ 32\pi \alpha |\la_L^i \la_L^j|^2 }{ m_X^2 s_W^2} \, f_W^{S_c} (x,y) \\
\left| {\cal M}_h^{S_c} \right|^2 
   & = \frac{ |\la_L^i \la_L^j|^2 }{ m_X^2 } 
       \left[ \left(\frac{m_i}{v}\right)^2 \frac{ 1 - x + \xi_h/4 }{ ( 3 + \mu - 2x - 2y )^2 } + \left(\frac{m_j}{v}\right)^2 \frac{ 1 - x + \xi_h/4 }{ ( 1 - \mu - 2y )^2 } \right] ,
\end{split}
\end{equation}
where we mean $\lp^i \bar{\lp}^j W = e^i \bar{\nu}^j W^+ + \bar{e}^i \nu^j W^-$ and 
\begin{equation}
f_V^{S_c} (x,y) = \frac{ (1-x) ( 2 - 2x + x^2 - 4y + 2xy + 2y^2 ) + \frac{\xi_V}{4} ( x^2 - 2x + 2 ) }{ ( 1 - \mu - 2 y )^2 ( 3 + \mu - 2x - 2y )^2 } .
\end{equation}

\paragraph{\it Differential cross sections:}
\begin{align}
\frac{d(\sigma v)_{\lp^i\bar{\lp}^j\gamma}^{S_c}}{dx} 
   & = \frac{\alpha Q_\lp^2 |\la_L^i \la_L^j|^2}{16\pi^2 m_X^2} (1-x) \left[ \frac{2x}{(1+\mu)(1+\mu-2x)} - \frac{x}{(1+\mu-x)^2} \right. \nonumber\\
   &\quad \left. - \, \frac{(1+\mu)(1+\mu-2x)}{2(1+\mu-x)^3} \log\left(\frac{1+\mu}{1+\mu-2x}\right) \right] ,\\
\frac{ d (\sigma v)_{\lp^i \bar{\lp}^j Z}^{S_c} }{ dx }
   & = \frac{\alpha |\la_L^i \la_L^j|^2}{16\pi^2 m_X^2} \frac{((T_3)_{\lp_L} - Q_\lp s_W^2)^2}{c_W^2 s_W^2} \, g_Z^{S_c} (x) ,\\
\frac{d(\sigma v)_{\lp^i\bar{\lp}^jW}^{S_c}}{dx}
   & = \frac{\alpha |\la_L^i \la_L^j|^2}{16\pi^2 m_X^2 s_W^2} \, g_W^{S_c} (x) ,\\
\frac{d(\sigma v)_{\lp^i\bar{\lp}^j h}}{dx}
   & = \frac{|\la_L^i \la_L^j|^2}{128\pi^3 m_X^2} \frac{m_i^2 + m_j^2}{v^2} \frac{(1-x+\xi_h/4) \sqrt{x^2-\xi_h} }{(1+\mu)(1+\mu-2x)+\xi_h} ,
\end{align}
where 
\begin{align}
g_V^{S_c} (x) 
   & = \frac{ 2 (1-x) \sqrt{x^2 - \xi_V} }{ (1+\mu)(1+\mu-2x) + \xi_V } 
   - \frac{ (1-x) \sqrt{x^2 - \xi_V} }{ (1+\mu-x)^2 } \nonumber\\
   & \quad - \frac{ (1-x) (1+\mu) (1+\mu-2x) }{(1+\mu-x)^3} \tanh^{-1} \left( \frac{\sqrt{x^2-\xi_V}}{1+\mu-x} \right) + \frac{ \xi_V }{ 2(1+\mu-x)^2 } \nonumber\\
   & \quad \times \left\{ \frac{ (x-2)^2 \sqrt{x^2 - \xi_V} }{ (1+\mu)(1+\mu-2x) + \xi_V } + \frac{ x^2 - 2x + 2 }{ 1 + \mu - x } \tanh^{-1} \left( \frac{\sqrt{x^2-\xi_V}}{1+\mu-x} \right) \right\} .
\end{align}
The results are consistent with \cite{Toma:2013bka,Giacchino:2013bta,Ibarra:2014qma}.

\paragraph{\it Cross sections:}
\begin{align}
(\sigma v )_{\lp^i \bar{\lp}^j \gamma}^{S_c} 
& = \frac{ \alpha Q_\lp^2 | \la_L^i \la_L^j |^2 }{ 32 \pi^2 m_X^2 } 
\left[ (1+\mu) \left\{ \frac{\pi^2}{6} - \log^2\left(\frac{1+\mu}{2\mu}\right) 
- 2 {\rm Li}_2 \left(\frac{1+\mu}{2\mu}\right) \right\}\right. \nonumber\\
&\quad \left. + \, \frac{4\mu+3}{\mu+1} + \frac{(4\mu+1)(\mu-1)}{2\mu} \log\left(\frac{\mu-1}{\mu+1}\right) \right] ,\\
\nonumber\\
(\sigma v)_{\lp^i \bar{\lp}^j h}^{S_c} 
& = \frac{ |\la_L^i \la_L^j|^2 }{ 128 \pi^3 m_X^2 } 
\frac{ m_i^2 + m_j^2 }{ v^2 } \frac{1}{64(1+\mu)^3} \nonumber\\
& \times \bigg[ (1+\mu) (4-\xi_h) \left\{ 4\mu^2 + \mu (4-\xi_h) + 3 r_h \right\} \\
& \quad - 4 (\mu-1) (2+2\mu-\xi_h) \left\{ (1+\mu)^2 - \xi_h \right\} \log \left( \frac{2+2\mu-\xi_h}{2(\mu-1)} \right) \nonumber\\
& \quad + 4 \xi_h \left\{ \mu (4 + \xi_h) + 4 - \xi_h \right\} \log \left(\frac{\xi_h}{4} \right) 
\bigg] ,\nonumber
\end{align}
where ${\rm Li}_2 (z) = - \int^1_0 dt \log(1-zt)/t$ is the dilogarithm function. 
The cross sections for the $Z$ and $W$ emissions are obtained by numerical integrations. 
\\
\begin{center}
{3. $XX^\dagger \to VV'$}
\end{center}
\begin{align}
\label{eq;XXVV}
(\sigma v)_{\gamma \gamma}^{S_c} 
& = \frac{\alpha^2 Q_e^4}{128\pi^3 m_X^2} \left( \sum_i |\la_L^i|^2 \right)^2
\left[ 2 + {\rm Li}_2 \left(\frac{1}{\mu}\right) - {\rm Li}_2 \left(-\frac{1}{\mu}\right) - 2 \mu \arcsin^2\left(\frac{1}{\sqrt{\mu}}\right) \right]^2 , \nonumber\\
(\sigma v)_{\gamma Z}^{S_c} 
& = \frac{\alpha^2 Q_e^2 \left( (T_3)_{e_L} - Q_e  s_W^2 \right)^2}{64\pi^3 m_X^2 c_W^2 s_W^2} 
\left( \sum_i |\la_L^i|^2 \right)^2 \left( 1 - \frac{\xi_Z}{4} \right) 
\left| A^S_{\gamma Z} \right|^2 ,\\
(\sigma v)_{ZZ}^{S_c} 
& = \frac{\alpha^2}{512\pi^3 m_X^2 c_W^4 s_W^4} 
\left(\sum_i \sum_{\lp=e,\nu_L} |\la_L^i|^2 \left( (T_3)_{\lp_L} - Q_\lp  s_W^2 \right)^2 \right)^2 \sqrt{1-\xi_Z}
\left| A^S_{ZZ} \right|^2 , \nonumber\\
(\sigma v)_{WW}^{S_c} 
& = \frac{\alpha^2}{1024\pi^3 m_X^2 s_W^4} \left(\sum_i |\la_L^i|^2\right)^2 \sqrt{1-\xi_W}
\left| A^S_{WW} \right|^2 , \nonumber
\end{align}
where $A^S_{\gamma Z}$ is  
\begin{equation}
\begin{split}
A^S_{\gamma Z} & 
= 2 - \frac{\xi_Z}{4-\xi_Z} \bigg[ \frac{4\mu}{\sqrt{\mu-1}} \arccot\left(\sqrt{\mu-1}\right) 
    + 2 \log\left(\frac{\mu-1}{\mu}\right) \\
& \quad - 2 \sqrt{ \frac{4\mu}{\xi_Z} - 1 } \arccot\left(\sqrt{\frac{4\mu}{\xi_Z} - 1}\right) 
       - \log\left(\frac{\xi_Z}{\mu}\right) + i \pi \bigg] \\
& \quad + m_X^2 \bigg\{ \frac{\mu(4-4\mu-\xi_Z)}{1-\mu} \,\,
       C_0 \left( m_Z^2, 4m_X^2, 0; m_L^2, m_L^2, m_L^2 \right) \\
& \quad + \frac{\mu(4+\xi_Z)(-2+2\mu+\xi_Z)}{2(1-\mu)(4\mu+\xi_Z)} \,\,
       C_0 \left( \frac{m_Z^2}{2}-m_X^2, m_X^2, 0; m_L^2, 0, m_L^2 \right) \\
& \quad + \left[ \frac{\xi_Z}{4} - \frac{4(1+\mu)}{4-\xi_Z} + \frac{4\mu(1+\mu)}{4\mu+\xi_Z} \right]
       C_0 \left( \frac{m_Z^2}{2}-m_X^2, m_X^2, m_Z^2; 0, m_L^2, 0 \right) \\
& \quad + \left[ \mu + \frac{\xi_Z}{2(1-\mu)} - \frac{4(1+\mu)}{4-\xi_Z} \right] 
       C_0 \left( \frac{m_Z^2}{2}-m_X^2, m_X^2, m_Z^2; m_L^2, 0, m_L^2 \right) \bigg\} ,\\
\end{split}
\end{equation}
where $C_0$ is the Passarino-Veltman function defined by 
\begin{equation}
C_0 (p_1^2,(p_1-p_2)^2,p_2^2;m_1^2,m_2^2,m_3^2) 
= \int \frac{d^4l}{i\pi^2} \frac{1}{l^2-m_1^2} \frac{1}{(l+p_1)^2-m_2^2} \frac{1}{(l+p_2)^2-m_3^2} .
\end{equation}
The above results are consistent with Ref.~\cite{Ibarra:2014qma}.

The loop function $A^S_{ZZ}$ is more complicated. 
To the best of our knowledge, the explicit expression is given for the first time, which is described by 
\begin{equation}
\left| A_{ZZ}^{S} \right|^2 = |A|^2 + 2 |B|^2 ,
\end{equation}
where 
\begin{align}
A & = 4 - \frac{\xi_Z }{1-\xi_Z} \bigg[ \frac{4\mu}{\sqrt{\mu-1}} \arccot\left(\sqrt{\mu-1}\right) + 2 \log\left(\frac{\mu-1}{\mu}\right) \nonumber\\
& \quad - 2\sqrt{\frac{4\mu}{\xi_Z}-1} \arccot\left(\sqrt{\frac{4\mu}{\xi_Z}-1}\right) - \log\left(\frac{\xi_Z}{\mu}\right) + i\pi \bigg] ,\nonumber\\
&\quad + \frac{\xi_Z}{1-\xi_Z} m_X^2 \bigg\{ 2 \left[ \frac{4\mu}{\xi_Z} - \frac{\mu(4\mu-6+\xi_Z)}{(\mu-1)} \right] C_0 \left( 4m_X^2, m_Z^2, m_Z^2; m_L^2, m_L^2, m_L^2 \right) \\
& \quad - \left[ \frac{4}{\xi_Z} + \frac{(\mu-1)^3+(\mu+1)(2-\xi_Z)}{\mu(\mu-1)} \right] C_0 \left( m_X^2, m_Z^2-m_X^2, m_Z^2; m_L^2, 0, m_L^2 \right)  \nonumber\\ 
& \quad - \frac{1+\mu^2-\xi_Z}{\mu} \, C_0 \left( m_X^2, m_Z^2-m_X^2, m_Z^2; 0, m_L^2, 0 \right) \bigg\} , \nonumber
\end{align}
and
\begin{align}
B & = \frac{\xi_Z}{1-\xi_Z} 
\bigg[ \frac{4\mu}{\sqrt{\mu-1}} \arccot\left(\sqrt{\mu-1}\right) + 2 \log\left(\frac{\mu-1}{\mu}\right) \nonumber\\
&\quad - 2 \sqrt{\frac{4\mu}{\xi_Z}-1} \arccot\left(\sqrt{\frac{4\mu}{\xi_Z}-1}\right) - \log\left(\frac{\xi_Z}{\mu}\right) + i\pi \bigg] \nonumber\\
& \quad + \frac{\xi_Z}{1-\xi_Z} m_X^2 
\bigg\{ \frac{2\mu(-2+\xi_Z)}{\mu-1} \, C_0 \left( 4m_X^2, m_Z^2, m_Z^2; m_L^2, m_L^2, m_L^2 \right) \\
& \quad + \frac{\mu^2-1+\xi_Z}{\mu} \, C_0 \left( m_X^2, m_Z^2-m_X^2, m_Z^2; 0, m_L^2, 0 \right) \nonumber\\
& \quad + \left[ \frac{(\mu-1)^2}{\mu} + \frac{4\mu}{\mu-1} + \frac{\xi_Z(1-3\mu)}{\mu(\mu-1)} \right] C_0 \left( m_X^2, m_Z^2-m_X^2, m_Z^2; m_L^2, 0, m_L^2 \right) \bigg\}. \nonumber
\end{align}
In the limit of $m_Z \to 0$, the latter function is not contributing $B \approx {\cal O}(m_Z^2)$, while the former function is approximate to 
\begin{equation}
A \approx 2 \left[ 2 + {\rm Li}_2 \left(\frac{1}{\mu}\right) - {\rm Li}_2 \left(-\frac{1}{\mu}\right) - 2 \mu \arcsin^2\left(\frac{1}{\sqrt{\mu}}\right) \right] + {\cal O}(m_Z^2) ,
\end{equation}
which is equivalent to the loop function appearing in the photon contribution Eq.~(\ref{eq;XXVV}), as it should be. 
For $WW$, the amplitude is obtained by the replacement $A^S_{WW}=\left.A^S_{ZZ}\right|_{m_Z\to m_W}$.

\subsubsection{Direct detection}

\paragraph{\it Charge radius operator:}
The photon penguin diagram induces the so-called DM charge radius operator, 
\begin{equation}
{\cal L}_{\rm eff}^S \supset b_X (i X^\dagger \overleftrightarrow{\partial_\mu} X) \partial_\nu F^{\mu\nu} ,
\end{equation}
which in turn provides the DM coupling to the quark vector current, $C_{V,q} = - e Q_q b_X$, through the equation of motion for photon. 
The penguin contribution involving $L$ and $\lp^i$ is given by 
\begin{equation}
(b_X)_{\lp^i} = \frac{eQ_\lp |\la_L^i|^2}{16\pi^2 m_X^2} \, \hat{b}_X (\mu,\eps_i) .
\end{equation}
Loop functions are 
\begin{equation}
\begin{split}
\hat{b}_X (\mu,\eps) & = 
- \frac{1}{3}
 \left[ 
    \frac{\mu-\eps}{ \Delta^{3/2} } 
       \left\{ ( \mu + \eps + 1 ) \Delta - 4 \mu \eps \right\} 
       \tanh^{-1} \left( \frac{ \Delta^{1/2} }{ \mu + \eps - 1 } \right) \right. \\
  & \quad \left. + \, \frac{ 2 + \mu + \eps }{2} \log \left( \frac{\eps}{\mu} \right) 
       + \frac{(\mu-\eps)(\mu+\eps-1)}{\Delta}
 \right] ,
\end{split}
\end{equation}
with $\Delta \equiv \mu^2 + (\eps-1)^2 - 2 \mu ( \eps + 1)$.

\paragraph{\it $Z$-penguin:}
The $Z$ penguin diagram also induces the DM coupling to the quark vector current, 
\begin{equation}
{\cal L}_{\rm eff}^S \supset C_{V,q} (i X^\dagger \overleftrightarrow{\partial_\mu} X) (\ol{q}\gamma^\mu q) .
\end{equation}
The contribution is expressed by $C_{V,q}^Z = \sum_{\lp^i} (C_{V,q}^Z)_{\lp_L^i}$ with 
\begin{align}
(C_{V,q}^Z)_{e_L^i} &= \frac{2\sqrt{2}G_F g_{V,q}}{16\pi^2 m_X^2} \, g_{A,e^i} |\lambda_L^i|^2 \, \hat{a}_Z^S (\mu,\eps_i) ,\\
(C_{V,q}^Z)_{\nu_L^i} &=0 ,
\end{align}
with $g_{V,q} = (T_3)_q - Q_q s_W^2$ and $g_{A,\lp} = (T_3)_\lp$, and the loop function is 
\begin{equation}
\hat{a}_Z^S (\mu,\eps) = \eps \left[ 1 + \frac{ 1 + \mu - \eps }{ 2 } \log \left( \frac{\eps}{\mu} \right)
+ \frac{ \Delta + 2 \mu }{ \Delta^{1/2} } \tanh^{-1} \left( \frac{ \Delta^{1/2} }{ \mu + \eps - 1 } \right) \right] .
\end{equation}
It should be noted that, if DM couples to the singlet leptons $e_R^i$, the sign of the $Z$ penguin contribution is flipped: 
\begin{equation}
(C_{V,q}^Z)_{e_R^i} = - \frac{2\sqrt{2}G_F g_{V,q}}{16\pi^2 m_X^2} \, g_{A,e^i} 
|\lambda_R^i|^2 \, \hat{a}_Z^S (\mu,\eps_i) ,
\end{equation}
with $g_{A,e^i}=-1/2$.

\subsection{Real scalar DM}
\label{app:Sr}
The notation of Yukawa interaction is the same as the complex case, but DM is self-conjugate in this case ($X^\dagger=X$).

\subsubsection{Annihilation}

\begin{center}
{1. $XX \to \lp^i \bar{\lp}^j$}
\end{center}
Similarly, for the following velocity expansion, 
\[
(\sigma v)_{XX \to \lp^i \bar{\lp}^j} = a_{ij} + b_{ij} v^2 + c_{ij} v^4 ,
\]
the coefficients are 
\begin{equation}
\begin{split}
a_{ij} & = \frac{(\eps_i+\eps_j)}{8\pi m_X^2 (1+\mu)^2} |\la_L^i \la_L^j|^2 ,\\
b_{ij} & = \frac{ -(1+2\mu) (\eps_i+\eps_j)}{12\pi m_X^2 (1 + \mu)^4} |\la_L^i \la_L^j|^2 ,\\
c_{ij} & = \frac{|\lambda_L^i \lambda_L^j|^2}{60\pi m_X^2 (1+\mu)^4} .
\end{split}
\end{equation}
\\
\begin{center}
{2. $XX \to \lp^i \bar{\lp}^j V$ and $VV'$}
\end{center}
Cross sections for these processes are just four times larger than the corresponding cross sections in the complex DM. The spectrum of vector boson in $XX\to \lp^i \bar{\lp}^j V$ and the dependence on $\mu$ and $\xi_V$ in both processes are completely same.

\subsection{Dirac DM}
\label{app:Fc}
Relevant interaction is given by 
\begin{equation}
{\cal L}_F = - \la_L^{*i} \, \wt{L}^\dagger \ov{\chi}_R \, \lp_L^i + h.c. .
\end{equation}
All results given here are consistent with Ref.~\cite{Ibarra:2015fqa}.

\subsubsection{Annihilation}

\begin{center}
{1. $\chi\ov{\chi} \to \lp^i \bar{\lp}^j$}
\end{center}
\begin{equation}
(\sigma v)_{\chi\ov{\chi} \to \lp^i \bar{\lp}^j} = \frac{|\la_L^i \la_L^j|^2}{32\pi m_\chi^2 (1+\mu)^2} .
\end{equation}

\subsubsection{Direct detection}

Effective DM interactions to photon are described by 
\begin{equation}
{\cal L}_{\rm eff}^{F} = b_\chi \ol{\chi} \gamma^\mu \chi \partial^\nu F_{\mu\nu} 
+ \frac{\mu_\chi}{2} \ol{\chi} \sigma^{\mu\nu} \chi F_{\mu\nu} 
+ a_\chi \ol{\chi} \gamma^\mu \gamma_5 \chi \partial^\nu F_{\mu\nu} 
+ i \frac{d_\chi}{2} \ol{\chi} \sigma^{\mu\nu} \gamma_5 \chi F_{\mu\nu} . 
\label{eq;multipoleDM}
\end{equation}
The contributions involving $\wt{L}$ and $\lp^i$ are written as follows: 
\begin{align}
(b_\chi)_{\lp^i} & = \frac{eQ_\lp |\la_L^i|^2}{16\pi^2m_\chi^2} \, \hat{b}_\chi (\mu,\eps_i) ,\\
(\mu_\chi)_{\lp^i} & = \frac{eQ_\lp |\la_L^i|^2}{16\pi^2m_\chi} \, \hat{\mu}_\chi (\mu,\eps_i) ,\\
(a_\chi)_{\lp^i} & = \frac{eQ_\lp |\la_L^i|^2}{16\pi^2m_\chi^2} \, \hat{a}_\chi (\mu,\eps_i) ,\\
(d_\chi)_{\lp^i} & = 0.
\end{align}

\paragraph{\it Charge radius operator:}
\begin{align}
\hat{b}_\chi (\mu,\eps) & = - \frac{1}{24} 
\left[ ( 8 \mu - 8 \eps + 1 ) \log \left( \frac{ \eps }{ \mu } \right) 
+ 4 \left( 4 + \frac{ \mu + 3 \eps - 1 }{\Delta} \right) \right. \\
& \left. + \, \frac{ 2 }{ \Delta^{3/2} } 
    \left\{ 8 \Delta^2 + ( 9 \mu + 7 \eps - 5 ) \Delta - 4 \eps ( 3 \mu + \eps -1 ) \right\} 
    \tanh^{-1} \left( \frac{ \Delta^{1/2} }{ \mu + \eps - 1 } \right) 
\right] ,\nonumber
\end{align}

\paragraph{\it Magnetic dipole operator:}
\begin{align}
\hat{\mu}_\chi (\mu,\eps) & = - \frac{1}{2} \left[ \frac{1}{2} ( \eps - \mu ) \log \left( \frac{\eps}{\mu} \right) - 1 
- \frac{ \Delta + \mu + \eps - 1 }{ \Delta^{1/2} } \tanh^{-1} \left( \frac{ \Delta^{1/2} }{ \mu + \eps - 1 } \right) 
\right] ,
\end{align}

\paragraph{\it Anapole operator:}
\begin{equation}
\hat{a}_\chi (\mu,\eps) = \frac{1}{12} \left[ \frac{3}{2} \log \left( \frac{\eps}{\mu} \right) 
+ \frac{3\mu-3\eps+1}{\Delta^{1/2}} \tanh^{-1} \left( \frac{ \Delta^{1/2} }{ \mu + \eps - 1 } \right) 
\right] 
\end{equation}

\paragraph{\it $Z$-penguin:}
The couplings to vector and axial vector currents, 
\begin{equation}
{\cal L}_{\rm eff}^F \supset C_{V,q}^Z \, \ov{\chi} \gamma_\mu \chi \, \ov{q} \gamma^\mu  q 
+ C_{A,q}^Z \, \ov{\chi} \gamma_\mu \gamma_5 \chi \, \ov{q} \gamma^\mu \gamma_5 q ,
\end{equation}
are induced. 
The $Z$ penguin contribution involving $\wt{L}$ and $\lp^i$ is 
\begin{align}
(C_{V,q}^Z)_{e_L^i} &= \frac{ \sqrt{2} G_F g_{V,q} }{16\pi^2} \, g_{A,e^i} | \lambda_L^i |^2 \, \hat{a}_Z^F (\mu,\eps_i) ,\\
(C_{A,q}^Z)_{e_L^i} &= - \frac{ \sqrt{2} G_F g_{A,q} }{16\pi^2} \,  g_{A,e^i} | \lambda_L^i |^2 \, \hat{a}_Z^F (\mu,\eps_i) ,
\end{align}
and $(C_{V,q}^Z)_{\nu_L^i} = (C_{A,q}^Z)_{\nu_L^i}=0$, where the loop function is  
\begin{equation}
\hat{a}_Z^F (\mu,\eps) = \eps \left[ \frac{1+\mu-\eps}{\Delta^{1/2}} \tanh^{-1} \left( \frac{\Delta^{1/2}}{\mu+\eps-1} \right) 
+ \frac{1}{2} \log\left(\frac{\eps}{\mu}\right)  \right] .
\end{equation}
Like the real DM case, if DM couples to the singlet leptons $e_R^i$, the sign of the contribution to $C_{V,q}$ should be flipped, while the contribution to $C_{A,q}$ is unchanged: 
\begin{align}
(C_{V,q}^Z)_{e_R^i} &= - \frac{ \sqrt{2} G_F g_{V,q} }{16\pi^2} \, g_{A,e^i} | \lambda_R^i |^2 \, \hat{a}_Z^F (\mu,\eps_i) ,\\
(C_{A,q}^Z)_{e_R^i} &= - \frac{ \sqrt{2} G_F g_{A,q} }{16\pi^2} \,  g_{A,e^i} | \lambda_R^i |^2 \, \hat{a}_Z^F (\mu,\eps_i) .
\end{align}

\paragraph{\it Higgs exchanging:}
\begin{equation}
{\cal L}_{\rm eff}^F \supset C_{S, q} \, m_q \, \ov{\chi} \chi \, \ov{q} q ,
\end{equation}

\begin{equation}
(C_{S,q}^{\rm Higgs})_{f^i} = \frac{ - \sqrt{2} G_F m_\chi }{ 16\pi^2 m_h^2 } \, |\la_L^i|^2 \, \hat{c}_H (\mu,\eps_i) ,
\end{equation}
with 
\begin{equation}
\hat{c}_H (r,\eps) = \eps \left[ \frac{ \Delta + \mu + \eps - 1 }{\Delta^{1/2}} \tanh^{-1} \left( \frac{\Delta^{1/2}}{\mu+\eps-1} \right) + 1 + \frac{ \mu - \eps }{2} \log\left(\frac{\eps}{\mu}\right) \right] .
\end{equation}

\subsection{Majorana DM}
\label{app:Fr}
The notation of Yukawa interaction is the same as the Dirac case.

\subsubsection{Annihilation}

\begin{center}
{1. $\chi\chi \to \lp^i \bar{\lp}^j$}
\end{center}

\[
(\sigma v)_{\chi\chi \to \lp^i \bar{\lp}^j} = a_{ij} + b_{ij} v^2 ,
\]
with
\begin{equation}
\begin{split}
a_{ij} & = \frac{(\eps_i+\eps_j)}{64\pi m_\chi^2 (1+\mu)^2} |\la_L^i \la_L^j |^2 ,\\
b_{ij} & = \frac{(1+\mu^2)}{48\pi m_\chi^2 (1+\mu)^4} |\la_L^i \la_L^j |^2 .
\end{split}
\end{equation}
\\
\begin{center}
{2. $\chi\chi \to \lp^i \bar{\lp}^j V$}
\end{center}

\paragraph{\it Squared amplitudes:}
\begin{align}
\left| {\cal M}_\gamma^{F_r} \right|^2 
   & = \frac{1}{2} \left| {\cal M}_\gamma^{S_c} \right|^2 ,\\
\left| {\cal M}_Z^{F_r} \right|^2 
   & = \frac{16\pi \alpha |\la_L^i \la_L^j|^2}{m_\chi^2} \frac{ ( (T_3)_{\lp_L} - Q_\lp s_W^2 )^2 }{ c_W^2 s_W^2 } \, f_Z^{F_r} (x,y) ,\\
\left| {\cal M}_W^{F_r} \right|^2 
   & = \frac{16\pi \alpha |\la_L^i \la_L^j|^2}{m_\chi^2 s_W^2} \, f_W^{F_r} (x,y) ,\\
\left| {\cal M}^{F_r}_h \right|^2 
   & = \frac{1}{2} \left| {\cal M}^{S_c}_h \right|^2 ,
\end{align}
where 
\begin{equation}
f_V^{F_r} (x,y) = \frac{ ( 1 - x + \frac{\xi_V}{4} )( 2 - 2x + x^2 - 4y + 2xy + 2y^2 - \frac{\xi_V}{2} ) }
{ ( 1 - \mu - 2 y )^2 ( 3 + \mu - 2x - 2y )^2 } .
\end{equation}

\paragraph{\it Differential cross sections:}
\begin{align}
\frac{ d (\sigma v)_{\lp^i \bar{\lp}^j Z}^{F_r} }{ dx }
& = \frac{ \alpha | \la_L^i \la_L^j |^2 }{ 32 \pi^2 m_\chi^2 } 
\frac{ ( (T_3)_{\lp_L} - Q_\lp s_W^2 )^2 }{ c_W^2 s_W^2 } \, g_Z^{F_r} (x) ,\\
\frac{ d (\sigma v)_{\lp^i \bar{\lp}^j W}^{F_r} }{ dx }
& = \frac{ \alpha | \la_L^i \la_L^j |^2 }{ 32 \pi^2 m_\chi^2 s_W^2 } \, g_W^{F_r} (x) ,
\end{align}
where
\begin{align}
g_V^{F_r}(x) & = \left( 1 - x + \frac{\xi_V}{4} \right) 
\bigg[ \frac{ 2 \sqrt{x^2-\xi_Z} }{ (1+\mu) (1+\mu-2x) + \xi_Z } - \frac{ \sqrt{ x^2 - \xi_Z } }{ (1+\mu-x)^2 } \nonumber\\
&\quad - \frac{ (1+\mu) (1+\mu-2x) + \xi_Z }{ (1+\mu-x)^3 } \, \tanh^{-1} \left( \frac{\sqrt{x^2-\xi_Z}}{1+\mu-x} \right) \bigg] .
\end{align}
\\
\begin{center}
{3. $\chi\chi \to VV'$}
\end{center}

\begin{align}
(\sigma v)_{\gamma\gamma}^{F_r} 
   & = \frac{ \alpha^2  Q_e^2 }{256\pi^3 m_\chi^2} \left( \sum_i |\la_L^i|^2 \right)^2 
\left[ {\rm Li}_2 \left(\frac{1}{\mu}\right) - {\rm Li}_2 \left(-\frac{1}{\mu}\right) \right]^2, \\
(\sigma v)_{\gamma Z}^{F_r} 
   & = \frac{\alpha^2 Q_e^2 \left( (T_3)_{e_L} - Q_e  s_W^2 \right)^2}{512\pi^3 m_\chi^2 c_W^2 s_W^2} \left( \sum_i |\la_L^i|^2  \right)^2 
   \frac{ \left| A^F_{\gamma Z} \right|^2 }{ \mu^2 (1-\frac{\xi_Z}{4} ) ( 1 - \frac{\xi_Z^2}{16 \mu^2} )^2 } , \\
(\sigma v)_{ZZ}^{F_r} 
   & =  \frac{\alpha^2}{1024\pi^3 m_\chi^2 c_W^4 s_W^4} 
\left( \sum_i \sum_{\lp=e,\nu_L} |\la_L^i|^2 \left( (T_3)_{\lp_L} - Q_\lp  s_W^2 \right)^2 \right)^2 
\frac{ \left| A^F_{ZZ} \right|^2 }{ \mu^2 \sqrt{ 1 - \xi_Z } } , \\
(\sigma v)_{WW}^{F_r} 
   & =  \frac{\alpha^2}{2048\pi^3 m_\chi^2 s_W^4} \left( \sum_i |\la_L^i|^2 \right)^2 
\frac{ \left| A^F_{WW} \right|^2 }{ \mu^2 \sqrt{ 1 - \xi_W } } , 
\end{align}
where $A^F_{\gamma Z}$, $A^F_{ZZ}$ and $A^F_{WW}$ are
\begin{equation}
\begin{split}
A^F_{\gamma Z} & 
  = m_\chi^2 \left( 1 - \frac{\xi_Z}{4\mu} \right) 
  \bigg\{
    2 \mu \left( 1 - \frac{\xi_Z}{4} \right)^2 
       C_0 \left( m_\chi^2 , 0 , \frac{m_Z^2}{2} - m_\chi^2 ; 0 , m_{\wt{L}}^2 , m_{\wt{L}}^2 \right) \\
& + \xi_Z \left( \frac{1+\mu^2}{2} + \frac{\xi_Z(\mu-1)}{4} + \frac{\xi_Z^2}{16} \right)
       C_0 \left( m_\chi^2 , m_Z^2 , \frac{m_Z^2}{2} - m_\chi^2 ; m_{\wt{L}}^2 , 0 , 0 \right) \\
& + \left( \mu + \frac{\xi_Z}{4} \right) \left( 2 - \xi_Z + \frac{\mu \xi_Z}{2} \right) 
       C_0 \left( m_\chi^2 , m_Z^2 , \frac{m_Z^2}{2} - m_\chi^2 ; 0 , m_{\wt{L}}^2 , m_{\wt{L}}^2 \right) 
   \bigg\} \\
& + \frac{\mu \xi_Z}{2} \left( 1 - \frac{\xi_Z^2}{16\mu^2} \right) 
       \left[ 2 \sqrt{\frac{4\mu}{\xi_Z} -1} \, \arccot \left(\sqrt{\frac{4\mu}{\xi_Z} -1}\right) - \log\left(\frac{\xi_Z}{\mu}\right) + i \pi \right] ,
\end{split}
\end{equation}
and 
\begin{align}
A^F_{ZZ} & = m_\chi^2 
  \bigg\{ 
     \xi_Z \left( 1 + \mu^2 - \xi_Z \right)
       C_0 \left( m_\chi^2 , m_Z^2 , m_Z^2 - m_\chi^2 ; m_{\wt{L}}^2 , 0 , 0 \right) \nonumber\\
&\quad + \left[ 4 \mu + ( \mu^2 - 4 \mu - 1 ) \xi_Z + \xi_Z^2 \right] 
       C_0 \left( m_\chi^2 , m_Z^2 , m_Z^2 - m_\chi^2 ; 0 , m_{\wt{L}}^2 , m_{\wt{L}}^2 \right) 
  \bigg\} \\
&\quad + \mu \xi_Z \left[ 2 \sqrt{\frac{4\mu}{\xi_Z} -1} \, \arccot \left(\sqrt{\frac{4\mu}{\xi_Z} -1}\right) - \log\left(\frac{\xi_Z}{\mu}\right) + i \pi \right] ,\nonumber\\
A^F_{WW} & = \left. A^F_{ZZ} \right|_{m_Z\to m_W} .
\end{align}

\subsubsection{Direct detection}
In Majorana case, $a_\chi$, $C_{A,q}^Z$ and $C_{S,q}^{\rm Higgs}$ are non-vanishing. With the notation of Eq.~(\ref{eq;multipoleDM}), the coefficients are are exactly same as the Dirac case.

\section{Direct detection limit with non-contact interactions}
\label{dipole}
In this Appendix, we briefly explain how we have calculated the number of DM-nuclei scattering events in presence of non-contact interaction, and recast the public limit~\cite{Aprile:2018dbl}, that assumes contract-type DM-nucleon interactions, to our cases. Dedicated studies of this subject can be found in the context of mutipolar dark matter~\cite{Sigurdson:2004zp,Masso:2009mu,Chang:2010en,DelNobile:2012tx,Ho:2012bg,DelNobile:2014eta}, where they make some likelihood analysis to calculate the number of events. 
In this paper, we have performed simpler analysis. 

The total expected scattering rate (per target mass) is expressed in terms of integrals of the differential rate over nuclear recoil energy $E_R$:
\begin{equation}
R = \int^\infty_{E_{R,{\rm th}}} \, dE_R \, \eps(E_R) \, \sum_i \eta_i \, \frac{dR_i}{dE_R},
\label{eq;eventrate}
\end{equation}
where $\eta_i$ is an isotope fraction of target nuclei (see {\em e.g.} Ref.~\cite{Banks:2010eh}) and 
$\eps(E_R)$ an efficiency function for a given recoil energy $E_R$, 
for which we used the black solid line in Fig.~1 of \cite{Aprile:2018dbl}, 
which is a best-fit total efficiency after taking the energy region-of-interest (ROI) selection into account. 
The differential scattering rate (per target mass) is given by 
\begin{equation}
\frac{dR_i}{dE_R} = \frac{\rho_0}{m_\chi m_{N_i}} \int_{|\vec{v}|>v_{\rm min}(E_R)} d^3\vec{v} \, v f_E(\vec{v}) \frac{d\sigma_i(v,E_R)}{dE_R},
\end{equation}
where $v=|\vec{v}|$ and $v_{\rm min}=\sqrt{m_NE_R/(2\mu_{\rm red}^2)}$ is the minimum velocity with $\mu_{\rm red}$ the DM-nucleus reduced mass. 
The differential cross section for an elastic DM-nucleus scattering, $d\sigma/dE_R$, is given by Eq.~(\ref{eq;dsigma}), which involves two nuclear form factors $F(E_R)$ and $F_{\rm spin}(E_R)$, for which we use the Helm form factor normalized to $F(0)=1$, and a spin form factor in \cite{Lewin:1995rx} with a thin-shell approximation, respectively. 
The DM velocity distribution in the earth frame $f_E(\vec{v})$ is obtained by the Galilean transformation of a velocity distribution in the galactic rest frame $f_G(\vec{v})$, for which we use an isotropic Maxwellian distribution with 
a smooth cutoff~\cite{Kopp:2009et,Kopp:2009qt,Kopp:2011yr,Fitzpatrick:2010br,Gresham:2013mua},
\begin{equation}
f_G(\vec{v})=\frac{N}{(v_0 \sqrt{\pi})^3} \left(e^{-\vec{v}^2/v_0^2} - e^{-v_{\rm esc}^2/v_0^2}\right) \theta(v_{\rm esc}-v),
\end{equation}
where $N$ is a normalization constant such that $\int d^3\vec{v} \, f_G(\vec{v})=1$. In our analysis, we take $\rho_0=0.3$\,GeV/cm$^3$, $v_0=220$\,km/s and $v_{\rm esc}=544$\,km/s~\cite{Fitzpatrick:2010br,Gresham:2013mua}. 

Combining all pieces above, we can calculate the expected number of DM-nuclei scattering events detected at experiments, given masses of DM and vectorlike lepton and Yukawa coupling. 
Then, we can translate the null results at direct detection into limits on our models by comparing our calculation with experimental results at scattering event level. 
To pull the number of exclusion events out of the public limit (the solid black line in Fig.~5 of \cite{Aprile:2018dbl}), we required that contact-type scattering reproduces the exclusion curve. 
The number of exclusion events extracted in this way is $N_{\rm exc.}\simeq 5.6,\,12,\,14$ for $m_{DM}=30$\,GeV, 100\,GeV, 1\,TeV, respectively. 
This is the way we have drawn the red lines in Figs.~\ref{fig;plots1} and \ref{fig;plots2} in the text. 
We have confirmed that our approach produces consistent results with \cite{Ibarra:2015fqa} which studied the same type of DM candidate before\footnote{The definition of our signal region differs by a factor 2 from the definition in Ref.~\cite{Ibarra:2015fqa}, in which the authors defined their signal region as the lower half of the mean nuclear recoil band ({\em e.g.} the red solid line in Fig.~4 of \cite{Akerib:2013tjd}). Thus, our bound is aggressive by the factor. Once we use their definition instead of ours, we can find practically the same exclusion line as what they gave in \cite{Ibarra:2015fqa}.}.

We would like to note that even if we take statistical methods more carefully, the results are not affected so much. 
Taking the statistical fluctuations into account, the number of DM-nuclei scattering events is expressed by 
\begin{equation}
R = \int_{S_1^{\rm min}}^{S_1^{\rm max}} dS_1
\sum_{n=0}^\infty {\rm Gauss}(S_1|n,\sqrt{n}\sigma_{\rm PMT}) 
\int_{E_{\rm min}}^{\infty} dE_R \, \tilde{\eps}(E_R) \, {\rm Poiss}(n|\nu(E_R)) \sum_i \frac{dR_i}{dE_R}, 
\end{equation}
with $\tilde{\eps}(E_R)$ an efficiency function\footnote{This is a different function from Eq.~(\ref{eq;eventrate}).}, and $\nu(E_R)$ the expected number of photoelectrons (PEs) for a given recoil energy $E_R$, for which, respectively, we can take the green line in Fig.~1 of \cite{Aprile:2018dbl} and the black dotted line in Fig.~13 of \cite{Aprile:2015uzo}. The PMT efficiency can be taken to $\sigma_{\rm PMT}=0.4$~\cite{Aprile:2015lha,Barrow:2016doe}, for example.
With this expression and using $S_1^{\rm min} = 3$\,PEs and $S_1^{\rm max} = 70$\,PEs, we would be able to calculate the expected number of scattering events in the parameter space considered. We have checked that, for some reference points, this statistical treatment gives practically the same results as those of our simpler analysis.

\section{Renormalization Group Equations}
\label{sec-beta}
\subsection{Scalar DM}
\subsubsection{Complex Scalar DM}
The Yukawa interactions are given by  
\begin{align}
 -\Lcal  =  Y_e^{ij} \ol{\ell}^i_L H e^j_R + 
                  \kap\ol{L}_L H E_R + \tkap\ol{E}_L \tH L_R +\la_L^i \ol{\ell}^i_L X L_R + 
                   \la_R^i \ol{E}_L X^* e_R^i + h.c. .  
\end{align}
The RGEs of the Yukawa couplings are given by 
\begin{align}
 16\pi^2 \beta_{\la^i_L} =&\ 2 \tkap Y^{ij}_e \la_R^{*j} + \la_L^i \la_L^{*j} \la_L^j 
                                 + \frac{1}{2} \left(Y_e Y_e^\dagger\right)^{ij} \la^j_L
                          + \la_L^i \left(\frac{1}{2} \tkap^* \tkap + Y_2(X) - \frac{9}{4}(g_1^2 + g_2^2)  \right), \\
16\pi^2 \beta_{\la^i_R} =&\ 4\tkap \la_L^{*j} Y_e^{ji} + \la_R^j \left(Y_e^\dagger Y_e \right)^{ji}  
                    + \la_R^i \left(  \la_R^{*j} \la_R^j 
                                                 +\tkap^*\tkap  + Y_2(X) - \frac{9}{4}(g_1^2 + g_2^2)  \right), \\
16\pi^2 \beta_{\kap} =&\ \kap \left( \frac{3}{2} \kap^*\kap + Y_2(H) - \frac{9}{4} \left(g_1^2+g_2^2\right)  \right), \\
16\pi^2 \beta_{\tkap} =&\  2\la_R^i \left(Y_e^\dagger \right)_{ij} \la_L^j 
                                          + \tkap \left( \frac{3}{2}\tkap^*\tkap 
                                            + \frac{1}{2}\la_L^{*i} \la_L^i +\frac{1}{2} \la_R^{*i} \la_R^i  
                +   Y_2(H) - \frac{9}{4} \left(g_1^2+g_2^2\right)  \right), \\
16\pi^2 \beta_{Y_e^{ij}} =&\  \frac{3}{2} \left(Y_e Y_e^\dagger Y_e\right)^{ij} + 2 \tkap   \la^i_L \la^j_R 
                            + \frac{1}{2}\tkap \left(  \la_L^i  \left( \la^\dagger_L  Y_e\right)^j 
                                                     + \left(Y_e\la_R^\dagger \right)^i \la_R^j 
                                                      \right)   \notag \\ 
           &\quad      + Y^{ij}_e \left( Y_2(H) - \frac{9}{4} \left(g_1^2+g_2^2\right)  \right), 
\end{align}
where 
\begin{align}
 Y_2(X) =&\  2  \la^{*i}_L \la_L^i +    \la^{*i}_R \la_R^i , \\   
 Y_2(H) =&\   \text{Tr}\left(3Y_uY_u^\dagger + 3Y_dY_d^\dagger +Y_eY_e^\dagger \right) 
                         + \kappa^* \kappa + \tkap^* \tkap.  
\end{align}
Here, $Y_u$ and $Y_d$ are the up and down quark Yukawa matrices in the SM, respectively. 
We show the full 1-loop RGEs for completeness.  
We neglect the Yukawa couplings except  $\la_L^2$ and $\la^2_R$ in our numerical evaluation. 
Note that sizable $\kappa$ and/or $\tkap$ will make the Landau pole scale lower.

\subsubsection{Real Scalar DM}
The interaction is 
\begin{align}
 -\Lcal  = Y_e^{ij} \ol{\ell}_L^i H e^j_R + \kap\ol{L}_L H E_R + \tkap\ol{E}_L \tH L_R +\la_L^i \ol{\ell}^i_L X L_R + 
                   \la_R^i \ol{E}_L X e_R^i + h.c. , 
\end{align}
where $X$ is now a real scalar field. 

Compared with the complex scalar DM case, 
there are Yukawa coupling renormalization fro $\la_L, \la_R$, 
\begin{align}
 16\pi^2 \beta_{\la^i_L} =&\ 2 \tkap Y^{ij}_e \la_R^{*j} +3  \la_L^i \la_L^{*j} \la_L^j 
                                 + \frac{1}{2} \left(Y_e Y_e^\dagger\right)^{ij} \la^j_L
                          + \la_L^i \left(\frac{1}{2} \tkap^* \tkap + Y_2(X) - \frac{9}{4}(g_1^2 + g_2^2)  \right), \\
16\pi^2 \beta_{\la^i_R} =&\ 4\tkap \la_L^{*j} Y_e^{ji} + \la_R^j \left(Y_e^\dagger Y_e \right)^{ji}  
                    + \la_R^i \left(3 \la_R^{*j} \la_R^j  + \tkap^*\tkap  
                                                 +  Y_2(X) - \frac{9}{4}(g_1^2 + g_2^2)  \right), 
\end{align}
and the other parts are same as the complex case.

\subsection{Fermion DM}
The Yukawa interactions are 
\begin{align}
 -\Lcal = 
              \la_L^i  \ol{\ell}^i_L \chi  \tilde{L}_R + 
             \la_R^{i} \tilde{E}^*_L  \ol{\chi} e_R^i  + h.c.,
\end{align}
where the SM Yukawa matrix $Y_e$ is neglected.

\subsubsection{Dirac Fermion}
The RGEs are given by 
\begin{align}
16\pi^2  \beta_{\la_L^i} =&\ 
                                         \la_L^i \left[ \frac{5}{2} \left(\la_L^\dagger \la_L \right) 
                                                 +   \frac{1}{2} \left(\la_R^\dagger \la_R \right)
                                          -\frac{9}{20} g_1^2 -\frac{9}{4} g_2^2\right], \\
16\pi^2  \beta_{\la_R^i} =&\  
                                         \la_R^i \left[ 2 \left(\la_R^\dagger \la_R \right) + 2 \left(\la_L^\dagger \la_L \right) 
                                            -\frac{9}{5} g_1^2 \right].    
\end{align}

\subsubsection{Majorana Fermion}
The RGEs are given by 
\begin{align}
16\pi^2  \beta_{\la_L^i} =&\  \frac{1}{2} \left(Y_e Y_e^\dagger \la_L \right)^i  + 
                                         \la_L^i \left[ \frac{9}{2} \left(\la_L^\dagger \la_L \right) 
                                                 +   \frac{1}{2} \left(\la_R^\dagger \la_R \right)
                                          -\frac{9}{20} g_1^2 -\frac{9}{4} g_2^2\right], \\
16\pi^2  \beta_{\la_R^i} =&\  \frac{1}{2} \left(\la_R  Y_e^\dagger Y_e \right)^i  + 
                                         \la_R^i \left[ 4 \left(\la_R^\dagger \la_R \right) + 2 \left(\la_L^\dagger \la_L \right) 
                                            -\frac{9}{5} g_1^2 \right].   
\end{align}

{\small
\bibliographystyle{JHEP}
\bibliography{ref_fermionportal}
}

\end{document}